\documentclass[a4paper,11pt]{article}
\DeclareUnicodeCharacter{2212}{\ensuremath{-}}
\pdfoutput=1 
\usepackage{jheppub} 
\usepackage[numbers]{natbib}
\usepackage[T1]{fontenc} 
\usepackage[compat=1.1.0]{tikz-feynman}
\usepackage{xcolor}
\usepackage{multirow}
\usepackage{multicol}
\usepackage{float}
\usepackage{verbatim,slashed}
\usepackage{amsmath,mathtools}
\usepackage{relsize,float}
\usepackage{bm, nicefrac}
\usepackage{hyperref}
\usepackage[capitalise]{cleveref}
\usepackage{comment}
\usepackage{booktabs}
\usepackage[normalem]{ulem}
\usepackage{overpic}

\renewcommand{\phi}{\ensuremath{\varphi}}
\newcommand{\sss}{\scriptscriptstyle}

\newcommand{\mt}{m_{t}}

\newcommand{\dimreg}{dimensional regularisation}

\def\beq{\begin{equation}}
\def\eeq{\end{equation}}
\def\beqn{\begin{eqnarray}}
\def\eeqn{\end{eqnarray}}

\newcommand{\bqa}{\begin{eqnarray}}
\newcommand{\eqa}{\end{eqnarray}}
\chardef\MyArticleWithColor=\pdfcolorstackinit page direct{0 g}

\def\cCode#1{\begin{lstlisting}[mathescape,basicstyle=\small
\ttfamily,frame=leftline,aboveskip=4mm,belowskip=4mm,xleftmargin=20pt,framexleftmargin=10pt,
numbers=none,framerule=2pt,abovecaptionskip=0.0mm,belowcaptionskip=3.5mm #1]}

\newcommand\as{\alpha_{\sss S}}

\newcommand\aNLO{{\sc\small MadGraph5\_aMC@NLO}}

\newcommand{\LO}{{\rm LO}}

\newcommand{\NLO}{{\rm NLO}}

\newcommand{\ord}{{\cal O}}

\def\EWSL{\rm EWSL}

\def\beq{\begin{equation}}
\def\eeq{\end{equation}}
\def\beqar{\begin{eqnarray}}
\def\eeqar{\end{eqnarray}}
\def\barr#1{\begin{array}{#1}}
\def\earr{\end{array}}
\def\bfi{\begin{figure}}
\def\efi{\end{figure}}
\def\btab{\begin{table}}
\def\etab{\end{table}}
\def\bce{\begin{center}}
\def\ece{\end{center}}

\def\de{\delta}

\newcommand{\GeV}{\unskip\,\mathrm{GeV}}

\newcommand{\M}{{\cal{M}}}

\def\mathswitchr#1{#1}
\newcommand{\PW}{\mathswitchr W}

\newcommand{\PZ}{\mathswitchr Z}

\newcommand{\PH}{\mathswitchr H}

\newcommand{\Pt}{\mathswitchr t}

\def\mathswitch#1{\relax\ifmmode#1\else$#1$\fi}
\newcommand{\MW}{\mathswitch {M_\PW}}
\newcommand{\MZ}{\mathswitch {M_\PZ}}
\newcommand{\MH}{\mathswitch {M_\PH}}

\newcommand{\Mt}{\mathswitch {m_\Pt}}

\newcommand{\GF}{\mathswitch {G_\mu}}

\newcommand{\SC}{{\mathrm{LSC}}}
\renewcommand{\SS}{{\mathrm{SSC}}}
\newcommand{\cc}{{\mathrm{C}}}

\newcommand{\pre}{{\mathrm{PR}}}

\newcommand{\lrMwithabs}{l(|r_{kl}|,M^2)}

\newcommand{\LrM}{L(|r_{kl}|,M^2)}

\newcommand{\TO}{\rightarrow}

\newcommand{\denpoz}{{\sc\small DP}}

\newcommand{\MSbar}{{\rm \overline{MS}}}

\newcommand{\MzSM}{\mathcal{M}_0^{\rm SM}}
\newcommand{\MzSMp}{\mathcal{M}_{0'}^{\rm SM}}
\newcommand{\MoSM}{\mathcal{M}_{1, {\rm EW}}^{\rm SM}}
\newcommand{\dMSM}{\delta \mathcal{M}^{\rm SM}_{{\rm EW}}}
\newcommand{\MzNP}{\mathcal{M}_0^{\rm NP}}
\newcommand{\MzNPp}{\mathcal{M}_{0'}^{\rm NP}}
\newcommand{\MoNP}{\mathcal{M}_{1, {\rm EW}}^{\rm NP}}
\newcommand{\dMNP}{\delta \mathcal{M}^{\rm NP}}
\newcommand{\SM}{\rm SM}
\newcommand{\NP}{\rm NP}

\newcommand{\MoSMQCD}{\mathcal{M}_{1, {\rm QCD}}^{\rm SM}}
\newcommand{\MoNPQCD}{\mathcal{M}_{1, {\rm QCD}}^{\rm NP}}

\newcommand{\SDK}{\rm SDK}
\newcommand{\SDKw}{{\SDK}_{\rm weak}}
\newcommand{\SDKz}{{\SDK}_{0}}

\newcommand{\sigmaSM}{\sigma^{\rm SM}}
\newcommand{\sigmaINT}{\sigma^{\rm INT}}
\newcommand{\sigmaSQ}{\sigma^{\rm SQ}}

\title{Electroweak corrections in the SMEFT: four-fermion operators at high energies}

\author[a]{Hesham El Faham,}
\author[b]{Ken Mimasu,}
\author[c]{Davide Pagani,}
\author[a]{Claudio Severi,}
\author[a]{Eleni Vryonidou,}
\author[d,e]{Marco Zaro}

\affiliation[a]{Department of Physics and Astronomy, University of Manchester, Oxford Road, Manchester M13~9PL, United Kingdom}
\affiliation[b]{School of Physics and Astronomy, University of Southampton,
Highfield, Southampton S017 1BJ, United Kingdom}
\affiliation[c]{INFN, Sezione di Bologna, Via Irnerio 46, 40126 Bologna, Italy}
\affiliation[d]{TIFLab, Universit\`a degli Studi di Milano, Via Celoria 16, 20133 Milano, Italy}
\affiliation[e]{INFN, Sezione di Milano, Via Celoria 16, 20133 Milano, Italy}

\emailAdd{hesham.elfaham@manchester.ac.uk}
\emailAdd{ken.mimasu@soton.ac.uk}
\emailAdd{davide.pagani@bo.infn.it}
\emailAdd{claudio.severi@manchester.ac.uk}
\emailAdd{eleni.vryonidou@manchester.ac.uk}
\emailAdd{marco.zaro@mi.infn.it}

\preprint{
\begin{flushright}
TIF-UNIMI-2024-19
\end{flushright}
}

\abstract{
In the Standard Model (SM), electroweak (EW) corrections become significant at high energies, particularly at the tera-electronvolt scale and beyond, due to the presence of Sudakov logarithms. At these energy scales, the Standard Model Effective Field Theory (SMEFT) framework provides an enhanced sensitivity to potential new physics effects. This motivates the inclusion of EW corrections not only for SM predictions but also for analyses within SMEFT.
In this work, we compute EW corrections in the high-energy limit for a selected set of dimension-six operators, specifically the class of four-fermion contact interactions, in key hard-scattering processes relevant to both current and future colliders: top-quark pair production at the Large Hadron Collider (LHC) and in a muon collider scenario, as well as the Drell-Yan process at the LHC. We first discuss the technical details and challenges associated with evaluating EW Sudakov logarithms in SMEFT, contrasting them with the SM case. We then present phenomenological results for the aforementioned processes, highlighting the non-trivial effects introduced by EW corrections arising from the insertion of dimension-six, four-fermion operators. Importantly, the resulting \( K \)-factors exhibit significant deviations from their SM counterparts, with dependencies not only on the process but also on the specific operators considered.
Finally, we explore the potential to lift flat directions in the SMEFT parameter space by incorporating higher-order corrections, using Fisher information techniques.
}


\begin{document}
\maketitle
\flushbottom

\section{Introduction}
\label{sec:intro}
The most realistic approach to deepening our understanding of physics at short distances, in the absence of clear indications of resonant production of new particles, is through indirect methods. Following the discovery of the Higgs boson~\cite{ATLAS:2012yve, CMS:2012qbp} and the remarkable precision with which the Standard Model (SM) has been confirmed as the most complete theory at our disposal, a natural extension is to treat the SM as an Effective Field Theory (EFT). In this framework, potential effects of unknown particles can be accounted for by incorporating  higher-dimensional operators constructed from SM fields, suppressed by a large ultraviolet (UV) scale, \(\Lambda\), where new physics (NP) is assumed to reside. The Standard Model Effective Field Theory (SMEFT) consists of the SM and all such effective operators that respect the gauge symmetries and the electroweak symmetry-breaking structure of the SM. The construction of the SMEFT is general, provided that any additional degrees of freedom are kinematically decoupled due to their heavy masses~\cite{Appelquist:1974tg}.

Searches based on the SMEFT framework have the substantial advantage of being able to collect and correlate deviations across several analyses, potentially spanning various final states, and even considering data from multiple colliders. This approach results in a greatly improved sensitivity to NP. It is imperative, in this context, that the SM and SMEFT theoretical predictions used in these analyses are achieved at the highest standard of precision available. Whilst SM predictions at next-to-leading order (NLO), both in quantum chromodynamics (QCD) and in electroweak (EW) couplings, are available for arbitrary processes in public automated codes (see {\it e.g.}~Refs.~\cite{Alwall:2014hca, Kallweit:2014xda, Frixione:2015zaa, Chiesa:2015mya, Biedermann:2017yoi, Chiesa:2017gqx, Frederix:2018nkq, Pagani:2021iwa}), with most analyses nowadays relying on accuracies at the level of next-to-NLO (NNLO) or even higher, the case of SMEFT is not nearly as developed. For example, whilst the structure of UV poles at \(\mathcal{O}(\alpha)\) and \(\mathcal{O}(\alpha_S)\) is known in the literature~\cite{Jenkins:2013wua,Jenkins:2013zja,Alonso:2013hga}, a general automated framework for the evaluation of NLO cross-sections is only available for corrections of purely QCD origin~\cite{Degrande:2020evl}. The significantly more complicated structures stemming from effective operators have so far prevented an equivalent general evaluation of NLO EW corrections, and only a few calculations for selected \(2 \to 1\), \(2 \to 2\), and \(1 \to 3\) processes have been performed at this level of accuracy in EW couplings~\cite{Hartmann:2015aia,Hartmann:2015oia,Gauld:2015lmb,Hartmann:2016pil,Trott:2017yhn,Dawson:2018dxp,Dawson:2018jlg,Dawson:2018liq,Dawson:2018pyl,Dedes:2018seb,Dedes:2019bew,Dawson:2019clf,Cullen:2019nnr,Boughezal:2019xpp,Cullen:2020zof,Corbett:2021cil,Dawson:2021ofa,Dawson:2022bxd,Bellafronte:2023amz,Biekotter:2023xle,Maura:2024zxz,Biekotter:2025nln,Asteriadis:2024xuk,Asteriadis:2024xts,Dawson:2024pft}.

A naive power counting, 
\begin{equation}
    \frac{\text{NLO}_\text{QCD}}{\text{NLO}_\text{EW}} \sim \frac{\alpha_S}{\alpha} \sim 10,
\end{equation}
would suggest that EW corrections are one order of magnitude smaller than those stemming from QCD and thus probably negligible. However, this is not necessarily the case. EW corrections feature terms proportional to the so-called Electroweak Sudakov Logarithms (EWSL), {\it i.e.}, large logarithms resulting from loop diagrams featuring a massive EW boson, whose mass acts as a regulator of the soft/collinear divergence. 
It is interesting to note that such terms, which grow logarithmically with energy, are not cancelled even if the real emission of the heavy boson, yielding a finite contribution to the cross-section, is included (see {\it e.g.}, Refs.~\cite{Ciafaloni:2001vt,Manohar:2014vxa,Frixione:2015zaa,Czakon:2017wor, Ma:2024ayr}), and therefore represent a genuine feature of EW corrections at high energy. Since EWSL grow with energy, they become dominant in the tails of differential distributions, which, incidentally, are also the kinematic regions most sensitive to SMEFT operators. 

The evaluation of complete NLO corrections with all possible one-loop contributions from QCD and EW origins in the SMEFT is highly desirable and will likely represent the next key milestone in the quest for precise theoretical predictions in EFT studies. Leaving this major undertaking aside for the moment, our focus in this work is the extraction of NLO EW corrections in the high-energy limit,
\begin{equation}
    \MW^2, \MZ^2, \mt^2, v^2 \ll s < \Lambda^2,
\end{equation}
including only terms that grow with energy, namely the EWSL. Indeed, the extraction of EWSL is, in principle, much simpler than an exact calculation of NLO EW corrections. Their evaluation is expected to be relatively straightforward, and they are numerically dominant when \( s \gg \MW^2 \), which is also the region where the sensitivity to SMEFT is maximised.

Electroweak corrections in the high-energy limit have been studied for a long time. The pioneering works of Denner and Pozzorini~\cite{Denner:2000jv,Denner:2001gw} resulted in a general algorithm, which we will refer to as the {\denpoz} algorithm, for the extraction of EWSL in the SM. This algorithm has also been recently automated within general-purpose event generators~\cite{Bothmann:2020sxm,Bothmann:2021led,Pagani:2021vyk,Pagani:2023wgc,Lindert:2023fcu}. In particular, Refs.~\cite{Pagani:2021vyk,Pagani:2023wgc} address the automation within {\aNLO}~\cite{Alwall:2014hca}. The starting point of our study has been to determine whether the {\denpoz} algorithm for the calculation of EWSL could be minimally modified for SMEFT calculations, as an approach based on EWSL appears to be highly appealing.

The {\denpoz} algorithm can EWSL for arbitrary processes, as long as the corresponding tree-level amplitudes are {\it not mass-suppressed}, that is, not proportional to some positive power of \(\MW/\sqrt{s}\).\footnote{~Such amplitudes exactly vanish in the high-energy limit.} In the SM, mass suppression is exceedingly rare. However, in the SMEFT, terms containing the Higgs doublet \(\Phi\) entering effective operators can result in Feynman rules proportional to the Higgs vacuum expectation value (vev) \(v\) (and thus to the \(W\) boson mass \(\MW\), since \(\MW = \tfrac{1}{2} g v\)) when \(\Phi\) is not dynamical.
It is important to note that the {\denpoz} algorithm is not expected to be applicable in these cases, and we will discuss an example where the algorithm fails to capture all of the EWSL. On the other hand, dimensional analysis dictates that non-mass-suppressed amplitudes at a given order in the SMEFT expansion must grow maximally with energy compared to the SM, implying that the algorithm is effective in scenarios where the best sensitivity to new interactions is expected.

In this paper, we consider selected processes and SMEFT operators for which the computation of EWSL using the {\denpoz} algorithm is not only possible but also structurally similar to the SM case. However, as will become evident, this does \emph{not} imply that the numerical impact of EWSL in the SMEFT and the SM is identical. We will compute the EWSL for the case of four-fermion operators at dimension-six, applied to the Drell-Yan process and top-quark pair production, the latter both at the LHC and a future high-energy muon collider. In recent years, there has been renewed interest in muon colliders, as accelerating elementary particles to several tera-electronvolts (TeV)~\cite{Delahaye:2019omf,Bartosik:2020xwr,Schulte:2021hgo,Long:2020wfp,MuonCollider:2022nsa,MuonCollider:2022ded,MuonCollider:2022glg} could enable the probing of fundamental interactions at unprecedented scales~\cite{MuonCollider:2022xlm,Aime:2022flm,Black:2022cth,Maltoni:2022bqs,Belloni:2022due,Accettura:2023ked}.

Once the numerical impact of the higher-order EW corrections on the differential distributions is established, it becomes particularly interesting to investigate how these corrections affect the linear combinations of Wilson coefficients probed by the observables of interest. The presence of weakly constrained or flat directions often hinders SMEFT interpretations of data and motivates the consideration of additional observables to constrain the parameter space. Higher-order corrections are expected to both rotate and lift flat directions, and our aim is to determine the extent to which this occurs through the computation of first QCD corrections and then EWSL for the processes under consideration. A Fisher information analysis is particularly well-suited to quantify these effects, and we will employ this approach in the present work.

This paper is organised as follows. In~\cref{sec:theo}, we study in detail the theoretical framework for the extraction of NLO EW corrections in the SMEFT, both in general and in the high-energy limit via EWSL, and discuss the domain of applicability of the {\denpoz} algorithm. In~\cref{sec:phenomenology}, we explain our setup and the Monte Carlo (MC) implementation. \Cref{sec:results} presents the phenomenological results for the aforementioned processes involving four-fermion operators, starting with the impact of the corrections on the differential distributions, followed by an exploration of their influence on the flat directions in parameter space through a Fisher information analysis. Finally, we conclude in~\cref{sec:conclusions}.

\section{Theoretical framework}
\label{sec:theo}
In this section, we discuss the theoretical framework of our computation. We provide a general overview of the calculation of NLO EW corrections at high energies in the presence of SMEFT operators. As mentioned in the introduction, we consider three processes and the corresponding sets of dimension-six operators, summarised in~\cref{tab:processes}. These processes are selected as representative examples. The choice of SMEFT operators included in this study will be justified in the following discussion, where we demonstrate that four-fermion operators present a particularly favourable case for computing approximate NLO EW corrections via EWSL.
\begin{table}[ht]
\centering
\begin{tabular}{lll}
    \hline
    \textbf{Process} & \textbf{Collider} & \textbf{Operator Class} \\ 
    \hline
    Top-quark pair production & LHC & Colour-octet four-quark \\ 
    The Drell-Yan process       & LHC & Two-quark–two-lepton \\ 
    Top-quark pair production  & Muon collider & Two-quark–two-lepton \\ 
    \hline
\end{tabular}
\caption{Processes and SMEFT operators considered in this work, taken as a representative sample for the LHC and future collider studies. The definitions of the operators listed in the table can be found in~\cref{sec:OPandNOT}.}
\label{tab:processes}
\end{table}

As noted in the introduction, the {\denpoz} algorithm cannot generally be applied to the SMEFT. It is only expected to work when the amplitude of interest is not mass-suppressed, which, as we will discuss, is equivalent to the amplitude exhibiting maximal energy growth. Four-fermion operators belong precisely to this class of operators that exhibit maximal growth, and in this work, we will demonstrate how the {\denpoz} algorithm can be applied in such cases.

This section is structured as follows: In~\cref{sec:theoamp}, we discuss the evaluation of \(\mathcal{O}(\alpha)\) corrections in the high-energy limit, specifically the Sudakov approximation for amplitudes involving dimension-six SMEFT operators at tree-level. In particular, we explain how the {\denpoz} algorithm can be applied to such SMEFT amplitudes. In~\cref{sec:theoXS}, we address the associated physical cross-sections and their approximate NLO EW corrections, with a focus on the additional complexities introduced by the triple expansion in \(\alpha_s\), \(\alpha\), and \(1/\Lambda\).

\subsection{EWSL at the amplitude-level}
\label{sec:theoamp}
At high energies, N$^n$LO EW corrections are dominated by the EWSL, which correspond to terms of order: 
\begin{equation}
    \alpha^n \log^m \frac{|r_{kl}|}{M^2}, \quad \text{with} \quad m \leq 2n,\label{eq:generalsudakov}
\end{equation}
w.r.t.~the LO amplitude. The denominators in the argument of the EWSL contain \(M\), which represents the mass of any of the heavy particles in the SM:
\begin{equation}
    M \in \big\lbrace \MW, \MZ, \MH, \Mt \big\rbrace\, ,
\end{equation}
or, in the case of pure QED contributions involving photons, the IR-regularisation scale \(Q\). Following standard notation, we denote by \(r_{kl}\) the kinematic invariants that can be constructed from the momenta of a pair of external particles (all momenta defined as incoming),
\begin{equation}
    r_{kl} \equiv (p_k + p_l)^2.\label{eq:rkl}
\end{equation}

In a generic scattering process, one of the requirements for the {\denpoz} algorithm to work is that all invariants \(r_{kl}\) in~\cref{eq:rkl} are of the same order as \(s\), the total energy available in the centre-of-mass frame:
\begin{equation}
    |(p_k + p_l)^2| = |r_{kl}| \simeq s = (p_1 + p_2)^2,
\end{equation}
where \(p_1\) and \(p_2\) are the incoming momenta.

Here, we focus on the high-energy approximation of NLO EW corrections, \emph{i.e.}, the \(\ord(\alpha)\) corrections to the LO prediction, corresponding to \(n=1\) in~\cref{eq:generalsudakov}. The associated EWSL are:
\begin{equation}
\LrM \equiv \frac{\alpha}{4\pi}\log^2{\frac{|r_{kl}|}{M^2}} \quad \text{and} \quad 
\lrMwithabs \equiv \frac{\alpha}{4\pi}\log{\frac{|r_{kl}|}{M^2}}.\label{eq:generallogs}
\end{equation}

Using the {\denpoz} algorithm and its refinement presented in Ref.~\cite{Pagani:2021vyk}, it is possible to compute approximate one-loop EW corrections to a generic tree-level SM scattering amplitude. In the SM, the {\denpoz} algorithm states that, starting from a tree-level amplitude with particles \(i_1, \ldots, i_n\) with momenta \(p_1, \ldots, p_n\),
\begin{equation}
    (\MzSM)^{i_1 \ldots i_n}(p_1,\ldots, p_n), \label{eq:M0}
\end{equation}
the \(\ord(\alpha)\) one-loop EW corrections, denoted as \(\MoSM\), can be expressed in the high-energy limit as:
\begin{equation}
    \lim_{\MW^2/s\to 0} (\MoSM)^{i_1 \ldots i_n}(p_1,\ldots, p_n)  =  \sum_{k,l} (\MzSM)^{\ldots i^\prime_k \ldots i^\prime_l \ldots}(p_1, \ldots, p_n)\delta^{\SM}_{{\rm EW,}\,i^\prime_k i^{\phantom{\prime}}_k, i^\prime_l i^{\phantom{\prime}}_l}. \label{LAfactorization}
\end{equation}

In the summand, \emph{up to two} external states, generically denoted by \(i_k\) and \(i_l\), may be replaced by \(i_k^\prime\) and \(i_l^\prime\), corresponding, {\it e.g.}, to the other component in the same SU(2) doublet. For instance, in the computation of the EWSL for the process \(\mu^+\mu^- \to t \, \bar t\), one of the examples considered in this work, the tree-level amplitude for \(\mu^+\mu^- \to b \, \bar b\) also contributes to the sum. 

\Cref{LAfactorization} states that \(\MoSM\) in the high-energy limit can be written in a factorised form, involving Born amplitudes (both for the original process and those with particle replacements) and \(\delta^{\SM}_{\rm EW}\), which is a product of logarithms from~\cref{eq:generallogs} and couplings of fields to EW gauge bosons, or related quantities such as EW Casimir operators.

The contributions to \(\delta^{\SM}_{\rm EW}\) can be split as:
\begin{equation}
    \delta^{\SM}_{\rm EW} = \de_{\SC}^{\SM} + \de_{\SS}^{\SM} + \de_{\cc}^{\SM} + \de_\pre^{\SM}, \label{eq:deltatodeltas}
\end{equation}
where \(\de_\SC\) and \(\de_\SS\) are the leading and sub-leading soft-collinear (LSC and SSC) logarithms, respectively, both emerging from the eikonal approximation of one-loop diagrams with a gauge boson exchanged between external legs. The term \(\de_\cc\) includes the collinear (C) logarithms, originating from virtual gauge bosons collinear to external lines, as well as from field renormalisation constants. All these contributions have an IR origin, whilst the last term, \(\de_\pre\), arises from the UV parameter renormalisation (PR) of \(\MzSM\), which is determined from the running of the physical input parameters.

Further details on the {\denpoz} algorithm, its refinement, and its automation in {\aNLO} can be found in Refs.~\cite{Pagani:2021vyk,Pagani:2023wgc}. In particular, Ref.~\cite{Pagani:2021vyk} introduces several novel aspects, including the regularisation of IR divergences using dimensional regularisation, an imaginary term omitted in earlier literature, and an efficient approximation of logarithms among invariants (denoted as \(\Delta^{s\TO r_{kl}}\)), which we account for in the predictions of this work. For additional technical details, we refer the reader to these publications, and in the following, we focus on aspects relevant to the application to SMEFT. Whenever possible, we use notation consistent with Refs.~\cite{Denner:2000jv,Denner:2001gw}.

\subsubsection{Mass suppression and the validity of the {\denpoz} algorithm}
\label{sec:MSDP}
The brief introduction to EWSL in the previous section highlights why these contributions can be computed in a much faster and more stable manner than the exact NLO EW corrections. This has spurred renewed interest in EWSL and the {\denpoz} algorithm in recent years. Unlike exact NLO EW corrections, the {\denpoz} algorithm framework does not require the computation of loops, thereby avoiding the associated technical and numerical complexities.

It is natural to expect that this approach could be extended to the Beyond Standard Model (BSM) case in a relatively straightforward manner, capturing the leading EW corrections at high energies. However, there are a few crucial assumptions underpinning the derivation of the {\denpoz} algorithm, and one of these is particularly critical for the SMEFT program:

{\it  ``For the helicity configuration considered, in the high-energy limit, the tree-level amplitude $\M_0$ must {\it not} be mass-suppressed by powers of the form $(M/\sqrt{s})^k$ with $k>0
$. In other words, by dimensional analysis, a $2\TO n$ process requires that $[\M]=E^{2-n}$, where $E$ has units of energy, and therefore 
\begin{equation}
\M\propto s^{(2-n)/2}\, ,  \label{eq:masssuppr}
\end{equation}
with no extra $(M/\sqrt{s})^k$ powers.''
}

In the SM, there are processes where all helicity configurations are mass-suppressed, such as Higgs Vector-Boson Fusion production, but these are exceptions: it is usually the case that at least one helicity configuration is not mass-suppressed. At most, what can occur is that, in a specific corner of the phase space, a mass-suppressed contribution becomes numerically dominant, spoiling the reliability of the approximation (see Section 4.1.3 of Ref.~\cite{Ma:2024ayr}). In the SMEFT, the situation is quite the opposite: mass-suppressed contributions are ubiquitous.

To avoid mass suppression in an amplitude featuring a single insertion of a dimension-six operator, the amplitude must scale with energy as: 
\begin{equation}
    \M \propto \frac{s^{(4-n)/2}}{\Lambda^2}\, , \label{eq:masssupprSMEFT}
\end{equation}
The energy dependence in~\cref{eq:masssupprSMEFT} corresponds to cases where SMEFT predictions at dimension-six exhibit the maximal possible growth with energy. This is the case, for instance, for vertices involving four fermions, directly related to the four-fermion operators considered in this work. However, it is quite common to observe dependencies of the form:
\begin{equation}
    \M \propto \frac{v \, s^{(3-n)/2}}{\Lambda^2} \propto \frac{M}{\sqrt{s}} \times \frac{s^{(4-n)/2}}{\Lambda^2}\, , \label{eq:masssupprSMEFTb}
\end{equation}
or even:
\begin{equation}
    \M \propto \frac{v^2 s^{(2-n)/2}}{\Lambda^2} \propto \frac{M^2}{s} \times \frac{s^{(4-n)/2}}{\Lambda^2}\, , \label{eq:masssupprSMEFTc}
\end{equation}
which are suppressed by one or two powers of \(M/\sqrt{s}\), respectively, compared to~\cref{eq:masssupprSMEFT}.
The Higgs vacuum expectation value (vev) dependence in~\cref{eq:masssupprSMEFTb,eq:masssupprSMEFTc} arises from effective operators containing the Higgs doublet \(\Phi\) when the field is not dynamical. 

A simple and general example related to~\cref{eq:masssupprSMEFTc} occurs when an operator induces a vertex that can be rewritten as one already present in the SM, multiplied by a coupling modifier, with all diagrams contributing to \(\M\) featuring the same number of such vertex insertions. 
An instance of this is the modification of the top-quark Yukawa coupling in \(t \bar t H\) production, which corresponds to the SMEFT equivalent of the `kappa' framework widely discussed in the literature. Indeed, an SMEFT-induced modification that does not introduce any new Lorentz structure compared to the SM must be proportional to \(v^2/\Lambda^2\) due to dimensional analysis. 

The case described in~\cref{eq:masssupprSMEFTb} is also common. One relevant example for some of the processes considered in this work is the \(t \bar t g\) vertex induced by the top-quark chromomagnetic operator \(\mathcal{O}_{tG}\), defined as:
\begin{equation}
    \mathcal{O}_{tG} \equiv \overline{Q}_L \widetilde{\Phi} \, \sigma^{\mu \nu} G_{\mu \nu} \, t_R\,, \label{eq:ctg_lagrangian}
\end{equation}
where \(\sigma^{\mu \nu} = \frac{i}{2} [\gamma^\mu, \gamma^\nu]\), \(\widetilde{\Phi} = i \sigma_2 \Phi^\dagger\) is the conjugated Higgs field, \(G_{\mu \nu} = G_{\mu \nu}^A T_A\) is the gluon field strength tensor, and \(Q_{L}\) and \(t_{R}\) represent the third-generation left-handed quark doublet and the right-handed top quark, respectively. 
After electroweak symmetry breaking, the new Lorentz structure induces modifications to the tree-level \(q \bar q \to t \bar t\) amplitude that scale as described in~\cref{eq:masssupprSMEFTb}.

For both energy scalings described by~\cref{eq:masssupprSMEFTb,eq:masssupprSMEFTc}, the {\denpoz} algorithm is not generally expected to work, as these cases lie outside its domain of applicability. We do not claim that it is never possible to use the algorithm in the presence of mass-suppressed tree-level amplitudes; rather, we emphasise that there {\it can} be cases where the {\denpoz} algorithm fails to capture all of the EWSL contributions.
We have explicitly verified that this is the case for NLO EW corrections to the dimension-six amplitude \(q \bar q \to t \bar t\) induced by \(\mathcal{O}_{tG}\). The details of this calculation are provided in~\cref{sec:ctg}, where we demonstrate how the explicit computation of virtual corrections in the high-energy limit leads to a contribution different from the naive application of the {\denpoz} algorithm to the SMEFT scenario.

It is straightforward to understand, simply by examining the Feynman diagrams in~\cref{fig:ctg_diagsinmain}, why mass-suppressed terms can cause problems when the {\denpoz} algorithm is applied:  
\begin{figure}[ht]
    \centering
    \includegraphics[width=0.5\textwidth]{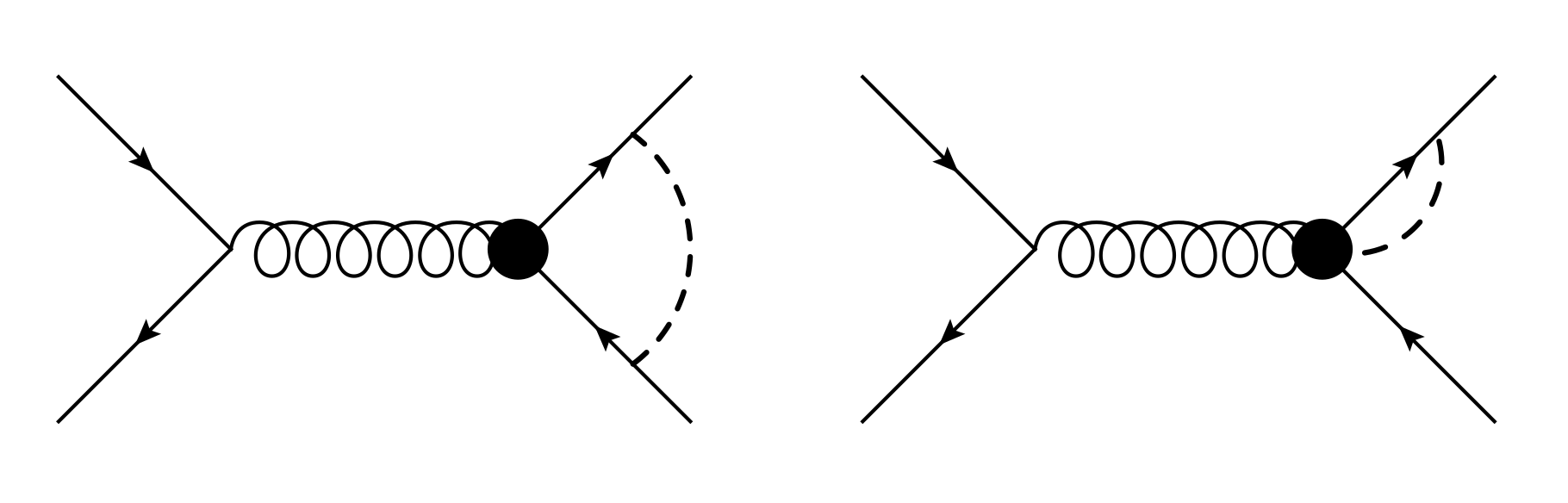}
    \caption{Representative diagrams for one-loop EW corrections to \(q \bar q \to t \bar t\) at \(\ord{(y_t^2)}\), with the insertion of \(\mathcal{O}_{tG}\) denoted by the black blob. In Feynman gauge, similar diagrams involving the Goldstone bosons, \(G^0\) and \(G^\pm\), are also present.}
    \label{fig:ctg_diagsinmain}
\end{figure}
Apart from corrections to the top-quark wave function\footnote{~See Eqs.~(B.6)-(B.8) of Ref.~\cite{Denner:1991kt}.}, the contributions of loops induced by Higgs boson exchange are always neglected in the {\denpoz} algorithm, as they result in mass-suppressed terms at high energies. For example, the diagram on the left in~\cref{fig:ctg_diagsinmain} contributes at \(\ord{(v^2/s)}\) relative to the tree-level amplitude involving \(\mathcal{O}_{tG}\). As in the SM case, this contribution can be safely neglected in the high-energy limit. In contrast, the diagram on the right does not have an equivalent in the SM and cannot be ignored. The vertex involving the \(t \bar t g H\) interaction from SMEFT, unlike the \(t \bar t g\) vertex stemming from the same operator, is {\it not} mass-suppressed because the Higgs field is dynamical and the Feynman rule substitutes \(v \to h\). Whilst this diagram is essential in the Sudakov approximation, it cannot be captured by a straightforward application of the {\denpoz} algorithm to the SMEFT case.

This result carries an important implication. Whilst SM amplitudes are rarely mass-suppressed, the opposite is true in SMEFT. However, several classes of operators, notably {\it all} four-fermion operators, never lead to mass-suppressed contributions, \emph{i.e.}, they always lead to amplitudes that grow maximally with energy, as shown in~\cref{eq:masssupprSMEFT}. For these operators, we can validate the applicability of the {\denpoz} algorithm and subsequently study the phenomenology of NLO EW corrections at high energies. The three cases listed at the beginning of Sec.~\ref{sec:theo} belong precisely to this category.

\subsubsection{Multi-coupling expansion in the {\denpoz} algorithm} \label{sec:multicoupling} \label{sec:DPsmeft}
We begin this discussion by examining the multi-coupling expansion within the {\denpoz} algorithm in the SM case, before extending the analysis to the SMEFT framework. 

A scattering amplitude at tree-level can involve multiple coupling combinations in QCD and EW. In the context of our discussion on EWSL, we identify \(\MzSM\) as the amplitude with the highest power of \(\as\). However, as discussed in Ref.~\cite{Frederix:2018nkq}, when multiple coupling combinations are present, \(\ord(\alpha)\) virtual corrections become more intricate. Specifically, the amplitude \(\MoSM\) receives contributions not only from `EW loops' on top of \(\MzSM\), but also from `QCD loops' on top of \(\MzSMp\), {\it i.e.}, the tree-level amplitude with one power more in \(\alpha\) and one power less in \(\as\).

A simple example is top-quark pair production from light quarks, where:
\begin{equation}
    \MzSM = \mathcal{M}(q \bar q \to g \to t \bar t), \label{eq:ttbarMzSM}
\end{equation}
is the amplitude with the highest power of \(\as\), and:
\begin{equation}
    \MzSMp = \mathcal{M}(q \bar q \to Z/\gamma \to t \bar t), \label{eq:ttbarMzSMp}
\end{equation}
features one power more in \(\alpha\) and one power less in \(\as\). 

Indeed, NLO corrections at \(\ord(\alpha)\) involve EW perturbations applied to \(\MzSM\), as well as QCD perturbations applied to \(\MzSMp\), both resulting in an \(\ord(\as \alpha)\) amplitude. It is important to note that the distinction between `EW' and `QCD' loops is a useful mnemonic but is not well-defined on a diagram-by-diagram basis. Certain Feynman diagrams cannot be unambiguously categorised into one group or the other (see, for example, the detailed discussion in Section 2 of Ref.~\cite{Frixione:2014qaa}). 

However, since this categorisation aligns with the structure of IR and UV limits, one can extend~\cref{LAfactorization} as:
\begin{align}
     \label{LAfactorizationQCD}
\lim_{\MW^2/s\to 0} \MoSM  &= \dMSM \nonumber \\
    &= \MzSM \delta^{\SM}_{\rm EW} + \MzSMp \delta^{\SM}_{\rm QCD}\,,
\end{align}
where, for brevity, we have omitted the explicit dependence on momenta and the summation over indices. In the case of \(\delta^{\SM}_{\rm QCD}\), the summation would refer to colour indices rather than SU(2) partners. The explicit expression for \(\delta^{\SM}_{\rm QCD}\) is available in the literature\footnote{~An expression for \(\delta^{\SM}_{\rm QCD}\) can be found in Ref.~\cite{Pagani:2021vyk} at the level of {\it squared} matrix elements and in the so-called \(\SDKw\) scheme introduced later in~\cref{sec:theoXS}. For a general case, this quantity can be derived from, {\it e.g.}, the high-energy limit of Eq.~(B.2) in Ref.~\cite{Frederix:2009yq}.}. We emphasise that whenever \(\MzSMp \neq 0\), these EW corrections of QCD origin arise; see also the discussion in Ref.~\cite{Frixione:2014qaa}.

Having discussed the SM case above, we use it as a guideline to examine how the {\denpoz} algorithm can be applied to SMEFT amplitudes, provided that the leading contribution at tree-level is not mass-suppressed. Our discussion here is general, and we take the case of four-fermion operators entering a \(2 \to 2\) scattering process as a relevant example.

Before starting our discussion, we clarify our conventions. We assign auxiliary powers of SM couplings to the SMEFT interactions such that their amplitudes carry the same powers, \((n, m)\), as the corresponding SM amplitudes. For the case of contact interactions, the associated vertices are assigned a power of \(g_s^{n-2}/\Lambda^2\) or \(g^{n-2}/\Lambda^2\), depending on whether the coupling order is attributed to QCD or EW interactions, respectively. Here, \(n\) denotes the number of particles entering the vertex.

It is important to emphasise that this is an arbitrary choice, as SMEFT operators do not inherently need to carry any SM couplings. The decision to assign coupling powers is a matter of convention. In our case, these assignments are auxiliary and are introduced purely for bookkeeping purposes to facilitate the mixed counting of loop and EFT contributions. We consider the QCD and EW corrections as if the SMEFT operators were assigned these underlying coupling powers. However, we do not actually rescale the numerical values of the Wilson coefficients by powers of \(g_s\) or \(g\) in our simulations.

Starting from a tree-level amplitude involving contributions from dimension-six operators, denoted as \(\MzNP\), the structure of \(\ord(\alpha)\) corrections, which we denote as \(\MoNP\), is rather intricate. To appreciate this complexity, we first briefly discuss the simpler case of QCD, or \(\ord(\as)\), corrections to \(\MzNP\), which we will refer to as \(\MoNPQCD\). At one loop, as illustrated in~\cref{fig:example_multicouplingSMEFT_QCD}, there are two distinct contributions to \(\MoNPQCD\).
\begin{figure}[ht]
    \centering
    \includegraphics[width=.4\textwidth]{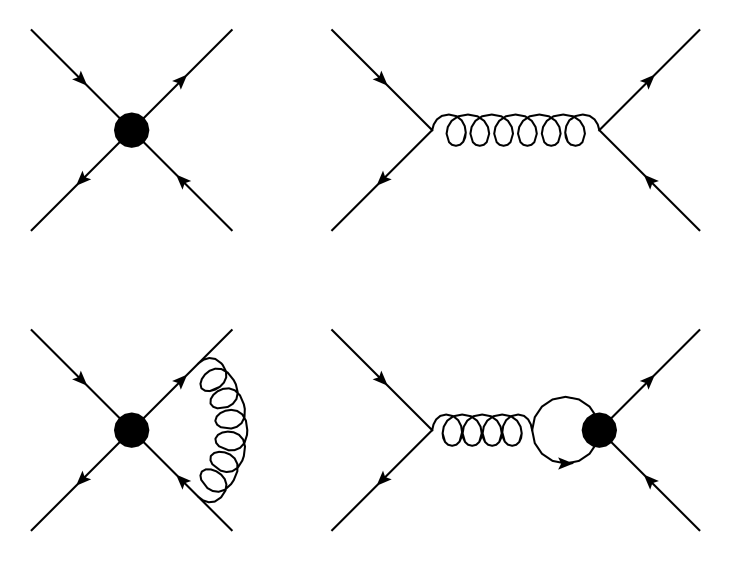}
    \caption{Illustration of the structure of \(\ord(\as)\) corrections in the SMEFT for an arbitrary \(2 \to 2\) process with the inclusion of colour-octet four-fermion operators (black blob). Top row, from left to right: \(\MzNP\) at \(\ord(\as / \Lambda^2)\) and \(\MzSM\) at \(\ord(\as)\). Bottom row, from left to right: one-loop perturbations to the corresponding amplitudes contributing to \(\MoNPQCD\).}
    \label{fig:example_multicouplingSMEFT_QCD}
\end{figure}

The first contribution to \(\MoNPQCD\) arises from adding a purely QCD perturbation to the SMEFT amplitude \(\MzNP\). The second contribution originates from starting with the SM amplitude \(\MzSM\) for the same process and adding a QCD correction involving a SMEFT insertion. Both procedures yield a contribution to \(\MoNPQCD\) that includes one additional power of \(\as\) relative to \(\MzNP\) and one additional factor of \(\as/\Lambda^2\) compared to \(\MzSM\).

Now moving on to EW corrections, the general concept is similar to the QCD case, but due to the multi-coupling expansion in three quantities—\(\as\), \(\alpha\), and \(1/\Lambda^2\)—rather than two (\(\as\), \(1/\Lambda^2\)), the number of terms proliferates compared to the purely QCD scenario. 

First, to categorise loop corrections, as in the SM case discussed in~\cref{sec:multicoupling}, \(\MzNP\) should factor a unique combination of powers of \(\as\) and \(\alpha\). However, at tree-level, several such combinations are often possible. We denote \(\MzNP\) as the combination with the highest power of \(\as\) (this is the only one needed for the computation of purely QCD corrections), whilst \(\MzNPp\) represents the tree-level amplitude with one power more in \(\alpha\) and one power less in \(\as\) compared to \(\MzNP\). Thus, \(\MoNP\) receives contributions from:
\begin{itemize}
    \item `SM EW loops' on top of \(\MzNP\),
    \item `SM QCD loops' on top of \(\MzNPp\),
    \item `NP loops' on top of \(\MzSM\) (loop corrections to SM amplitudes involving a SMEFT insertion).
\end{itemize}

The `NP loops' further subdivide into either `NP EW' or `NP QCD' contributions, depending on whether they involve additional EW or QCD interactions, acting on \(\MzSM\) and \(\MzSMp\), respectively. An illustrative example of this structure, for an arbitrary \(2 \to 2\) scattering process mediated by both gluons and EW bosons, is shown in~\cref{fig:example_multicouplingSMEFT}.
\begin{figure}[ht]
    \centering
    \includegraphics[width=.7\textwidth]{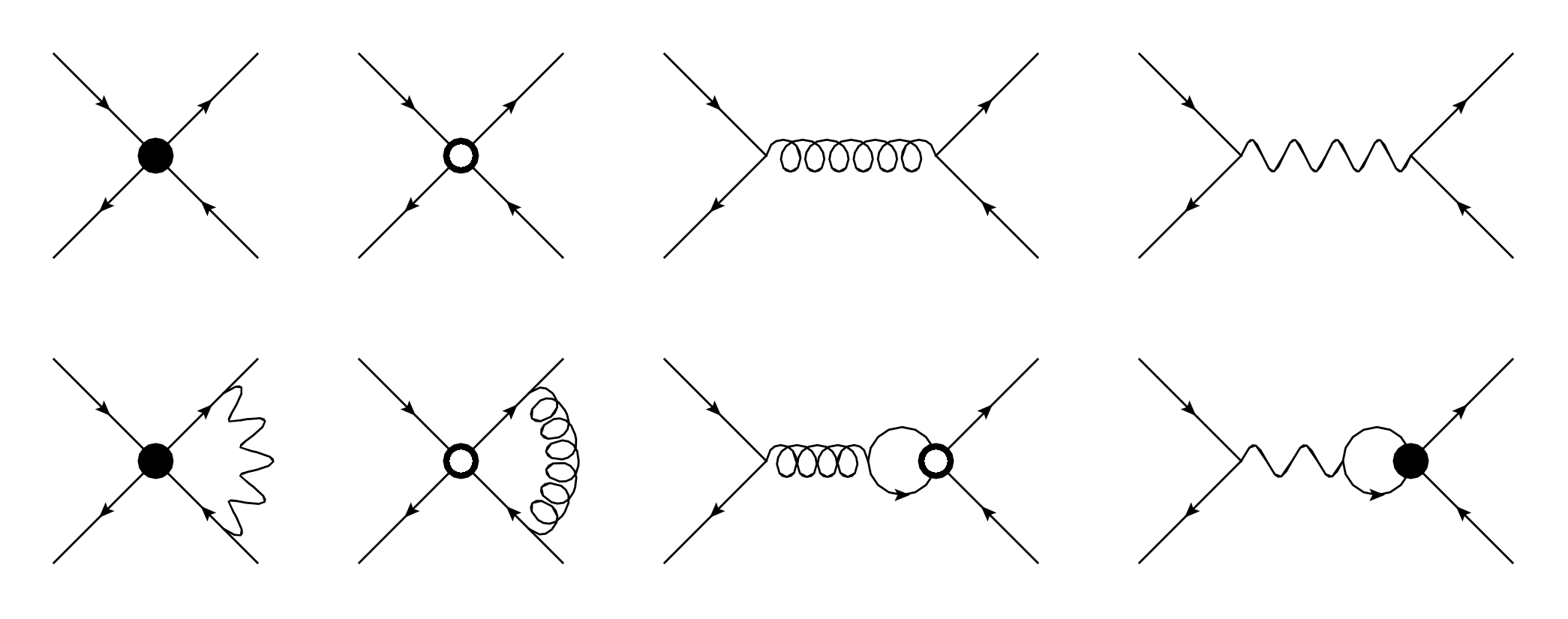}
    \caption{Illustration of the structure of \(\ord(\alpha)\) corrections in the SMEFT for an arbitrary \(2 \to 2\) process with the inclusion of colour-octet (black blob) and colour-singlet (white blob) four-fermion operators. Top row, from left to right: \(\MzNP\) at \(\ord(\as / \Lambda^2)\), \(\MzNPp\) at \(\ord(\alpha / \Lambda^2)\), \(\MzSM\) at \(\ord(\as)\), and \(\MzSMp\) at \(\ord(\alpha)\). Bottom row, from left to right: one-loop perturbations to the corresponding amplitudes contributing to \(\MoNP\).}
    \label{fig:example_multicouplingSMEFT}
\end{figure}
In the example shown in~\cref{fig:example_multicouplingSMEFT}, the third and fourth diagrams in the bottom row vanish due to colour configurations. However, processes such as four-top-quark production provide examples where equivalent contributions do not vanish.

As in the SM, diagrams in SMEFT cannot always be unambiguously associated with a single category. Nevertheless, this categorisation matches the structure of IR and UV limits. Thus, in the dimension-six case, the analogue of~\cref{LAfactorizationQCD} reads:
\begin{align}
    \label{LAfactorizationNP}
    \lim_{\MW^2/s \to 0} \MoNP  &= \dMNP \nonumber \\
    &= \MzNP \delta^{\SM}_{\rm EW} + \MzNPp \delta^{\SM}_{\rm QCD} + \MzSM \delta^{\NP}_{\rm EW} + \MzSMp \delta^{\NP}_{\rm QCD}\,,
\end{align}
where, for simplicity, indices and momenta are omitted.

We emphasise that the term \(\delta^{\SM}_{\rm EW}\) can always be computed using the {\denpoz} algorithm, which has been formally derived and proven valid for non-mass-suppressed amplitudes. The additional terms discussed here naturally arise in the generalisation of the multi-coupling expansion to include \(1/\Lambda^2\), in a manner analogous to the role of \(\as\) in \(\delta^{\SM}_{\rm QCD}\) in the SM.

To summarise, the underlying assumptions in the power expansion and factorisation are as follows: if
\begin{equation}
    \MzSM \propto \as^n \alpha^m \Lambda^0\,,
\end{equation}
with \(n\) and \(m\) being positive integers, then:
\begin{align}
    \MzSMp &\propto \as^{(n-1)} \alpha^{(m+1)} \Lambda^0\,, \\
    \MzNP &\propto \as^n \alpha^m / \Lambda^2\,, \\
    \MzNPp &\propto \as^{(n-1)} \alpha^{(m+1)} / \Lambda^2\,,
\end{align}
and:
\begin{equation}
    \delta^{\SM}_{\rm EW} \propto \alpha \Lambda^0\,, \quad 
    \delta^{\SM}_{\rm QCD} \propto \as \Lambda^0\,, \quad 
    \delta^{\NP}_{\rm EW} \propto \alpha / \Lambda^2\,, \quad 
    \delta^{\NP}_{\rm QCD} \propto \as / \Lambda^2\,,
\end{equation}
such that \(\MoSM\) factors one additional power of \(\alpha\) compared to \(\MzSM\), and the same holds true for \(\MoNP\) relative to \(\MzNP\).

\subsubsection{Processes considered in this work} \label{sec:ourprocesses}
So far, our argument has been general. In this section, we demonstrate that significant simplifications occur for the class of processes considered in this paper presented in ~\cref{tab:processes}. 

We associate an auxiliary power of \(\as / \Lambda^{2}\) (\(\alpha / \Lambda^{2}\)) to the four-fermion vector vertex arising from a colour-octet (colour-singlet) dimension-six four-fermion operator. This follows the convention discussed above, where octet operators are categorised as QCD order, whilst singlet operators are deemed to be of EW order. Under this convention, it becomes evident that, in the case of Drell-Yan and top-quark production in lepton collisions, only one combination of \(\as\) and \(\alpha\) is possible for tree-level amplitudes: \(\as^0 \alpha^1\). This leads to the conclusion that:
\begin{equation} \label{eq:zeroforDYandttlep} 
\text{for} \quad p \, p \to e^+ \, e^- \quad \text{and} \quad \mu^- \mu^+ \to t \, \bar t: \qquad  \MzNPp = \MzSMp = 0.
\end{equation}
Moreover, \(\delta^{\NP}_{\rm QCD}\) must also vanish, as no \(\ord(\as)\) corrections from NP are possible with the operators under consideration. The only tree-level diagrams contributing to~\cref{eq:zeroforDYandttlep}, both in the SM and the SMEFT, are shown in~\cref{fig:diags_DYandttlep}.
\begin{figure}[ht]
    \centering
    \begin{tikzpicture}
    \draw (0, 0) node[inner sep= 0] {\includegraphics[width=.4\textwidth]{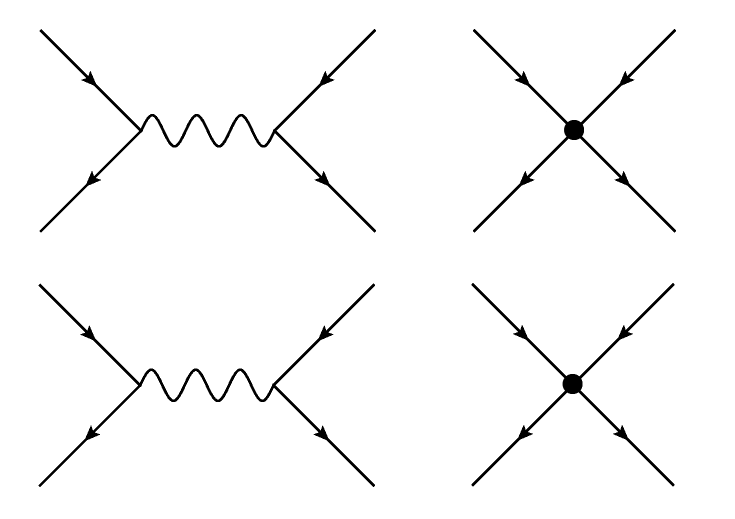}};
    \draw (-2.3, 1.8) node {$q$};
    \draw (-2.3, 0.3) node {$\bar q$}; 
    \draw (-0.35, 1.85) node {$e^-$};
    \draw (-0.35, 0.3) node {$e^+$};
    \draw (1.3, 1.8) node {$q$};
    \draw (1.3, 0.3) node {$\bar q$}; 
    \draw (2.1, 1.85) node {$e^-$};
    \draw (2.1, 0.3) node {$e^+$};
    \draw (-2.2, -0.2) node {$\mu^-$};
    \draw (-2.2, -1.7) node {$\mu^+$}; 
    \draw (-0.4, -0.3) node {$t$};
    \draw (-0.45, -1.7) node {$\bar t$};
    \draw (1.35, -0.2) node {$\mu^-$};
    \draw (1.35, -1.7) node {$\mu^+$}; 
    \draw (2.15, -0.2) node {$t$};
    \draw (2.1, -1.75) node {$\bar t$};
    \end{tikzpicture}
    \caption{Tree-level diagrams contributing to \(p \, p \to e^- \, e^+\) and \(\mu^- \mu^+ \to t \, \bar t\), both in the SM and the SMEFT. The black blob represents the insertion of a dimension-six four-fermion operator.}
\label{fig:diags_DYandttlep}
\end{figure}
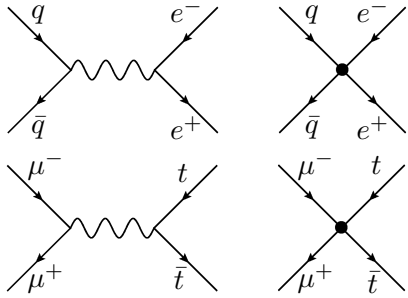

Furthermore, the four-fermion vertices do not induce any additional soft and/or collinear singularities. Therefore, in the high-energy limit, the IR logarithms in \(\delta^{\NP}_{\rm EW}\) must vanish. The UV high-energy logarithms within \(\delta^{\NP}_{\rm EW}\) can, in principle, be present, as \(\MzSM\) can depend on the SM parameter \(\mt\), which is the only relevant parameter renormalised at one loop by four-fermion operators. However, all helicity configurations of \(\MzSM\) that explicitly depend on \(\mt\) are mass-suppressed and are therefore negligible in the high-energy limit. The non-mass-suppressed helicity configurations are independent of \(\mt\), and consequently, \(\delta^{\NP}_{\rm EW} = 0\). Given these arguments, when colour-singlet four-fermion operators are considered,~\cref{LAfactorizationNP} simplifies to:
\begin{equation}
    \label{LAfactorizationNPDYandttlep}
    \text{for} \quad p \, p \to e^+ \, e^- \quad \text{and} \quad \mu^- \mu^+ \to t \, \bar t: \qquad \lim_{\MW^2/s \to 0} \MoNP =  \MzNP \delta^{\SM}_{\rm EW} \,.
\end{equation}

Moving on to the process \(q \bar q \to t \bar t\) with colour-octet four-fermion operators, the situation is slightly different. Since \(\MzSM\) is of \(\ord(\as)\), \(\MzSMp\) can, in principle, be non-vanishing, and corrections of QCD origin may be present. However, as we shall see in~\cref{sec:sdkweak}, these terms vanish due to colour configurations.

At this point, we note that with our choice of operators in~\cref{tab:processes}—driven by simplicity—we also have \(\MzNPp = 0\). Furthermore, since no NP effects of \(\ord(\alpha)\) are assumed to be present, \(\delta^{\NP}_{\rm EW}\) also vanishes. Similarly, due to colour considerations, \(\delta^{\NP}_{\rm QCD} = 0\) in all cases. In conclusion:
\begin{equation}
    \label{LAfactorizationNPqqttbar}
    \text{for} \quad p \, p \to t \bar t: \qquad \lim_{\MW^2/s \to 0} \MoNP = \MzNP \delta^{\SM}_{\rm EW} \,.
\end{equation}

From~\cref{LAfactorizationNPDYandttlep,LAfactorizationNPqqttbar}, it is evident that, in the cases studied here, the only inputs required for the computation of \(\ord(\alpha)\) corrections to the \(\MzNP\) amplitude are the tree-level amplitudes involving NP effects and the quantity \(\delta^{\SM}_{\rm EW}\). 

We remind the reader that in~\cref{LAfactorizationNPDYandttlep,LAfactorizationNPqqttbar}, a summation over SU(2) partners (\emph{e.g.}, \(b_L\) for the \(t_L\) in the final state) is implicitly understood. The computation of tree-level amplitudes is straightforward, and in the following, we will discuss how to extract \(\delta^{\SM}_{\rm EW}\).

\subsubsection{Evaluation of $\delta^{\SM}_{\rm EW} $ in the SMEFT}
\label{sec:dSMEWinSMEFT}
Since the quantity \(\delta^{\SM}_{\rm EW}\) appears in the main formula,~\cref{LAfactorization}, of the {\denpoz} algorithm, which was originally derived for the SM, a relevant question arises: ``Is \(\delta^{\SM}_{\rm EW}\) different in the SMEFT?'' 
In other words, moving from~\cref{LAfactorization} for a general SM process to the simplified cases in SMEFT described in~\cref{LAfactorizationNPDYandttlep,LAfactorizationNPqqttbar}, we ask: ``Can the {\denpoz} algorithm be used for calculating \(\delta^{\SM}_{\rm EW}\) in the SMEFT?''. To address this, we first separate the discussion into two components: the EWSL of IR origin and those of UV origin.

In the IR, the SMEFT shares the same particle and field content as the SM. It is invariant under the \({\rm SU(3)}\times{\rm SU(2)}\times{\rm U(1)}\) gauge symmetry and corresponds to the SM supplemented with, in our case, dimension-six operators. Importantly, the SMEFT does not modify the IR structure of the SM. Consequently, there is nothing in the derivation of the IR logarithms within the {\denpoz} algorithm that can be altered or invalidated by the inclusion of additional operators. Thus, the answer to our question regarding the IR contributions is affirmative.

The case of EWSL of UV origin is different. The SMEFT includes the same input parameters as the SM, along with the Wilson coefficients of the higher-dimension operators. Both the SM parameters and the Wilson coefficients, in general, receive \(\ord(\alpha)\) corrections and EWSL of UV origin. These logarithms account for the running of parameters from their renormalisation scale to the scale of the process under consideration.

The parameters already present in the SM (\(m_t\), \(\MW\), \emph{etc.}) are renormalised using the `on-shell' scheme, whilst \(\alpha\) is renormalised in the so-called `\(\alpha(\MZ)\)' or `\(G_\mu\)' scheme~\cite{Brivio:2017bnu, Brivio:2021yjb}. For all these SM parameters, the associated scales are of the order of \(\MW \ll \sqrt{s}\). Consequently, the corresponding EWSL take the same form as in the original formulation of the {\denpoz} algorithm\footnote{~We remind the reader that \(\ord(\alpha / \Lambda^{2})\) corrections are part of \(\delta_{\NP}^{\rm EW}\), so NP effects are not present in this case.}, and they induce non-negligible corrections.

The situation for the Wilson coefficients is, however, different. Typically, Wilson coefficients are renormalised in the \(\MSbar\) scheme, where a renormalisation scale, usually denoted as \(\mu_{\rm EFT}\)~\cite{Aoude:2022aro}, must be specified and is generally set equal to the typical scale of the process.

First, we note that as long as one is interested only in the high-energy tails of distributions, the EFT scale can be safely set to be of the order of \(\sqrt{s}\). In this case, the EWSL associated with the parameter renormalisation of Wilson coefficients takes the form:
\begin{equation}
    l(s, \mu^2_{\rm EFT}) = \frac{\alpha}{2 \pi} \log \frac{s}{\mu^2_{\rm EFT}} \sim \frac{\alpha}{2 \pi} \log \frac{s}{s} = 0,
\end{equation}
and is therefore subdominant, or strictly speaking, vanishing in our approximation. This is also the reason why, when NLO QCD corrections are computed in~\cref{sec:phenomenology}, the QCD renormalisation group (RG) running of the Wilson coefficients is neglected\footnote{~We emphasise that in our phenomenological analysis, the kinematic distributions start well above threshold. If one is instead interested in evaluating Wilson coefficients near the EW scale, \(\mu_{\rm EFT} \sim \MW\), RG effects must be considered~\cite{Aoude:2022aro, Maltoni:2024dpn}.}. Additionally, even when the full spectrum from the threshold up to the very boosted regime is considered, if one chooses a dynamical \(\mu_{\rm EFT}\), the \(\ord(\alpha)\) effects from RG running are exactly cancelled by the EWSL of UV origin in \(\delta^{\SM}_{\rm EW}\). In fact, at all orders, the dependence on \(\mu_{\rm EFT}\) must cancel, and in an NLO EW calculation, it can appear only at \(\ord(\alpha^2)\).

In conclusion, within our calculation set-up, we can neglect the EWSL of UV origin for the Wilson coefficients. As a result, \(\delta^{\SM}_{\rm EW}\) corresponds to its SM counterpart for both IR and UV terms.

\subsection{EWSL at the level of cross-sections}
\label{sec:theoXS} \label{sec:fullNLOsmeft}
So far, we have kept the discussion at the level of amplitudes. However, since our primary interest lies in phenomenological predictions, in this section, we extend the discussion to the level of squared matrix elements and cross-sections.

In fixed-order predictions, several contributions associated with different perturbative orders can arise. Expanding in powers of \(\as\) and \(\alpha\), and adopting the notation from Refs.~\cite{Frixione:2014qaa, Frixione:2015zaa, Pagani:2016caq, Frederix:2016ost, Czakon:2017wor, Frederix:2017wme, Frederix:2018nkq, Broggio:2019ewu, Frederix:2019ubd, Pagani:2020rsg, Pagani:2020mov, Pagani:2021iwa}, the contributions to the differential or inclusive cross-section, up to NLO, can be expressed as:
\begin{align}
    \LO(\as,\alpha) &= \LO_1 + \cdots + \LO_k\, , \label{eq:blobs_LO_general} \\
    \NLO(\as,\alpha) &= \NLO_1 + \cdots + \NLO_{k+1}\, , \label{eq:blobs_NLO_general}
\end{align}
where \(k \geq 1\) and is process-dependent. Here, \(\LO\) refers to contributions from squared amplitudes or interferences of tree-level diagrams only, whilst \(\NLO\) refers to contributions from interferences of tree-level and one-loop amplitudes or squared tree-level amplitudes with an additional massless parton. Using the notation established above:
\begin{align}
    \LO_1 &\supset |\MzSM|^2 \, , \\
    \NLO_1 &\supset 2\Re\left(\MzSM\left(\MoSMQCD\right)^*\right) \, , \\
    \NLO_2 &\supset 2\Re\left(\MzSM\left(\MoSM\right)^* + \MzSMp\left(\MoSMQCD\right)^*\right), \label{eq:NLO2supsets}
\end{align}
where the symbol `\(\supset\)' denotes ``receives contributions from''. Taking \(\LO_1\) as a reference, \(\NLO_1\) factors in one additional power of \(\as\), corresponding to the QCD corrections, whilst \(\NLO_2\) factors in one additional power of \(\alpha\), corresponding to the EW corrections.

Extending the discussion to the SMEFT framework, we restrict our analysis to amplitudes containing at most \emph{one} insertion of a dimension-six operator\footnote{~In all processes discussed in~\cref{sec:ourprocesses}, double insertions are not possible.}, and neglect operators of dimension eight or higher. The notation in~\cref{eq:blobs_LO_general,eq:blobs_NLO_general} can be generalised as:
\begin{align}
    \LO^{(j)}(\as,\alpha) &= \LO^{(j)}_1 + \cdots + \LO^{(j)}_k\, , \label{eq:blobs_LO_generalSMEFT} \\
    \NLO^{(j)}(\as,\alpha) &= \NLO^{(j)}_1 + \cdots + \NLO^{(j)}_{k+1}\, , \label{eq:blobs_NLO_generalSMEFT}
\end{align}
where the case with apex \((j) = (4)\), or simply no apex, corresponds to the SM as in~\cref{eq:blobs_LO_general,eq:blobs_NLO_general}. The case \((j) = (6)\) corresponds to one insertion of dimension-six EFT vertices and arises from the interferences of SM and NP amplitudes, proportional to \(\mathcal{O}(\Lambda^{-2})\). 
The case \((j) = (8)\) is proportional to \(\mathcal{O}(\Lambda^{-4})\) and, for our purposes, originates from the square of NP amplitudes with one insertion of dimension-six vertices. Since we restrict our study to four-fermion operators, double insertions are not allowed.

In the following, we base our discussion on the sample processes and operators outlined in~\cref{sec:ourprocesses}, which allow for significant simplifications in the structure of NLO corrections. Specifically, we restrict our analysis to the first LO contribution, \(\LO_1\), and the first two NLO contributions\footnote{~All the processes considered fall into this category: the Drell-Yan process and \(\mu^- \mu^+ \to t \bar t\) have only one LO and two NLO contributions, whilst top-quark hadroproduction features two LO and three NLO contributions. However, \(\LO_2\) and \(\NLO_3\) are largely subdominant.}. Under these assumptions, the corrections are expressed as follows for \((j) = (6)\):
\begin{align}
    \LO^{(6)}_1 &\supset 2\Re\left(\MzSM\left(\MzNP\right)^*\right) \, , \label{eq:LO1-6supsets} \\ 
    \NLO^{(6)}_1 &\supset 2\Re\left(\MzNP\left(\MoSMQCD\right)^* + \MzSM\left(\MoNPQCD\right)^*\right) \, , \label{eq:NLO1-6supsets} \\
    \NLO^{(6)}_2 &\supset 2\Re\left(\MzNP\left(\MoSM\right)^* + \MzSM\left(\MoNP\right)^* + \MzSMp\left(\MoNPQCD\right)^*\right) \, , \label{eq:NLO2-6supsets}
\end{align}
whilst for \((j) = (8)\), they read:
\begin{align}
    \LO^{(8)}_1 &\supset |\MzNP|^2 \, , \label{eq:LO1-8supsets} \\ 
    \NLO^{(8)}_1 &\supset 2\Re\left(\MzNP\left(\MoNPQCD\right)^*\right) \, , \label{eq:NLO1-8supsets} \\
    \NLO^{(8)}_2 &\supset 2\Re\left(\MzNP\left(\MoNP\right)^*\right) \, .  \label{eq:NLO2-8supsets}
\end{align}
The relative simplicity of~\cref{eq:LO1-8supsets,eq:NLO1-8supsets,eq:NLO2-8supsets} compared to~\cref{eq:LO1-6supsets,eq:NLO1-6supsets,eq:NLO2-6supsets} arises from our restriction to single insertions of dimension-six vertices and the omission of dimension-eight operators.

Our discussion so far assumes \(\MzNPp = 0\), as no colour-singlet operators are considered in this work. For completeness, if this assumption were relaxed—for example, by including additional SMEFT operators—the following terms would need to be added:
\begin{equation}
    \NLO^{(6)}_2 \supset 2\Re\left(\MzNPp\left(\MoSMQCD\right)^* \right), \qquad 
    \NLO^{(8)}_2 \supset 2\Re\left(\MzNPp\left(\MoNPQCD\right)^* \right). \label{eq:addtermsNLO6NLO8}
\end{equation}

The terms \(\LO^{(6)}_1\), \(\LO^{(8)}_1\), \(\NLO^{(6)}_1\), and \(\NLO^{(8)}_1\)—corresponding to LO and NLO corrections of QCD origin—can be computed in an automated way using {\aNLO}, as outlined in Ref.~\cite{Degrande:2020evl}. The new results presented in this paper are \(\NLO^{(6)}_2\) and \(\NLO^{(8)}_2\), which are calculated in the high-energy limit \(\MW^2/s \to 0\). Applying~\cref{LAfactorizationNPDYandttlep,LAfactorizationNPqqttbar}, the virtual contributions at the fully differential level are:
\begin{align}
    \lim_{\MW^2/s \to 0} \NLO^{(6)}_2\Big|_{\rm virt.} &\propto 2\Re\Big[\MzNP\left(\MzSM \delta^{\SM}_{\rm EW} \right)^* 
    + \MzSM\left(\MzNP \delta^{\SM}_{\rm EW} \right)^* \nonumber \\
    &\quad + \MzSMp\left(\MzSM  \delta^{\NP}_{\rm QCD} + \MzNP  \delta^{\SM}_{\rm QCD}\right)^*\Big], \label{eq:NLO2-6limit} \\
    \lim_{\MW^2/s \to 0} \NLO^{(8)}_2\Big|_{\rm virt.} &\propto 2\Re\left[\MzNP\left(\MzNP \delta^{\SM}_{\rm EW} \right)^*\right] \, . \label{eq:NLO2-8limit}
\end{align}
Integrating~\cref{eq:NLO2-6limit,eq:NLO2-8limit} over the relevant phase-space regions and dividing by the initial-state flux and symmetry factors yields the high-energy limit of the NLO EW virtual corrections at \(\mathcal{O}(\Lambda^{-2})\) and \(\mathcal{O}(\Lambda^{-4})\), respectively.

\subsubsection{Treatment of photons/gluons and the $\SDKw$ scheme} \label{sec:sdkweak}
In our previous discussion, we have interchangeably referred to cross-sections and interferences or squares of amplitudes. In doing so, we implicitly assumed that both tree-level and virtual contributions share the same kinematic and phase-space integration, whilst postponing the discussion of real-emission contributions, which also form part of the NLO corrections. Having introduced the relevant notation and theoretical background, we are now ready to address the inclusion of real-emission contributions.

The NLO EW contributions derived in the previous section stem solely from virtual corrections and cannot directly be used for phenomenological predictions. This is because they are IR divergent, as is typical for virtual corrections involving massless particles (in this case, photons). Refs.~\cite{Pagani:2021vyk, Pagani:2023wgc} discuss in detail how the {\denpoz} algorithm, originally derived for the high-energy approximation of \emph{virtual amplitudes}, can be adapted to obtain IR-safe predictions for physical observables through the proper inclusion of real-emission contributions. These modifications are referred to as the `\(\SDKw\)' scheme\footnote{~The \(\SDKw\) scheme is distinct from the `\(\SDK\)' scheme, which corresponds to the {\denpoz} algorithm applied solely at the amplitude level, and the `\(\SDKz\)' scheme, a widely used alternative method. The \(\SDKz\) scheme, unless supplemented by real-emission contributions, can yield less precise predictions and fails to capture part of the EWSL.}.

In Ref.~\cite{Pagani:2021vyk}, it was demonstrated that the \(\SDKw\) scheme, which retains only the purely weak component of the EWSL of IR origin, is superior to other prescriptions for obtaining physical predictions. This superiority is particularly evident in analyses where final-state charged particles, both massless and massive, are clustered together with photons and gluons. The \(\SDKw\) scheme ensures IR safety whilst maintaining the precision of the high-energy approximation.

For the remainder of this paper, we adopt the \(\SDKw\) scheme, which facilitates the simplification and practical use of~\cref{eq:NLO2-6limit,eq:NLO2-8limit} for phenomenological predictions. Under this scheme, the terms in~\cref{eq:NLO2-6limit,eq:NLO2-8limit} can be rewritten as:
\begin{align}
    \lim_{\MW^2/s \to 0} \NLO^{(6)}_2 &\propto 2\Re\Big[\MzNP\left(\MzSM \delta^{\SM}_{\rm EW}\Big|_{\SDKw} \right)^* + \MzSM\left(\MzNP \delta^{\SM}_{\rm EW}\Big|_{\SDKw} \right)^*\Big] \label{eq:NLO2-6limitSDKw}\, ,\\
    \lim_{\MW^2/s \to 0} \NLO^{(8)}_2 &\propto 2\Re\left[\MzNP\left(\MzNP \delta^{\SM}_{\rm EW}\Big|_{\SDKw} \right)^*\right] \, . \label{eq:NLO2-8limitSDKw}
\end{align}
We denote by \(\EWSL^{(6)}\) and \(\EWSL^{(8)}\) the contributions arising from~\cref{eq:NLO2-6limitSDKw,eq:NLO2-8limitSDKw}, as these form the logarithmically-growing components of \(\NLO^{(6)}_2\) and \(\NLO^{(8)}_2\), \emph{i.e.}, taking the \(\lim_{\MW^2/s \to 0}\) of \(\NLO^{(6)}_2\) and \(\NLO^{(8)}_2\), respectively.

In the transition from~\cref{eq:NLO2-6limit} to~\cref{eq:NLO2-6limitSDKw}, one might observe that the term in the second line of~\cref{eq:NLO2-6limit} appears to have been omitted. In a general case, an additional contribution:
\begin{align}
    2\Re\Big[
    &\MzSMp\left(\MzSM \delta^{\NP}_{\rm QCD}\Big|_{\SDKw} + \MzNP \delta^{\SM}_{\rm QCD}\Big|_{\SDKw}\right)^*
    \Big] \label{eq:NLO2-6limitSDKwmore}
\end{align}
can indeed appear in~\cref{eq:NLO2-6limitSDKw}. However, this term vanishes under the approximations applied for the processes and operators considered in this study. We elaborate on the reason for this omission below, as it is not trivial for the \(q \bar q \to t \bar t\) process. For the other two processes analysed in this work, \(\MzSMp\) is simply absent, making this term irrelevant.

First, we make a technical remark that has been somewhat obscured by our simplified notation. When writing terms such as \(\MzSM \delta^{\SM}_{\rm EW}\) or \(\MzNP \delta^{\SM}_{\rm EW}\), a summation over amplitudes with SU(2) partners as final states is implied. Similarly, terms like \(\MzSM \delta^{\NP}_{\rm QCD}\) and \(\MzNP \delta^{\SM}_{\rm QCD}\) involve a summation over SU(3)-rotated amplitudes, i.e., amplitudes with different external colour configurations. This is one reason why the terms within the square brackets in~\cref{eq:NLO2-6limitSDKw} cannot simply be factorised or simplified, as implicit summations over different \(\mathcal{M}_0\)'s are involved.

Next, we note that Ref.~\cite{Pagani:2021vyk} constructed the \(\SDKw\) scheme to handle contributions arising from QCD corrections applied to interferences of the form \(\MzSMp (\MzSM)^*\), such as the first term in~\cref{eq:NLO2-6limitSDKwmore}. It was shown that this scheme enables these contributions to be rewritten in the high-energy limit directly in terms of the original interference itself, without introducing colour-rotated terms, up to subleading logarithms such as \(\log(r_{kl}/r_{pq})\). The same reasoning applies to the second term, involving SMEFT amplitudes, in~\cref{eq:NLO2-6limitSDKwmore}. Consequently, these terms can be fully expressed as functions of \(\MzSMp (\MzSM)^*\) and \(\MzSMp (\MzNP)^*\), without implicit colour rotations or summations.
However, with the choice of operators considered in this work (colour-octet ones, as listed in~\cref{tab:processes}), these two terms vanish. This is due to the differing external \(t \bar t\) colour states: in both cases, a colour-singlet amplitude (\(\MzSMp\)) interferes with a colour-octet amplitude (\(\MzSM\) or \(\MzNP\)), leading to zero interference.

\subsubsection{Summary of the different terms in the SMEFT}
\label{sec:DP_summary_energy_behaviour}
In summary, taking the SM LO (\(\LO_1\), to be precise) as a reference, the LO quantities are:
\begin{equation}
\begin{aligned}
    \LO^{(6)}_1 / \LO &\propto s / \Lambda^2 \quad \text{and} \quad
    \LO^{(8)}_1 / \LO &\propto s^2 / \Lambda^4,
\end{aligned}
\end{equation}
and the NLO ones:
\begin{equation}
\begin{aligned}
    \NLO_1 / \LO &\propto \as \quad \text{and} \quad
    \NLO_2 / \LO &\propto \alpha \,, 
\end{aligned}
\end{equation}
\begin{equation}
\begin{aligned}
    \NLO_1^{(6)} / \LO &\propto \as \, s / \Lambda^2 \quad \text{and} \quad
    \NLO_1^{(8)} / \LO &\propto \as \, s^2 / \Lambda^4\,. \label{eq:NLOold}
\end{aligned}
\end{equation}
All these quantities have been evaluated and are already available in automated codes.

The novel results of this work are the extraction and automated evaluation of the following quantities:
\begin{align}
    \EWSL^{(6)} / \LO &\propto \alpha \, s / \Lambda^2 \log^m(s / \MW^2), \quad \label{eq:NLOnew1} \\
    \EWSL^{(8)} / \LO &\propto \alpha \, s^2 / \Lambda^4 \log^m(s / \MW^2), \quad \label{eq:NLOnew2}
\end{align}
where \(m = 1, 2\).

Having summarised the different SMEFT contributions examined in this section, we note that, when presenting our phenomenological predictions, we will refer to \(\LO_1\) simply as \(\LO\). For clarity, the following conventions will be used:
\begin{align}
    \begin{split}\label{eqn:QCDEWdef}
        \text{QCD} &\equiv \LO + \NLO_1\,, \\
        \text{EW}_{\SDK} &\equiv \LO + \EWSL\,, \\
        \text{QCD} + \text{EW}_{\SDK} &\equiv \LO + \NLO_1 + \EWSL\,.
    \end{split}
\end{align}
These definitions apply both to the SM case and to the contributions of order \(\mathcal{O}(\Lambda^{-2})\) (\(\LO^{(6)}\), \(\NLO_1^{(6)}\), and \(\EWSL^{(6)}\)) as well as \(\mathcal{O}(\Lambda^{-4})\) (\(\LO^{(8)}\), \(\NLO_1^{(8)}\), and \(\EWSL^{(8)}\)).

For simplicity, when referring to cross-sections associated with these three classes, we will adopt the following notations:
\begin{equation}
\begin{aligned}
    \label{eqn:lambdaINTSQ}
    \sigmaSM &\propto \Lambda^0 \,, \quad
    \sigmaINT &\propto 1 / \Lambda^2 \,, \quad
    \sigmaSQ &\propto 1 / \Lambda^4 \,,
\end{aligned}
\end{equation}
where `INT' and `SQ' indicate contributions arising from the interference between dimension-six EFT amplitudes and the SM amplitudes, and from the square of the EFT amplitudes, respectively.

\section{Implementation of SMEFT four-fermion operators} 
\label{sec:phenomenology}
As previously discussed, the {\denpoz} algorithm can consistently be used to compute the high-energy approximation of NLO EW corrections with the insertion of four-fermion operators, as those operators never lead to mass-suppressed $2\to2$ scattering amplitudes. Exploiting this capability, in this work we present the first phenomenological SMEFT study at $\text{QCD} + \text{EW}_{\text{SDK}}$ accuracy (exact NLO QCD corrections plus NLO EW corrections in the high-energy limit) including dimension-six four-fermion SMEFT operators. 

This section outlines our computational setup for the sample processes introduced in~\cref{tab:processes}: top-quark pair production, both at the LHC and at a 10~TeV muon collider, and electron-positron pair production at the LHC. After introducing the relevant SMEFT operators in~\cref{sec:OPandNOT},  in~\cref{sec:MC}, we discuss our MC implementation, based on {\aNLO}, which will be made public following this work. A validation of our implementation of the {\denpoz} algorithm is documented in~\cref{sec:validations}, showing through an analytical calculation that it reproduces the high-energy behaviour of exact NLO EW corrections and that it can be utilised in SMEFT computations when the amplitude is not mass-suppressed, as discussed at length in~\cref{sec:theo}. 

\subsection{Operators and notation}
\label{sec:OPandNOT}
In this section, we introduce the dimension-six four-fermion SMEFT operators that are the focus of this study, along with the corresponding notation. These four-fermion operators describe contact interactions that mediate $2 \to 2$ scattering processes of constituent fermion fields. We calculate the SMEFT contributions involving four-fermion operators under a specific flavour symmetry assumption that singles out the top-quark interactions:
\begin{equation}
    U(3)_l \times U(3)_e \times U(2)_q \times U(2)_u \times U(3)_d \equiv U(2)^2 \times U(3)^3,
    \label{eq:SMEFTatNLOflavor}
\end{equation}
where the subscripts correspond to the five fermionic representations in the SM. 
This minimal breaking of the $U(3)^5$ symmetry allows for chirality-flipping top-quark interactions, such as dipole operators and modifications to the top-Yukawa coupling. 
Throughout, we adopt the notation and operator conventions from Refs.~\cite{Aguilar-Saavedra:2018ksv,Brivio:2019ius,Degrande:2020evl}.
Unless otherwise mentioned, we use $q$, $u$, and $d$ to denote the left-handed quark doublet ($q$) and right-handed quarks ($u,d$) of the first two generations, and similarly, $Q$, $t$, and $b$ for the third generation. For the lepton fields, $e$ and $l$ represent right-handed singlets and left-handed doublets, respectively.

\paragraph{Top-quark pair production at the LHC} 
For this process, we focus on the four-fermion colour-octet operators defined as follows: 
\begin{align}
    \mathcal O_{tu}^8 &= \sum_{i = 1}^2 ~ (\overline t \gamma_\mu T^A t)(\overline u_{i} \gamma^\mu T_A u_{i}), & \mathcal O_{td}^8 &= \sum_{i = 1}^3 (\overline t \gamma_\mu T_A t)(\overline d_{i} \gamma^\mu T^A d_{i}), \nonumber \\ 
    \mathcal O_{tq}^8 &= \sum_{i = 1}^2 (\overline t \gamma^\mu T^A t)(\overline q_{i} \gamma_\mu T_A q_{i}), & \mathcal O_{Qu}^8 &= \sum_{i = 1}^2 (\overline Q \gamma_\mu T_A Q)(\overline u_{i} \gamma^\mu T^A u_{i}), \nonumber \\ 
    \mathcal O_{Qd}^8 &= \sum_{i = 1}^3 (\overline Q \gamma_\mu T_A Q)(\overline d_{i} \gamma^\mu T^A d_{i}), & \mathcal O_{Qq}^{1,8} &= \sum_{i = 1}^2 (\overline Q \gamma_\mu T^A Q)(\overline q_{i} \gamma^\mu T_A q_{i}), \nonumber \\
    \mathcal O_{Qq}^{3,8} &= \sum_{i = 1}^2 (\overline Q \gamma_\mu T^A \sigma_I Q)(\overline q_{i} \gamma^\mu T_A \sigma^I q_{i}), \label{eq:Qq38}
\end{align}
where $i$ represents the generation of the light fermion fields. Due to their colour structure, these operators interfere with the dominant contribution to this process, namely gluon-mediated $q\bar{q} \to t\bar{t}$ scattering. This is in contrast to colour-singlet operators, which only interfere with the subleading SM amplitude, mediated by EW interactions. 
It is worth noting that operators composed of four top-quark fields, either $Q$ or $t_R$, would also be compatible with our flavour assumption in~\cref{eq:SMEFTatNLOflavor} and, in principle, enter top-quark pair production. These operators are not considered in our study as their contribution to $t\bar{t}$ production is either suppressed by the $b$ quark parton distribution function (PDF) or entirely loop-induced.

\paragraph{Top-quark pair production at a lepton collider} 
For this process, we consider the following two-quark-two-lepton operators:
\begin{align}
    \mathcal{O}_{te} &= (\overline{t} \gamma^\mu t)(\overline{e}_{i} \gamma_\mu e_{i})\,, & \mathcal{O}_{Qe} &= (\overline{Q} \gamma^\mu Q)(\overline{e}_{i} \gamma_\mu e_{i})\,,  \nonumber \\
    \mathcal{O}_{tl} &= (\overline{t} \gamma^\mu t)(\overline{l}_{i} \gamma_\mu l_{i})\,, & \mathcal{O}_{Ql}^{(1)} &= (\overline{Q} \gamma^\mu Q)(\overline{l}_{i} \gamma_\mu l_{i})\,, \nonumber \\
    \mathcal{O}_{Ql}^{(3)} &= (\overline{Q} \gamma^\mu \tau^I Q)(\overline{l}_{i} \gamma_\mu \tau^I l_{i})\,, \label{eq:Qte}
\end{align}
with $i=1$ or $2$, corresponding to the first or second fermion generation (for electron or muon colliders, respectively). The operators involving $b$ quarks could in principle contribute but as they do not appear at LO in this production process, they are omitted. In line with the conventions of Refs.~\cite{Aguilar-Saavedra:2018ksv,Degrande:2020evl}, we make the following redefinitions of the Wilson coefficients:
\begin{align}  \label{eq:13rotation}
    C_{Ql}^- &= C_{Ql}^{(1)} - C_{Ql}^{(3)} & \quad C_{Ql}^3 &= C_{Ql}^{(3)}.
\end{align}
This rotation specifically applies to operators involving fermion neutral currents and charged leptons. Choosing such linear combinations makes explicit the number of degrees of freedom involved in top-quark pair production at a lepton collider. Specifically, the redefined coefficient $C_{Ql}^-$ mediates vertices with $t\bar{t}$, whilst the redefined $C_{Ql}^3$ is associated only with $b\bar{b}$ neutral currents. Consequently, the LO contributions to top-quark pair production in lepton colliders are governed by only four Wilson coefficients.

\paragraph{The Drell-Yan process at the LHC} 
For this process, we consider the following two-quark-two-lepton operators:
\begin{align}
    \mathcal{O}_{ue} &= \sum_{i = 1}^3 (\overline{u}_{i} \gamma^\mu u_{i})(\overline{e} \gamma_\mu e)\,, & \quad \mathcal{O}_{de} &= \sum_{i = 1}^3 (\overline{d}_{i} \gamma^\mu d_{i})(\overline{e} \gamma_\mu e)\,, \nonumber\\
    \mathcal{O}_{qe} &= \sum_{i = 1}^3 (\overline{q}_{i} \gamma^\mu q_{i})(\overline{e} \gamma^\mu e)\,, & \quad \mathcal{O}_{ul} &= \sum_{i = 1}^3 (\overline{u}_{i} \gamma^\mu u_{i})(\overline{l} \gamma_\mu l)\,, \nonumber\\
    \mathcal{O}_{dl} &= \sum_{i = 1}^3 (\overline{d}_{i} \gamma^\mu d_{i})(\overline{l} \gamma_\mu l)\,, & \quad \mathcal{O}_{ql}^{(3)} &= \sum_{i = 1}^3 (\overline{q}_{i} \gamma^\mu \tau^I q_{i})(\overline{l} \gamma_\mu \tau^I l)\,, \nonumber\\
    \mathcal{O}_{ql}^{(1)} &= \sum_{i = 1}^3 (\overline{q}_{i} \gamma^\mu q_{i})(\overline{l} \gamma_\mu l)\,, \label{eq:Que}
\end{align}
Following the change of basis in~\cref{eq:13rotation}, and for consistency, we will use the redefined coefficients:
\begin{align}  \label{eq:13rotation1}
    C_{ql}^- &= C_{ql}^{(1)} - C_{ql}^{(3)}\,, & \quad C_{ql}^3 &= C_{ql}^{(3)}\,.
\end{align}
We point out that in the results section, we will not present phenomenological results for operators involving right-handed $d$ quarks, {\it i.e.}, $\mathcal{O}_{dl}$ and $\mathcal{O}_{de}$. The reason is purely a technical one, relating to the present implementation of our UFO model, and they are planned for inclusion in future work. As detailed at the end of~\cref{sec:DPsmeft}, solely for the purposes of bookkeeping and ensuring a homogeneous power counting of SM couplings between different orders in $1/\Lambda$, the Wilson coefficients of~\cref{eq:Qq38} are considered to be of $\mathcal O(\alpha_s)$, whilst ones of~\cref{eq:Qte,,eq:Que} are of $\mathcal O(\alpha)$. 

\subsection{Monte Carlo implementation details} 
\label{sec:MC}
In this Section, we present our MC setup for extracting phenomenological results at NLO with the inclusion of SMEFT four-fermion operators. NLO QCD corrections in the SMEFT have been generally computed and automated~\cite{Degrande:2020evl} in {\aNLO}~\cite{Alwall:2014hca}. Our implementation is standard, with the notable difference that we cluster QCD real-radiation to nearby top quarks in the region of $\Delta R \le 0.4$, where $\Delta R$ is the separation between two objects in the detector\footnote{~\(\Delta R = \sqrt{(\Delta \eta)^2 + (\Delta \phi)^2}\), where \(\Delta \eta\) and \(\Delta \phi\) represent the differences in pseudorapidity and azimuthal angle, respectively.}. This clustering method ensures consistency with the treatment of QED real-radiation in the $\SDKw$ scheme. We remind the reader that for EW corrections, the $\SDKw$ approach outlined in~\cref{sec:sdkweak} is the phenomenologically relevant one, as it bypasses the inherent infrared sensitivity introduced by the DP algorithm.
We use a specialised version of {\aNLO} capable of computing full QCD corrections, as well as EW corrections in the Sudakov approximation, see Ref.~\cite{Pagani:2021vyk}. We also retain the logarithmic contributions among Mandelstam invariants via the $\Delta^{s\TO r_{kl}}$ prescription of Ref.~\cite{Pagani:2021vyk}. For SMEFT studies, we have prepared a \texttt{UFO} model based on~\cite{Frederix:2018nkq,Degrande:2020evl}, which contains the relevant EFT operators.

\paragraph{LHC} For LHC phenomenology ($t\bar t$ production and Drell-Yan) we consider proton-proton collisions at $\sqrt{s}=13~\text{TeV}$.
We use the PDF set 
\texttt{NNPDF3.1luxQED}~\cite{Bertone:2017bme,Manohar:2017eqh,Manohar:2016nzj} via the corresponding LHAPDF interface~\cite{Buckley:2014ana}. For $t\bar{t}$ production, the factorisation and renormalisation scales $\mu_F$ and $\mu_R$ are computed dynamically with a reference value defined as:
\begin{equation}
    \mu^{\text{dyn}} = \sum_{i \in \text{FS}} \frac{H_{\text{T},i}}{4}\,,
\end{equation}
where the sum runs over all final state particles, and $H_{\text{T},i}$ is the transverse energy of the final-state particle $i$. For Drell-Yan, $\mu_F$ and $\mu_R$ are set to the final-state dilepton invariant mass,
\begin{equation}
    \mu_{F} = \mu_{R} = m_{\ell\ell}.
\end{equation}
When computing NLO QCD corrections, the SMEFT is renormalised separately at the fixed scale $\mu_{\text{EFT}} = 1 \, \text{TeV}$, and since the SMEFT scale is kept fixed, there is no effect from RG running of the Wilson coefficients included, as discussed in detail in~\cref{sec:dSMEWinSMEFT}. We carry out the computations in the $G_\mu$~scheme~\cite{Denner:2000bj,Dittmaier:2001ay}, as it is considered a suitable choice for SMEFT analyses, according to the recent recommendations of the LHC EFT WG~\cite{Brivio:2021yjb}. Whilst EW parameters and fermion masses are renormalised on-shell, the QCD sector is renormalised in $\overline{\text{MS}}$, with $\alpha_s(m_Z)$ as input. Our complete choice of inputs is as follows:
\begin{table}[ht]
\centering
\begin{tabular}{ c  l }
\hline
 $\GF$ & $1.16637 \times 10^{-5} \, \GeV^{-2}$ \\
 $m_{W}$ & $80.419 \, \text{GeV}$ \\
 $m_{Z}$ & $91.18 \, \text{GeV}$ \\
$m_t$ & $172.5 \, \text{GeV}$ \\
$m_f$, $f \neq t$ & $0.0$ \\
  $\alpha_{s}(m_{Z})$ & $0.118$ \\ \hline
\end{tabular}
\caption{Common inputs used in the our MC simulations.}
\end{table}
\paragraph{Muon collider} For the muon collider, we consider $\sqrt{s} = 10 ~ \text{TeV}$, unpolarised beams, and the same choice of input scheme and parameters described above. For simplicity, we choose not to include effects stemming from initial-state radiation (typically encoded via PDFs of the lepton) as they are expected to be negligible in the regions where Sudakov enhancements are relevant (see {\it e.g.}~\cite{Ma:2024ayr}). The renormalisation scale is fixed to:
\begin{equation}
    \mu_R = \frac{\sqrt{s}}{2}\,.
\end{equation}
Similarly to the LHC case, in the NLO QCD corrections, the SMEFT is renormalised separately, but in this case the EFT scale is fixed to the muon beam energy, $\mu_{\text{EFT}} = \sqrt{s}/2$.

\section{Results}
\label{sec:results}
We now present the differential results for the three processes examined in this study, incorporating dimension-six four-fermion operators and focusing on the impact of high-energy approximations of NLO EW corrections compared to NLO QCD corrections. The former, when combined with the LO prediction, are referred to interchangeably as `EW$_{\text{SDK}}$' (or simply `EW'), and the latter as just `QCD' (See~\cref{eqn:QCDEWdef} in~\cref{sec:DPsmeft} for their precise definition). Predictions will be studied at the interference- and squared-level, generically referred to as `INT' and `SQ', as per~\cref{eqn:lambdaINTSQ}. The impact of corrections will be quantified using differential $K$-factors, defined as the ratio of the higher-order to LO predictions. These will be examined separately for the SM, INT and SQ contributions to provide the clearest picture of the effects of higher-order corrections.
In particular, we explore how the sensitivity of kinematic distributions to the SMEFT coefficients varies with the level of accuracy employed and how higher-order corrections can resolve flat directions that are indiscernible at LO. It is important to note that four-fermion operators are classified based on their chirality structure. Since \(\text{EW}_{\text{SDK}}\) corrections depend on the helicity configuration of the external states, the behaviour of these corrections is expected to be closely linked to the chirality structure of the corresponding SMEFT operator. We have chosen values for the Wilson coefficients primarily for illustrative purposes, ensuring that the SMEFT contributions are of $\mathcal{O}(1)$ in the kinematic ranges considered. We will highlight predictions for a subset of the relevant operators, with the remaining set presented in~\cref{app:AdditionalDistr}.

\subsection{Top-quark pair production at the LHC} 
\label{sec:LHCtt}
We begin with the phenomenology of EW corrections in top-quark pair production at the LHC,
\begin{equation}
    p p \to t \bar t,
\end{equation}
with the inclusion of the SMEFT four-quark operators in~\cref{eq:Qq38}.
As an example,~\cref{fig:tt_distr_Qd8} focuses on the operator $\mathcal O_{Qd}^8$, with a Wilson coefficient set to $C_{Qd}^8=0.25$ TeV$^{-2}$. 
The main plot in the left panel shows the differential cross-section with respect to the transverse momentum of the top quark, $d\sigma/dp^{t}_{\sss T}$, including both QCD and EW corrections. The insets of the left figure illustrate the relative impact of the linear and quadratic SMEFT contributions compared to the SM at different orders in the perturbative expansion. The solid, dotted, and dashed lines represent the ratios to the SM at LO, QCD, and EW$_{\text{SDK}}$ accuracy, respectively, whilst the solid black lines indicate the ratio of QCD+EW$_{\text{SDK}}$. 
The right plot of~\cref{fig:tt_distr_Qd8} displays the corresponding differential $K$-factors in the SM, and order by order in the SMEFT for the operator $\mathcal{O}_{Qd}^{8}$ indicated by the subscript on $K$.~\Cref{fig:tt_distr_Qq81} presents the same plots but for the operator $\mathcal{O}_{Qq}^{1,8}$. The remaining plots for the other four-quark operators in~\cref{eq:Qq38} are compiled in~\cref{sec:tt_distr_remaining}. 
\begin{figure}[ht]
    \centering
    \includegraphics[width=0.5\textwidth]{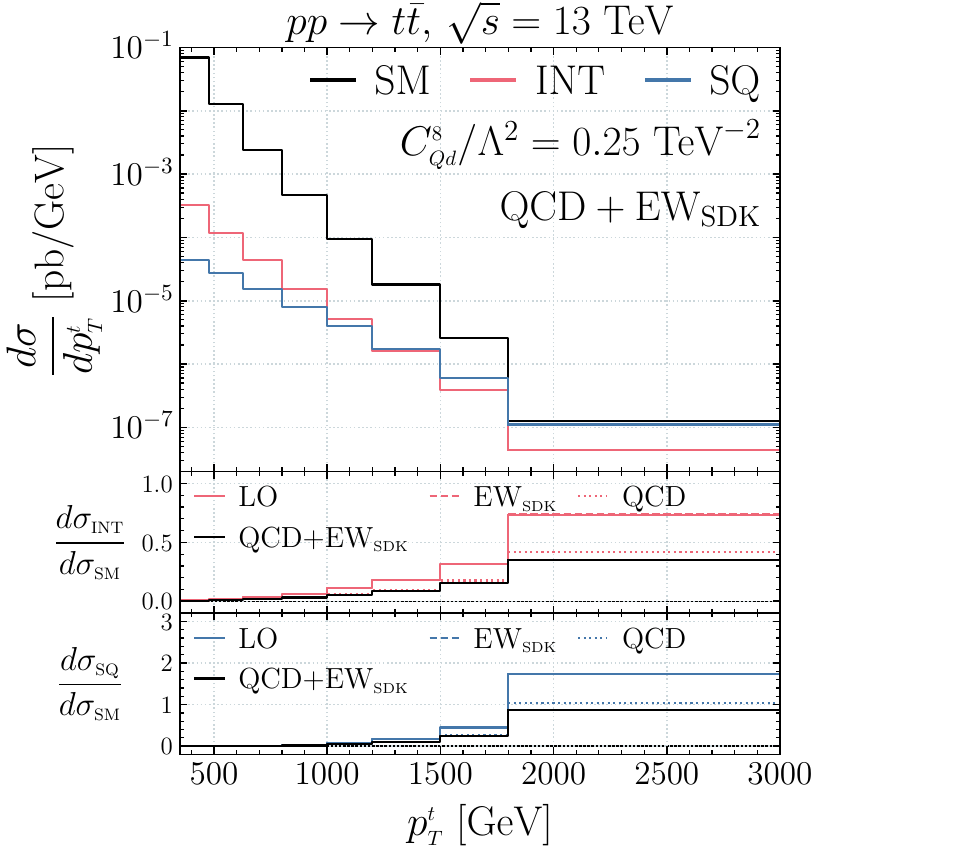}
    \hspace{-1.0cm}
    \includegraphics[width=0.5\textwidth]{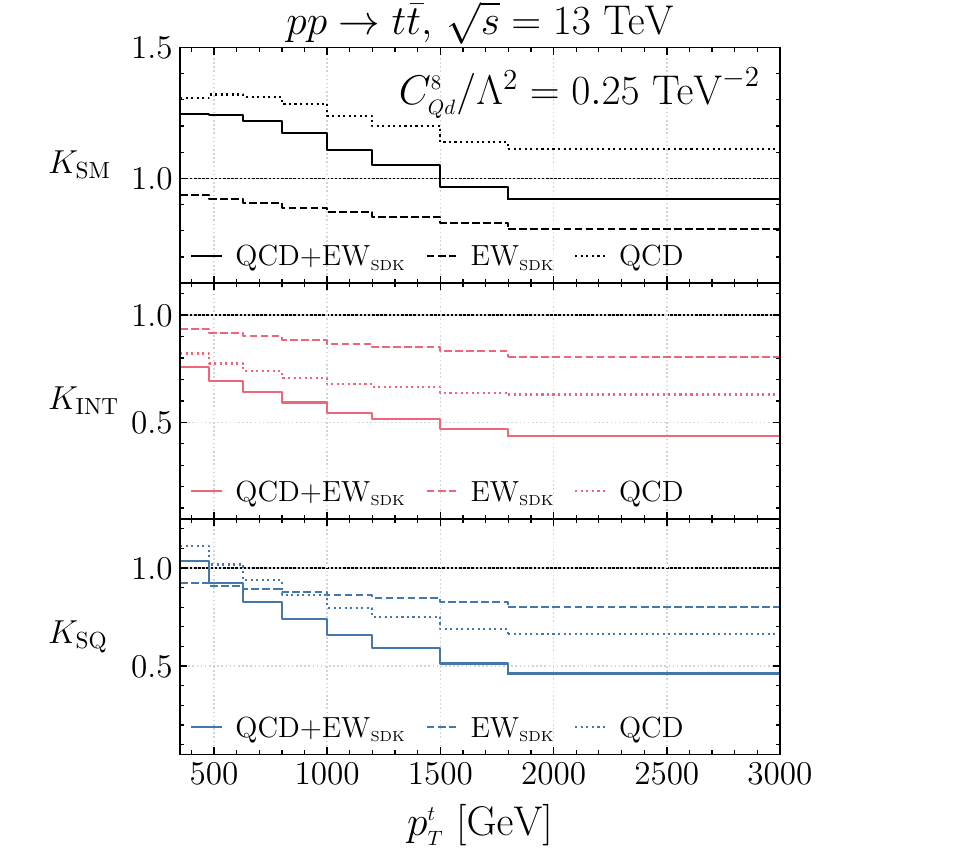}
    \vspace{-4mm}
   \caption{\textbf{Left:} Differential cross-section for top-quark pair production at the LHC, shown at QCD+EW$_{\text{SDK}}$ order as a function of the transverse momentum of the top quark, $p^{t}_{\sss T}$. The black line represents the SM, the red line indicates the linear term in $C_{Qd}^8$, and the blue line depicts the quadratic term. Insets illustrate the relative impact of the interference and quadratic terms compared to the SM at various perturbative orders. \textbf{Right:} Corresponding $K$-factors for $C_{Qd}^8$ at QCD order (dotted), EW$_{\text{SDK}}$ order (dashed), and the combined effect of the two (solid) at different orders in the EFT expansion.}
    \label{fig:tt_distr_Qd8}
\end{figure}
\begin{figure}[ht]
    \centering
    \includegraphics[width=0.5\textwidth]{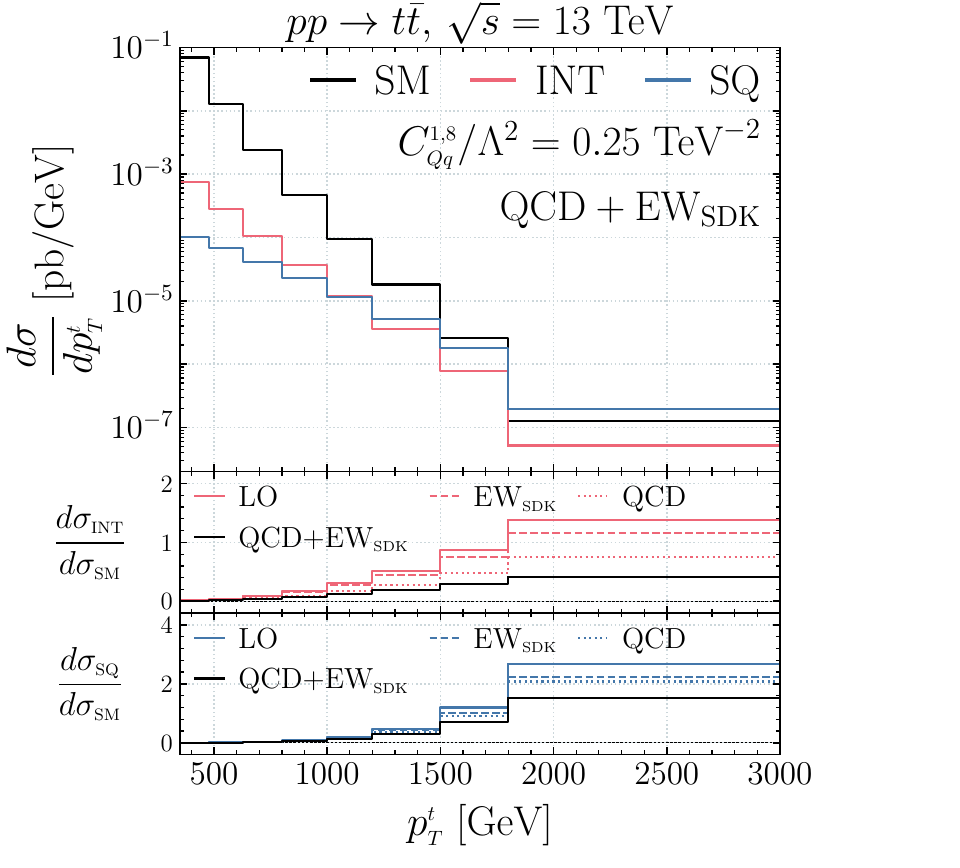}
    \hspace{-1.0cm}
    \includegraphics[width=0.5\textwidth]{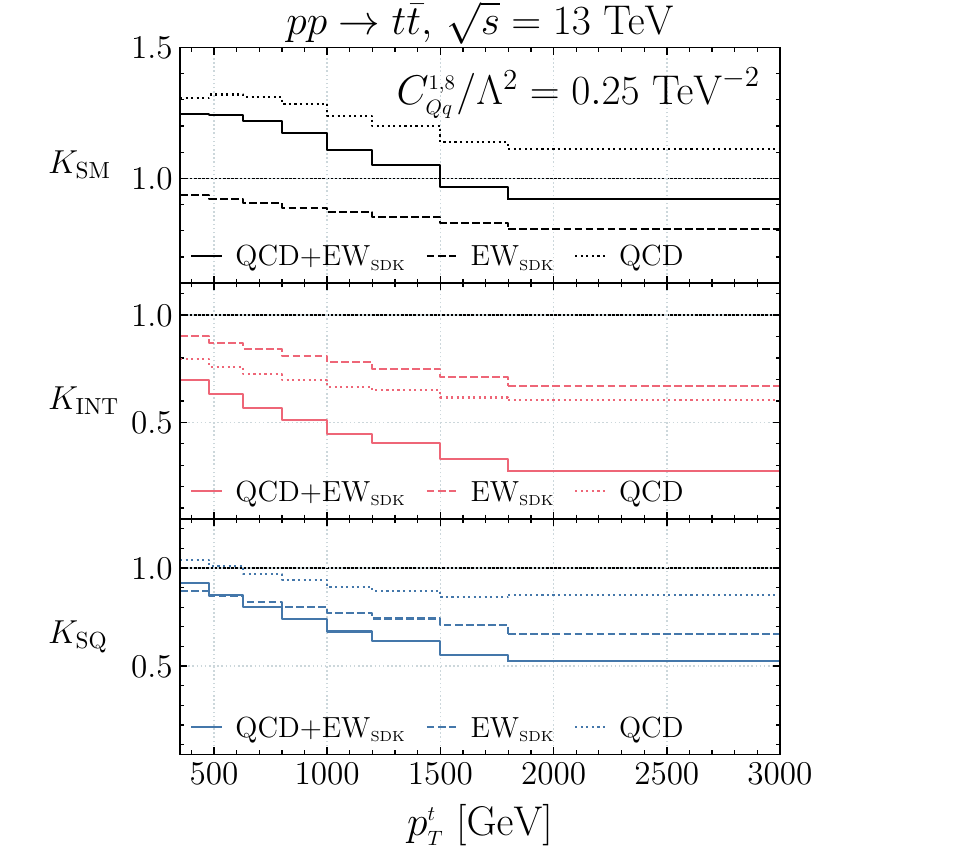}
    \vspace{-4mm}
   \caption{Same as~\cref{fig:tt_distr_Qd8} for the operator $\mathcal O_{Qq}^{1,8}$.}
    \label{fig:tt_distr_Qq81}
\end{figure}

Several interesting features can be seen when comparing the differential $K_{\text{SM}}, K_{\text{INT}}$ and $K_{\text{SQ}}$. Firstly, as expected, EW corrections are negative and grow with energy in absolute value, resulting in a downward shift in the tail of the distributions, a feature that underlines the importance of those contributions in precision SMEFT calculations. Secondly, whilst as expected, QCD corrections are positive in the SM, they are mostly negative in the SMEFT such that they combine constructively with the EW corrections leading to strong suppressions of the SMEFT effects in the tails. The negative QCD corrections in the SMEFT are likely associated with the reduction of the centre-of-mass energy of the fermion-fermion contact vertex due to QCD radiation.
This observation is valid for all the four-quark operators that we considered, both at linear and quadratic order. 
The different behaviour of the SM QCD $K$-factors, in comparison to the SMEFT case, is likely due to the important $gg$-initiated contribution, which is not impacted by the four-fermion operators. This contribution remains significant despite its decreasing relevance due to the sharp decline of the $gg$ luminosities with centre-of-mass energy.
Moreover, $K$-factors associated with different operators can differ, often significantly. This can already be seen by comparing~\cref{fig:tt_distr_Qd8,,fig:tt_distr_Qq81}, and is further confirmed by the additional~\cref{fig:tt_distr_tu8_td8_Qu8,,fig:tt_distr_tq8_Qq83} in~\cref{sec:tt_distr_remaining}. Therefore, we strongly advise against using a simple $K$-factor approach for including higher-orders, as it fails to represent the relevant underlying physical effects. This is particularly true when the $K$-factors evaluated for the SM are applied naively to the SMEFT predictions.
Such application results in an almost 100\% error in predictions, a conclusion that can be inferred for example from~\cref{fig:tt_distr_Qq81}, in which the QCD+EW$_{\text{SDK}}$ $K$-factor for the SM lies around $0.75$ in the highest energy bin, whilst the corresponding linear SMEFT one is close to $0.3$.

Overall, the sensitivity to four-fermion operators in $t\bar{t}$ production at the LHC exhibits significant variation at NLO compared to LO, as previously demonstrated in studies focusing on QCD corrections~\cite{Brivio:2019ius,Degrande:2020evl}. We observe that the numerical effect of the EW corrections is comparable to that of the QCD corrections. Furthermore, the disparity in $K$-factors between the SM and the SMEFT results in a  significant reduction in the relative impact compared to LO predictions. This significant change is evident in the insets of the left plot in~\cref{fig:tt_distr_Qd8,,fig:tt_distr_Qq81}, {\it i.e.}~the difference between the solid black and coloured lines where the LO impacts are strongly reduced when including the QCD+EW$_{\text{SDK}}$ corrections. 

\subsection{The Drell-Yan process at the LHC} 
\label{sec:LHCDY}
We continue our examination of key LHC processes by considering electron-positron pair production,
\begin{equation}
    p p \to e^+ e^-,
\end{equation}
with the inclusion of the SMEFT two-quark-two-lepton operators in~\cref{eq:Qte}, and considering the redefined coefficients in~\cref{eq:13rotation}.
In~\cref{fig:DY_distr_qlm,,fig:DY_distr_qe}, similarly to~\cref{fig:tt_distr_Qd8,,fig:tt_distr_Qq81}, we plot the differential cross-section in the dilepton invariant mass $m_{\ell \ell}$, $d\sigma/dm_{\ell \ell}$, for the SM, and in SMEFT at linear and quadratic order for the operators $\mathcal{O}_{ql}^{\sss(-)}$ and $\mathcal{O}_{qe}$, when $C/\Lambda^2=0.02$ TeV$^{-2}$.
For completeness, the differential results for the remaining four-fermion operators are shown in~\cref{sec:DY_distr_remaining}.
\begin{figure}[ht]
    \centering
    \includegraphics[width=0.5\textwidth]{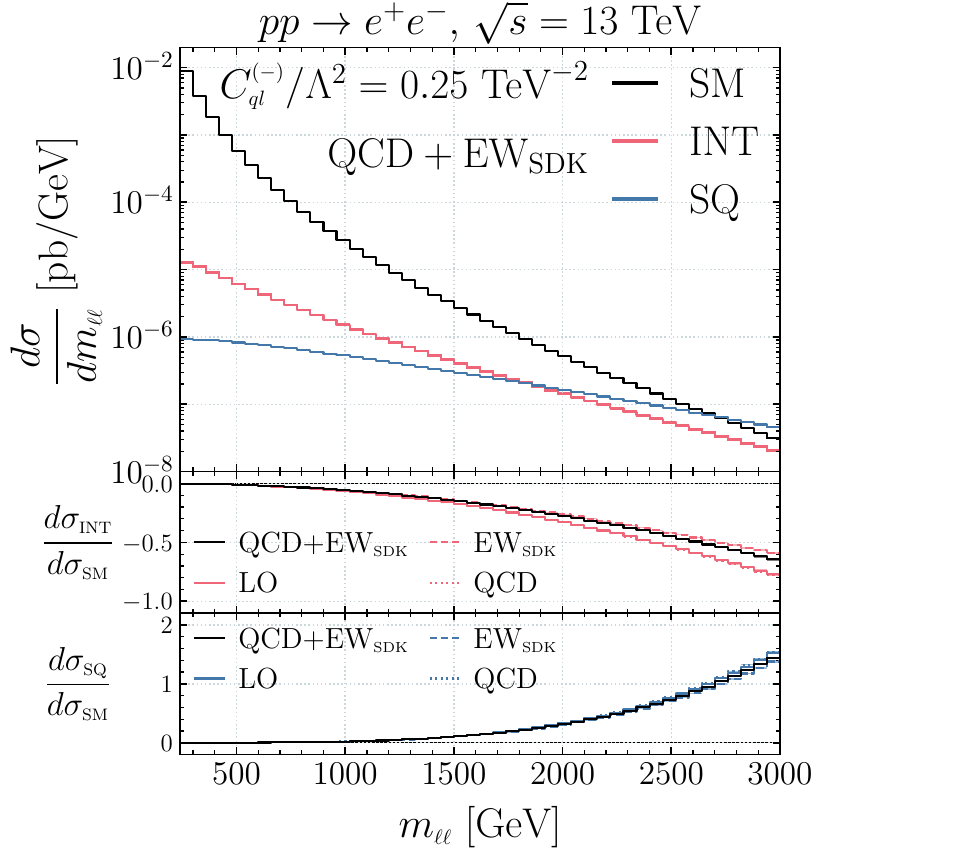}
    \hspace{-1.0cm}
    \includegraphics[width=0.5\textwidth]{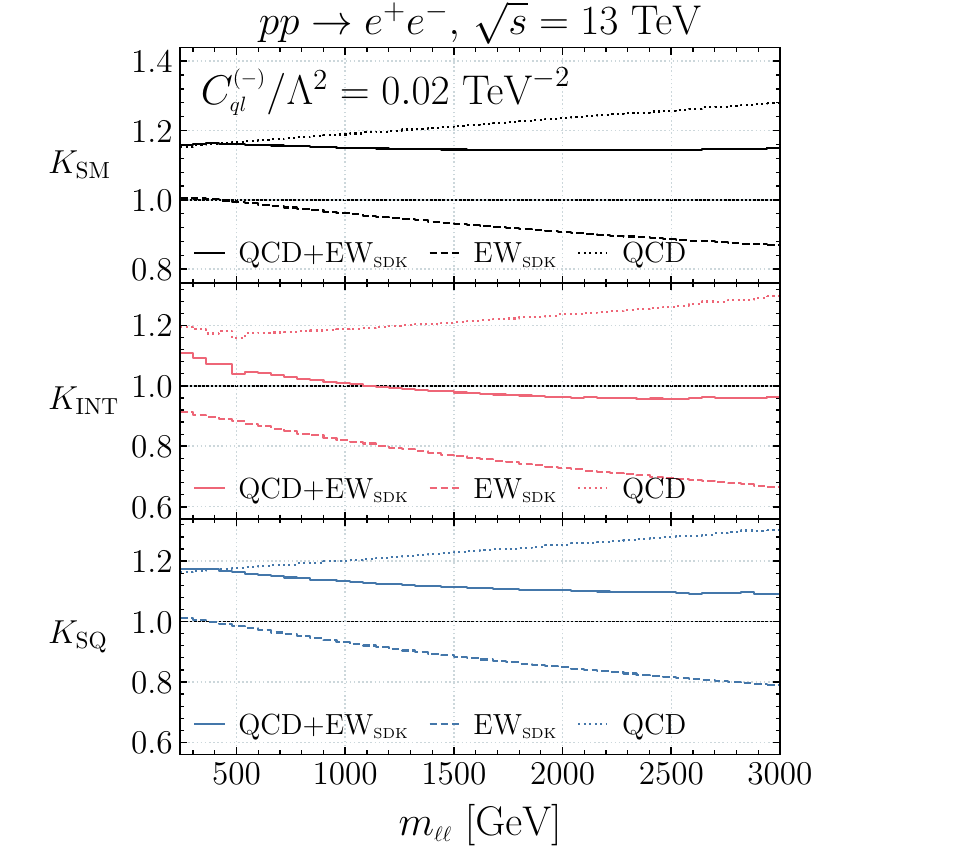}
    \vspace{-4mm}
   \caption{Differential cross-section in $m_{\ell\ell}$ (left) and the corresponding $K$-factors (right) for electron-positron pair production at the LHC, with the same notation and structure of~\cref{fig:tt_distr_Qd8}, for the operator $\mathcal O_{ql}^{\sss(-)}$.}
   \label{fig:DY_distr_qlm}
\end{figure}
\begin{figure}[ht]
    \centering
    \includegraphics[width=0.5\textwidth]{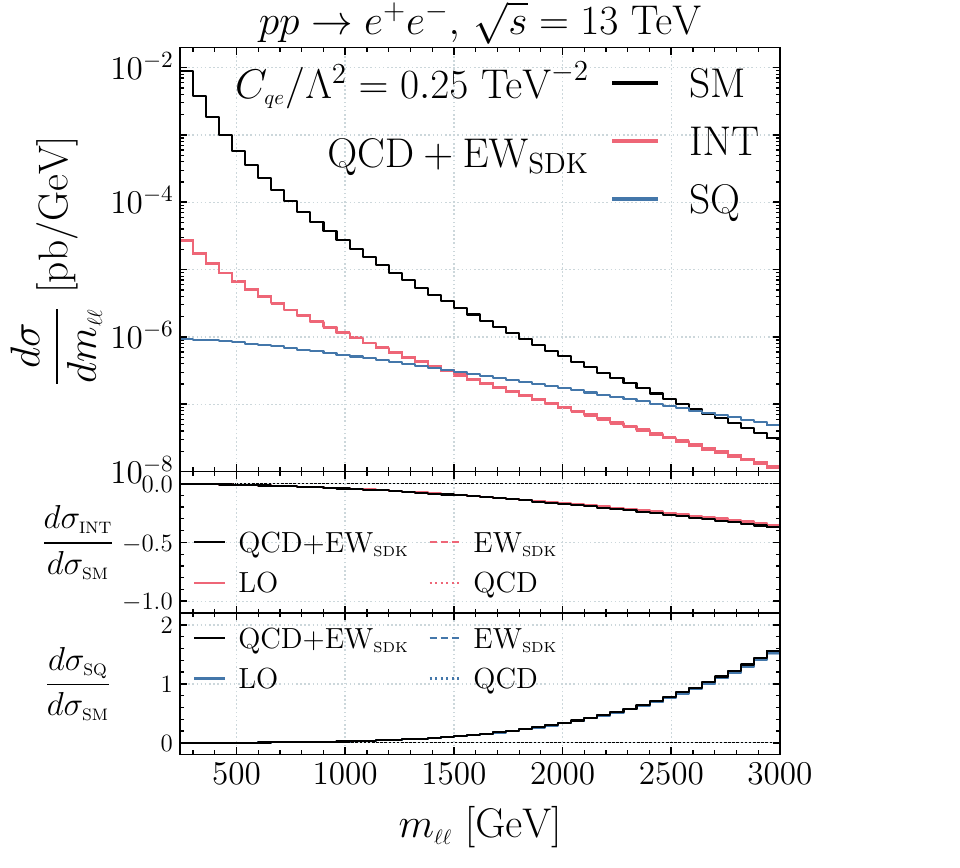}
    \hspace{-1.0cm}
    \includegraphics[width=0.5\textwidth]{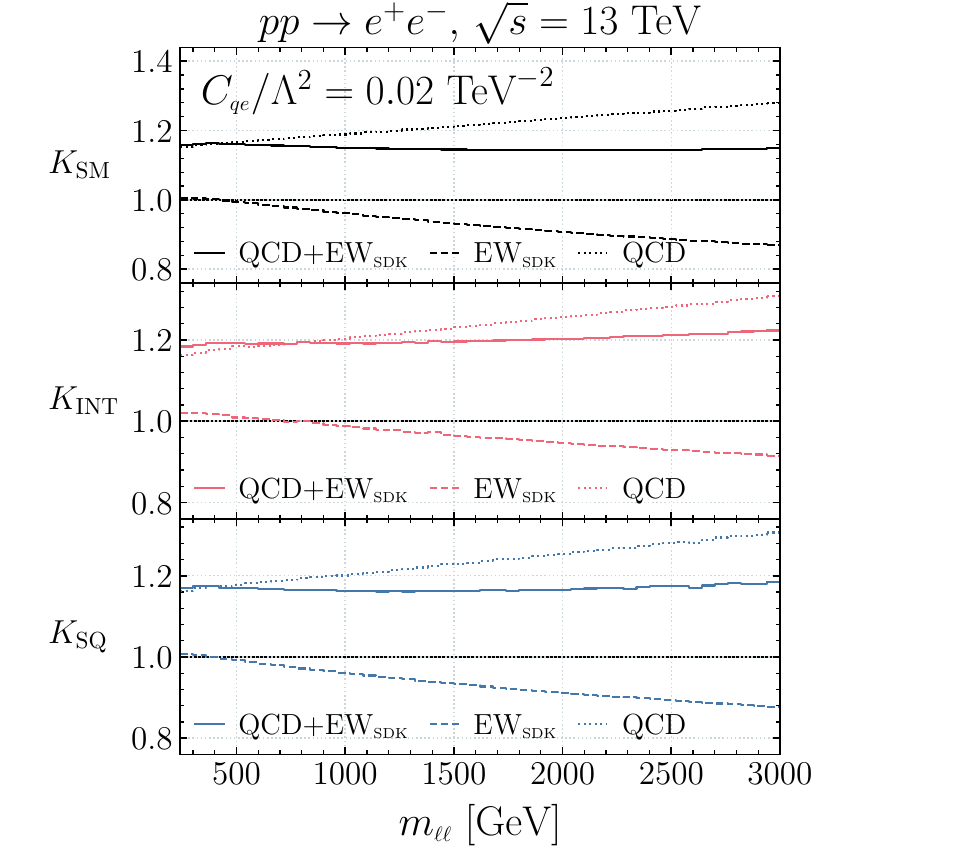}
    \vspace{-4mm}
   \caption{Same as~\cref{fig:DY_distr_qlm} for the operator $\mathcal O_{qe}$.}
   \label{fig:DY_distr_qe}
\end{figure}

Similarly to the $t \bar t$ case, and as expected, at high energies, EW$_{\text{SDK}}$ corrections are generally negative and grow in size with energy, with the characteristic $\alpha \log^2 (s/M^2)$ behaviour of Sudakov logarithms. The QCD corrections, instead, are positive and grow with energy, which leads to an interesting accidental cancellations between the two. In the SM, this results in an essentially flat $K$-factor. Looking, for example at the interference contribution of $C_{ql}^{\sss(-)}$ in the right panel of~\cref{fig:DY_distr_qlm}, we see that the cancellation leads to a QCD+EW$_{\text{SDK}}$ prediction lying very close to the LO one. 

Moving to the contributions from $C_{qe}$ in~\cref{fig:DY_distr_qe}, the aforementioned cancellation is less evident, leading to a flatter but non-zero result. This case is more representative of the remaining operators in~\cref{sec:DY_distr_remaining}, where QCD corrections are similar across the board but EW corrections are relatively less important.
The QCD corrections to SMEFT contributions  are known to be approximately SM-like. Indeed, the similarity of the differential QCD $K$-factors between the SM and the SMEFT is evident in the right plots of~\cref{fig:DY_distr_qlm,,fig:DY_distr_qe} as well as the ones in~\cref{sec:DY_distr_remaining}. This explains the stability under QCD corrections of the relative impacts w.r.t.~the SM, as depicted in the insets of left plots in~\cref{fig:DY_distr_qlm,,fig:DY_distr_qe}, as well as the additional~\cref{fig:DY_distr_ue_ul_ql3} in~\cref{sec:DY_distr_remaining}. 
In the case of $\mathcal O_{q\ell}^{\sss(-)}$, the relative impact is considerably modified by the inclusion of EW corrections, especially at the linear level, where around 30\% difference is predicted at high energies. However, for $\mathcal O_{qe}$, the overall impact is not significantly modified when including higher orders QCD or EW corrections. A similar range of sensitivity to EW$_{\text{SDK}}$ corrections is observed across all of the operators that we considered, leading to effects on the relative impacts at the level of a few tens of percent in the tails. 

The key takeaway is that, in contrast to QCD, these corrections do not have a universal behaviour, differing across operators and even between individual interference and squared contributions. Our results once again demonstrate the importance of including NLO corrections of both QCD and EW origin in precision SM and SMEFT predictions, particularly for the tails of differential distributions.

\subsection{Top-quark pair production at a muon collider}
\label{sec:muCtt}
To conclude our phenomenological studies, we now consider the relevance of EW corrections in the SMEFT within a muon collider scenario. We consider a circular muon collider operating at $\sqrt{s} =  10 \, \text{TeV}$, and focus on top-quark pair production,
\begin{equation}
    \mu^+ \mu^- \to t \bar{t}.
\end{equation}
Our analysis is similar to the one of the LHC presented in~\cref{sec:LHCtt}, apart from the difference in the relevant four-fermion operators, that here are of the two-muon-two-quark type, as given in~\cref{eq:Que} and considering the redefinition in~\cref{eq:13rotation1}. The SM-only component of this analysis has been carried out also in Ref.~\cite{Ma:2024ayr}, where many more details have been discussed.

It is important to note that in a muon collider, and in lepton colliders in general, the partonic energy available in the centre-of-mass frame is effectively constant (up to small effects from lepton PDFs, see {\it e.g.}~the discussion in Ref.~\cite{Ma:2024ayr}). Therefore, logarithms of the form $\log^2(s/M^2)$ and $\log(s/M^2)$ are constant throughout the phase space. Only logarithms depending explicitly on angles, such as $\log(-t/M^2)$ or $\log(-t/s)$ are capable of inducing an angular dependence from the Sudakov approximations.
\begin{figure}[ht]
    \centering
    \includegraphics[width=0.5\textwidth]{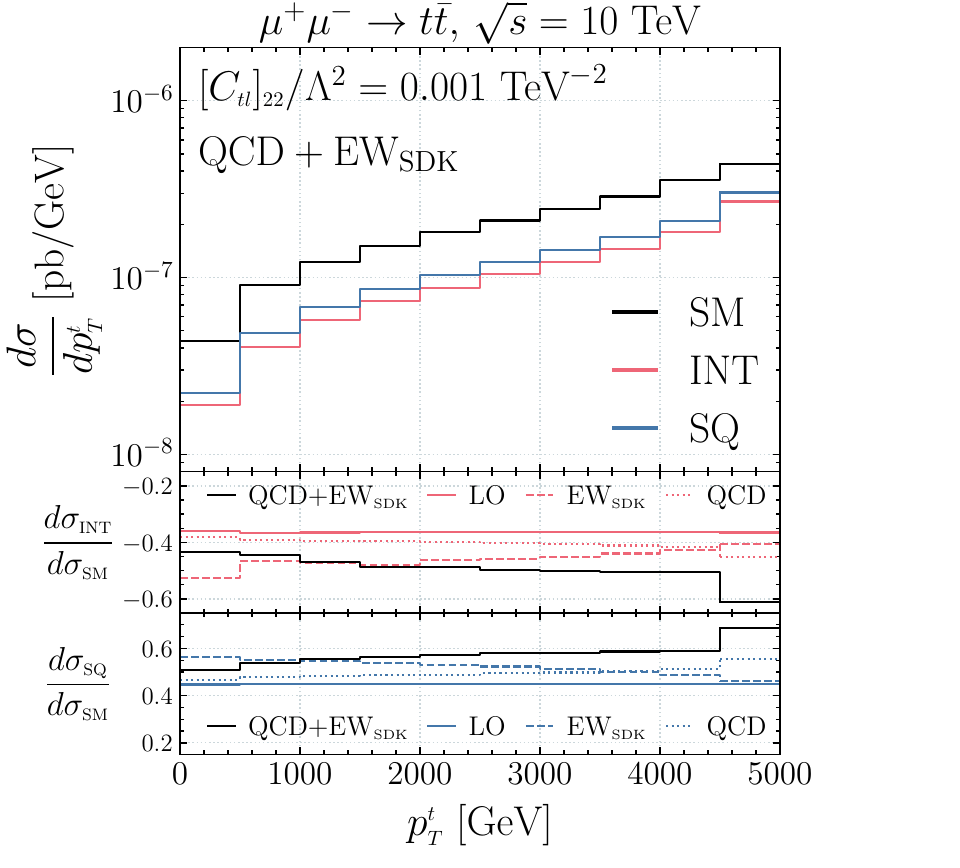}
    \hspace{-1.0cm}
    \includegraphics[width=0.5\textwidth]{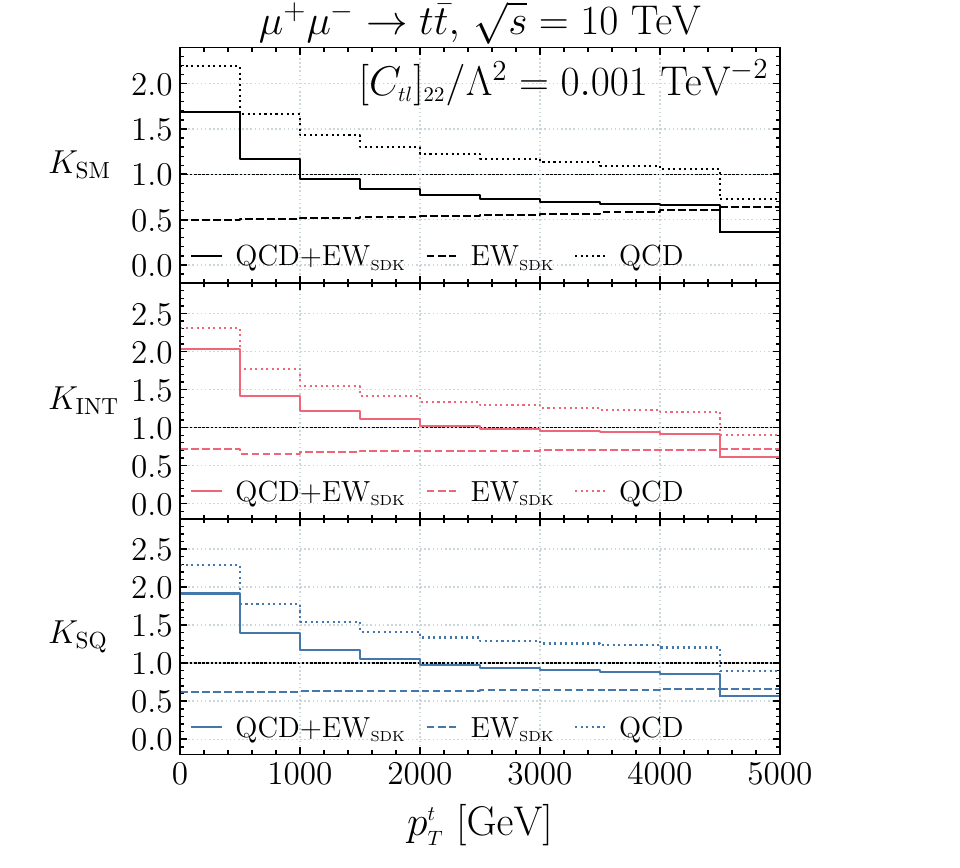}
    \vspace{-4mm}
   \caption{Differential distribution and corresponding $K$-factors for top-quark pair production in a $10 \, \text{TeV}$ muon collider, with the same notation and structure of~\cref{fig:tt_distr_Qd8}, for the operator $\mathcal{O}_{tl}$.}
    \label{fig:muC_distr_tl}
\end{figure}

~\Cref{fig:muC_distr_tl}, along with the additional Figures in~\cref{sec:muC_distr_remaining}, showcase our results for the differential cross-section $d\sigma/dp_{T}^{t}$, for the SM and at the linear and quadratic levels for the relevant four-fermion Wilson coefficients, with $C/\Lambda^2=0.001$ TeV$^{-2}$. The relatively small value of the Wilson coefficients reflects the fact that the sensitivity of this process far exceeds that of the LHC due to the larger partonic centre-of-mass energy.

Concerning QCD corrections, we find that they are generally positive and sizeable, with the exception of the rightmost bin, where they become negative. This behaviour is primarily due to QCD radiation escaping recombination, decreasing the invariant mass of the \(t\bar{t}\) pair and, consequently, reducing the value of \(p_T^t\). Given the shape of the distribution at LO, which peaks at large \(p_T^t\), this dynamic results in the characteristic shape of the \(K\)-factor observed.

Moving to EW corrections, we find, as expected, that they are consistently negative, but that they grow in absolute value at smaller $p_T^t$, especially in the case of the SM. This effect is counter-intuitive, but it has already been observed in Ref.~\cite{Ma:2024ayr} for the SM case, where it is also discussed in detail. The origin is the aforementioned independence of double logarithms on angular distributions, at variance with the case of single logarithms. In fact, the $p_T^t$-dependence on the relative EW corrections is milder in the SMEFT, in particular for the operator considered in~\cref{fig:muC_distr_tl} ({\it cf.} results in~\cref{sec:muC_distr_remaining}).

In summary, the inclusion of EW corrections is important at high energies, and is therefore highly recommended when precise predictions are needed. This is especially the case for a TeV-scale lepton collider, where EW corrections, for specific operators, can be of order $-50\%$ of the LO. We note that such large negative corrections may have implications beyond precision studies, and may even lead to a sizeable impact on projections for statistical uncertainties in experimental analyses. Indeed, it is important to note that, in contrast to the LHC case, the region at high $p_T^t$ corresponds to the bulk of the cross-section.
Furthermore, the SM, the linear, and the quadratic SMEFT contributions generally exhibit different $K$-factors, both in magnitude and energy dependence. Therefore, the point made in the previous two sections—that the naive
inclusion of higher-order corrections via a simplistic $K$-factor approach does not provide an accurate physical description—also applies to this process and is even more pertinent.  

\subsection{Lifting flat directions in the SMEFT parameter space}
Given a specific process, it is well known that EW corrections can induce sensitivity to additional parameters or interactions that were not present at the LO. For instance, in the case of $p p \to t \bar t$, these effects have already been studied in, {\it e.g.}, Refs.~\cite{Kuhn:2013zoa,Martini:2019lsi, Martini:2021uey, Blasi:2023hvb, Maltoni:2024wyh}, showing that they enable probes of the EW and Yukawa couplings of the top-quark, or new top-philic scalars. The situation for the processes and the operators that we have considered is analogous, with the minor complication that the newly probed directions correspond to linear combinations of Wilson coefficients.

Hence, the improved accuracy of higher-order predictions can lead to new conclusions on how data can indirectly constrain new physics. In the event that evidence for non-zero Wilson coefficients is observed, higher-order corrections will also be essential to accurately and precisely pinpoint their new physics origin. Independently of any discovery, in a highly multidimensional approach such as the SMEFT, correlations among Wilson coefficients in a given dataset are key in order to achieve a robust sensitivity.

It is also important to identify the presence of weakly-constrained or flat directions in order to motivate future measurements aimed at globally closing in on the parameter space. These features can be sensitive to higher-order corrections, given the non-trivial $K$-factors that we have highlighted in the previous section. In order to highlight this fact, we will mostly consider $pp\to t\bar{t}$ in~\cref{sec:flattt} as a simple example. We choose to focus on this process because it has a relatively simple dependence on the four-fermion operators, allowing for an intuitive understanding of the flat directions at LO. We also briefly comment on the Drell-Yan process and on $\mu^+\mu^-\to t\bar{t}$ production at the end of the section. Afterwards, in~\cref{sec:fisher}, a more systematic study based on Fisher information will be carried out for all the three processes considered in this work.

\subsubsection{Benchmarks and parameter subspaces}
\label{sec:flattt}
At LO, the partonic total rate for $u \bar{u} \to t \bar{t}$ at $\mathcal{O}(\Lambda^{-2}$) in the SMEFT (which enters in $\sigma_{\rm INT}$) only receives contributions from the purely vectorial combination of the four-fermion Wilson coefficients, $C_{VV}^u$. This combination is characterised by its equal coupling to left- and right-handed fermions~\cite{Brivio:2019ius}:
\begin{equation}
    C_{VV}^u = \frac{1}{4} \big( C_{Qq}^{1,8} + C_{Qq}^{3,8} + C_{tu}^{8} + C_{tq}^{8} + C_{Qu}^{8} \big)\,.\label{eq:cvvup}
\end{equation}
Similarly, the total rate for $d \bar{d} \to t \bar{t}$ only receives contributions from the down-type vectorial combination:
\begin{equation}
    C_{VV}^d = \frac{1}{4} \big( C_{Qq}^{1,8} - C_{Qq}^{3,8} + C_{td}^{8} + C_{tq}^{8} + C_{Qd}^{8} \big)\,. \label{eq:cvvdown}
\end{equation}
The other five linear combinations (7 four-fermion operators minus $C_{VV}^u$ and $C_{VV}^d$) do not contribute at this order when integrated over the angular phase space. 

Coupling combinations involving the axial current\footnote{~Fermion currents involving $\gamma^\mu\gamma^5$ Dirac structures are sometimes denoted as `axial-vector' currents. We refer to them as just `axial' for simplicity.}, such as those for the up-type quark,
\begin{align}
     C_{AA}^u &= \frac{1}{4} \big( C_{Qq}^{1,8} + C_{Qq}^{3,8} + C_{tu}^{8} - C_{tq}^{8} - C_{Qu}^{8}  \big)\,, \\
     C_{AV}^u &= \frac{1}{4} \big( - C_{Qq}^{1,8} - C_{Qq}^{3,8} + C_{tu}^{8} + C_{tq}^{8} - C_{Qu}^{8}  \big)\,, \\    
     C_{VA}^u &= \frac{1}{4} \big( - C_{Qq}^{1,8} - C_{Qq}^{3,8} + C_{tu}^{8} - C_{tq}^{8} + C_{Qu}^{8}  \big)\,,
\end{align}
can still manifest themselves in differential quantities constructed to sample the negative parity of the corresponding interaction, for instance, in angular observables such as symmetries, polarisation and spin correlations~\cite{Bernreuther:2015yna,Severi:2022qjy}.
Equivalent combinations for the down-quark initial state can be obtained by replacing $u\to d$ and taking $C_{Qq}^{3,8}\to-C_{Qq}^{3,8}$. 

The LO dependence on only $C_{VV}^u$ and $C_{VV}^d$ therefore extends to all of the standard non-angular observables, such as top-quark $p_T$ and the top-quark pair invariant mass.
Here we investigate how the sensitivity to certain coupling directions changes as higher-order corrections are introduced, and demonstrate how such `conventional' observables become sensitive to directions that are flat, or insensitive, at LO. 

We begin by considering the impact of higher-order corrections on the differential predictions in $p_T^{t}$ from the vector ($V$) and axial ($A$) coupling combinations defined above. Of particular interest are those involving the axial
current, which do not contribute at LO. To do so, we examine four benchmark parameter points that separately select purely-vectorial ($VV$), purely-axial ($AA$), mixed vector-axial ($VA$) and mixed axial-vector ($AV$) four-fermion interactions, setting the corresponding combinations for up and down quarks equal  as follows:
\begin{center}
\begin{minipage}{0.36\textwidth}
  \begin{itemize}
    \item[$VV$:]$C_{VV}^u=C_{VV}^d=1$ TeV $^{-2}$,
    \item[$VA$:]$C_{VA}^u=C_{VA}^d=1$ TeV $^{-2}$,
\end{itemize}  
\end{minipage}\hspace{0.5cm}
\begin{minipage}{0.36\textwidth}
  \begin{itemize}
    \item[$AA$:]$C_{AA}^u=C_{AA}^d=1$ TeV $^{-2}$,
    \item[$AV$:]$C_{AV}^u=C_{AV}^d=1$ TeV $^{-2}$,
\end{itemize}  
\end{minipage}
\end{center}
with all other combinations vanishing in each case.~\Cref{fig:axial_vector_combinations} plots the interference contribution, $\sigma_{\rm INT}$, relative to the SM of these four benchmarks at different perturbative orders in bins of $p_T^{t}$, with MC uncertainties given by the coloured bands.
\begin{figure}[ht]
    \centering
    \includegraphics[width=0.9\textwidth]{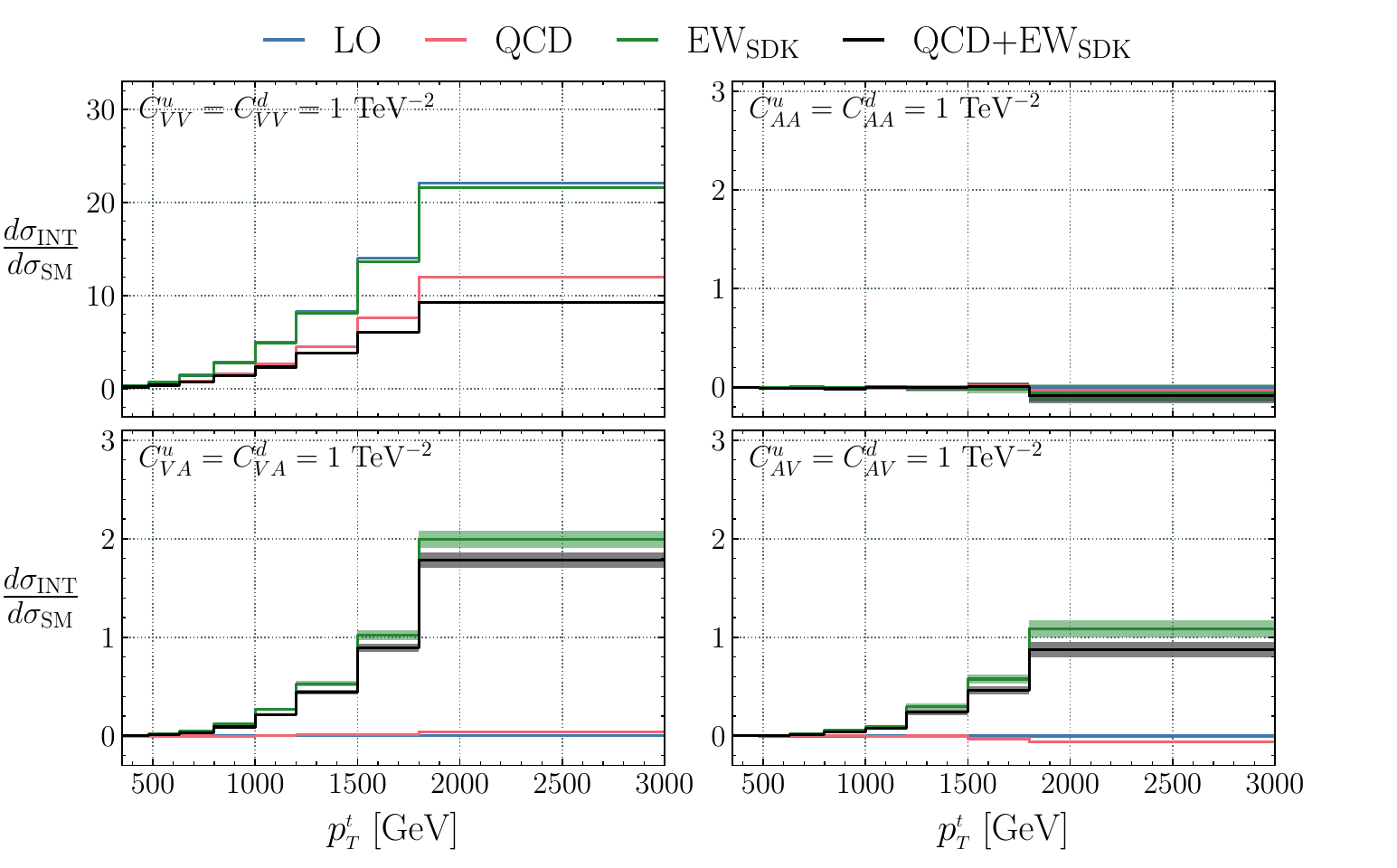}
    \vspace{-4mm}
    \caption{Impact of the SMEFT interference contributions to $pp\to t\bar{t}$, relative to the SM, of purely vectorial (upper left), purely axial-vector (upper right), mixed vector-axial (lower left) and mixed axial-vector (lower right) combinations of four-fermion operators, differentially in $p_T^{t}$. MC statistical uncertainties are represented by the bands around the central values.}
\label{fig:axial_vector_combinations}
\end{figure}
The purely vectorial case shown in the upper left panel is the only non-zero contribution to this observable at LO, and the higher-order corrections display a familiar pattern with respect to those discussed in~\cref{sec:LHCtt}. As expected, the interferences of all three other coupling combinations vanish at this order. 

When only QCD corrections are taken into account, the LO features are qualitatively unaltered. Only the $VV$ combination is not vanishing, although it receives non-negligible corrections.
In contrast, EW$_\text{SDK}$ predictions lead to non-vanishing results both for the $AV$ and $VA$ combinations and the relative corrections on the $VV$ combination is even larger than those induced by QCD corrections. The contribution of EW corrections to the $AV$ and $VA$ combinations is of the order of $10\%$ of the one induced to the $VV$ combination. Moreover, including EW corrections, the $p_T^t$ distribution does not depend on the  $AA$ combination.

It is clear that when EW and QCD corrections are combined, the results are qualitatively equivalent to those at EW accuracy\footnote{~In the last bin of the $AV$ combination the difference observed is due to statistical fluctuations.}, with only an additional effect from QCD for the $VV$ combination. It is important to note that the $AV$ and $VA$ directions at LO would be accessible only at $\mathcal{O}$($\Lambda^{-4}$) and/or via the measurement of the top-quark polarisation or spin correlations~\cite{Brivio:2019ius,Basan:2020btr}. In other words,  EW corrections introduce new sensitivity at the interference level in a more inclusive and easy-to-measure observables such as $p_{T}^{t}$. 

Beyond benchmarks for particular coupling structures, we can also consider flat directions at the linear-level characterised by relations between Wilson coefficients. At this $\mathcal{O}$($\Lambda^{-2}$) order, one can always define relations between pairs of coefficients that lead to a cancellation in a given cross-section.
In~\cref{fig:flat_tt}, we plot two such flat directions in selected 2-dimensional SMEFT parameter spaces at different orders in perturbation theory. They correspond to flat directions in the cross-section for the $1500 < p_T^{t}< 1800$ GeV bin of our kinematical distributions.
\begin{figure}[ht]
    \centering
    \includegraphics[width=.48\textwidth]{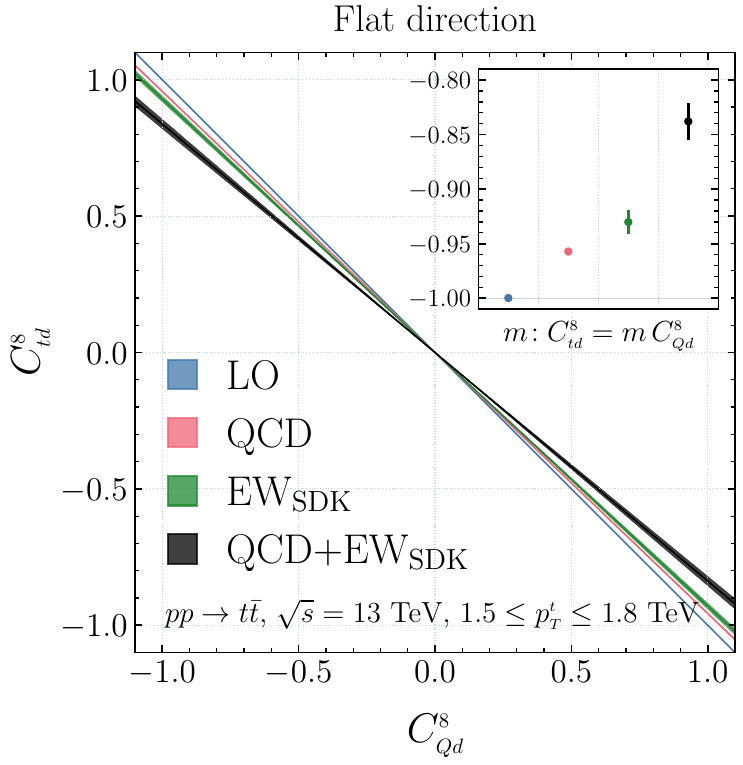}\hfill
    \includegraphics[width=.48\textwidth]{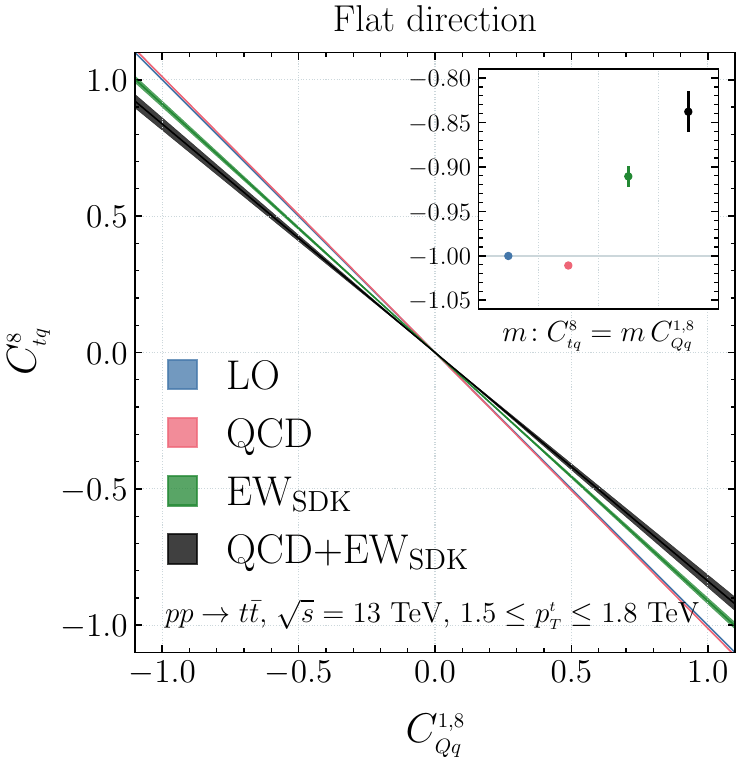}
    \vspace{-4mm}
    \caption{Flat directions in the $(C_{Qd}^8,C_{td}^8)$ and $(C_{Qq}^{1,8},C_{tq}^{8})$ planes for top-quark pair production in the region $1500 < p_{T}^{t} < 1800 \, \text{GeV}$, with thicknesses indicating their MC uncertainty. Blue: LO, Red: QCD, Green: EW$_{\text{SDK}}$, Black: QCD + EW$_{\text{SDK}}$. The insets show the gradients of the plotted lines, along with their MC uncertainties.
    The two flat directions of~\cref{flat_for_figure}, present exactly at LO and approximately at NLO QCD, are shifted by NLO EW corrections.}
\label{fig:flat_tt}
\end{figure}
The directions are defined by the relations:
\begin{align}
    C_{Qd}^8 &=-C_{td}^8\,, & \quad C_{Qq}^{1,8} &= -C_{tq}^{8}\,,
    \label{flat_for_figure}
\end{align}
where for both cases all the other coefficients are set equal to zero. Indeed, as can be seen from~\cref{eq:cvvup,,eq:cvvdown}, when either relation in~\cref{flat_for_figure} is satisfied and all the other coefficients are zero, both $C_{VV}^u$ and $C_{VV}^d$ vanish. In this case the interference contribution to the cross-section, $\sigma^{\rm INT}$, also vanishes when integrated over the angular phase space. This can also be seen in terms of the parity-conserving nature of strong interactions, noting for instance that $C_{Qd}^8$ and $C_{td}^8$ mediate the partonic reactions $d \bar d \to t_L \bar t_L$ and $d \bar d \to t_R \bar t_R$ respectively, two processes that QCD cannot distinguish. 

QCD corrections modify the correlations in the left plot of~\cref{fig:flat_tt}, $(C_{Qd}^8,C_{td}^8)$, whilst the picture is not significantly altered in the right plot for $(C_{Qq}^{1,8},C_{tq}^{8})$. EW corrections induce comparable shifts from the $-45^\circ$ lines, that are more important that the QCD corrections in the right plot. Moreover, in the case of $C_{Qd}^8$ against $C_{td}^8$ (left plot), both the QCD and EW corrections leads to modification to the anti-correlation in the same way. The most important result, however, is that in both plots, the QCD + EW$_{\text{SDK}}$ prediction leads to larger departures from the $-45^\circ$ lines than either the QCD or EW predictions individually.

To provide slightly more detail, in~\cref{tab:xseC_ppSM,,tab:xseC_ppEFT}, respectively, we present the cross-sections in the SM and in the SMEFT at linear and quadratic orders for the four-quark operators of~\cref{eq:Qq38}, in the  kinematic region $ 1.5 \, \text{TeV}<p_{\sss T}^{t}<1.8 \, \text{TeV}$. 
\begin{table}[ht]
\centering
\scalebox{1.0}{
\begin{tabular}{lccc}
\toprule
\textbf{Process} & \textbf{LO} & \textbf{EW}$_{\text{\bf SDK}}$ & \textbf{QCD} \\
\midrule
SM & $7.96 \cdot 10^{-4}$ & $6.61 \cdot 10^{-4}$ & $9.08 \cdot 10^{-4}$ \\
\bottomrule
\end{tabular}}
\caption{SM cross-section [pb] for $1.5~\text{TeV} < p_T^t < 1.8~\text{TeV}$ for the process $p p \to t \bar t$.}
\label{tab:xseC_ppSM}
\end{table}
\begin{table}[ht]
\centering
\scalebox{0.9}{
\begin{tabular}{lcccccc}
\toprule
\textbf{Operator} & \multicolumn{3}{c}{\textbf{Linear Contribution}} & \multicolumn{3}{c}{\textbf{Quadratic Contribution}} \\
\cmidrule(lr){2-4} \cmidrule(lr){5-7}
& \textbf{LO} & \textbf{EW}$_{\text{\bf SDK}}$ & \textbf{QCD} & \textbf{LO} & \textbf{EW}$_{\text{\bf SDK}}$ & \textbf{QCD} \\
\midrule
$C_{Qq}^{1,8}$ & $2.79 \cdot 10^{-3}$ & $1.98 \cdot 10^{-3}$ & $1.72 \cdot 10^{-3}$ & $1.52 \cdot 10^{-2}$ & $1.08 \cdot 10^{-2}$ & $1.29 \cdot 10^{-2}$ \\
$C_{Qq}^{3,8}$ & $7.81 \cdot 10^{-4}$ & $6.91 \cdot 10^{-4}$ & $4.08 \cdot 10^{-4}$ & $1.52 \cdot 10^{-2}$ & $1.26 \cdot 10^{-2}$ & $1.25 \cdot 10^{-2}$ \\
$C_{tq}^{8}$   & $2.79 \cdot 10^{-3}$ & $2.18 \cdot 10^{-3}$ & $1.70 \cdot 10^{-3}$ & $1.52 \cdot 10^{-2}$ & $1.18 \cdot 10^{-2}$ & $1.07 \cdot 10^{-2}$ \\
$C_{tu}^{8}$   & $1.79 \cdot 10^{-3}$ & $1.61 \cdot 10^{-3}$ & $1.04 \cdot 10^{-3}$ & $9.52 \cdot 10^{-3}$ & $8.56 \cdot 10^{-3}$ & $8.11 \cdot 10^{-3}$ \\
$C_{Qu}^{8}$   & $1.79 \cdot 10^{-3}$ & $1.49 \cdot 10^{-3}$ & $1.08 \cdot 10^{-3}$ & $9.52 \cdot 10^{-3}$ & $7.95 \cdot 10^{-3}$ & $6.74 \cdot 10^{-3}$ \\
$C_{td}^{8}$   & $1.00 \cdot 10^{-3}$ & $9.00 \cdot 10^{-4}$ & $6.67 \cdot 10^{-4}$ & $5.72 \cdot 10^{-3}$ & $5.14 \cdot 10^{-3}$ & $4.65 \cdot 10^{-3}$ \\
$C_{Qd}^{8}$   & $1.00 \cdot 10^{-3}$ & $8.37 \cdot 10^{-4}$ & $6.39 \cdot 10^{-4}$ & $5.72 \cdot 10^{-3}$ & $4.73 \cdot 10^{-3}$ & $3.93 \cdot 10^{-3}$ \\
\bottomrule
\end{tabular}}
\caption{SMEFT contributions at linear and quadratic orders to the cross-section [pb] for $1.5~\text{TeV} < p_T^t < 1.8~\text{TeV}$ for the process $p p \to t \bar t$ including the four-fermion operators of~\cref{eq:Qq38}, at LO, EW$_{\text{SDK}}$ and QCD accuracies. 
Wilson coefficients are set to unity, and the scale $\Lambda$ is set to $1 \, \text{TeV}$. The MC uncertainty is approximately $\pm 1$ on the least significant digit. The full QCD +EW$_{\text{SDK}}$ result can be obtained by combining the contributions additively whilst avoiding double-counting the LO contribution as: QCD$ + $EW$_{\text{SDK}} - $LO.}
\label{tab:xseC_ppEFT}
\end{table}
It is interesting to note that the pattern observed in the LO predictions of~\cref{tab:xseC_ppEFT} — same values for different operators — is in part altered by the QCD corrections and completely disrupted by the EW$_{\text{SDK}}$ ones. This picture is consistent with the discussion of~\cref{fig:flat_tt}, where we have shown that  for individual operators, such as $C_{Qq}^{1,8}$ and $C_{tq}^8$, the flat direction at LO  is rotated by the inclusion of higher-order corrections. We witness this effect at both the linear and quadratic levels of the SMEFT contributions. 

As shown in~\cref{sec:results}, EW corrections have a larger relative impact on the Drell-Yan process at the LHC and top-quark pair production in a muon collider scenario. We therefore anticipate even larger effects from these processes. To highlight this point,~\cref{fig:flat_other} illustrates a pair of selected two-dimensional planes in the Wilson coefficient space relevant for the Drell-Yan process at the LHC and top-quark pair production at a 10~TeV muon collider. In both cases, the highest energy bin of the differential distributions shown in~\cref{sec:LHCDY} and~\cref{sec:muCtt} is selected. We remind the reader that, in the case of muon collisions, this bin corresponds to the bulk of the cross-section.
\begin{figure}[ht]
    \centering
    \includegraphics[width=.48\textwidth]{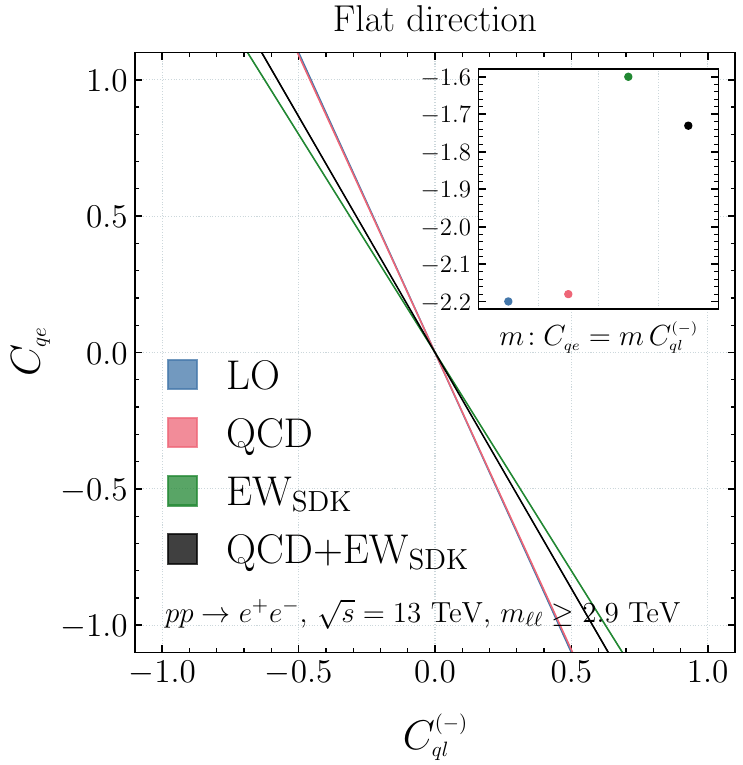}\hfill
    \includegraphics[width=.48\textwidth]{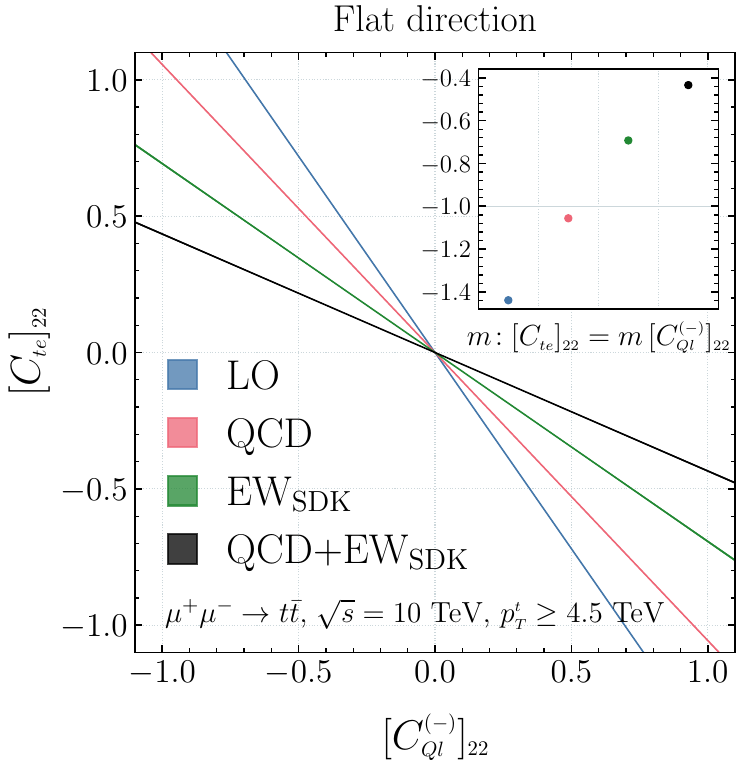}
    \vspace{-4mm}
    \caption{Same as \cref{fig:flat_tt} showing a selected 2D parameter planes for the Drell-Yan process at the LHC with $m_{\ell\ell}> 2900$ GeV (left), and for top-quark pair production at a 10~TeV muon collider with $p_{T}^{t} > 4500 \, \text{GeV}$ (right).}
\label{fig:flat_other}
\end{figure}
In the case of~\cref{fig:flat_tt}, we have chosen pairs of operators that lead to a vanishing LO cross-section when the values of the associated Wilson coefficients are anti-correlated, regardless of the bin considered, and therefore also at the inclusive-level. For the Drell-Yan process (left plot of~\cref{fig:flat_other}), and top-quark pair production at a 10~TeV muon collider (right plot of~\cref{fig:flat_other}), we simply select a pair of operators and identify the flat direction. This flat direction depends, at LO, on the bin selected.

For the bins considered, the impact of the EW corrections is particularly significant, especially for the top-quark pair production at a 10~TeV muon collider, compared to what was observed in~\cref{fig:flat_tt}, with large rotations induced by higher-order corrections. In the case of the Drell-Yan process at the LHC, QCD corrections have no impact on the flat direction and the entire effect is dominated by EW$_\text{SDK}$ contributions. Conversely, for $\mu^+\mu^-\to t\bar{t}$, QCD corrections already modify the correlations between the interference contributions of Wilson coefficients. However, the effects arising from EW$_\text{SDK}$ are more significant, with the combined result inducing the largest rotation.

The selected results discussed in this section underline the non-trivial nature of EW corrections in the SMEFT, again emphasising that a simplistic $K$-factor approach to estimating higher-order corrections will not capture the significant changes in the sensitivity of the data.
However, whilst it helps us to understand the relative impact of approximate EW corrections, whether such a `lifting' of blind directions represents the emergence of inherently new information in a global sense or a simpler shift/rotation in the parametric dependence of differential observables cannot be determined from this study alone.  

The model-independent spirit of the SMEFT approach motivates a global approach, accounting for all relevant Wilson coefficients at once. Flat directions should therefore be analysed in a global way, rather than looking at the individual benchmarks or lower-dimensional slices that have been shown so far. To address this, in the next section, we analyse the parameter space using the Fisher Information, and we will see the qualitative effects shown in this section be replicated at a more global level.

Before moving on, we note that, although the global approach is the right one for the SMEFT, matching concrete UV models often significantly reduces the number of free parameter imposing relations among the Wilson coefficients. For instance,~\cref{fig:axial_vector_combinations} demonstrates that if potential NP were to predominantly affect the colour octet $AV$ or $VA$ coupling combinations, EW corrections would induce an $\mathcal{O}(\Lambda^{-2})$ sensitivity in $p_{T}^{t}$ distributions of $t \bar t$ production at the LHC, which would not be the case at LO. 

\subsubsection{Global approach using the Fisher Information} 
\label{sec:fisher}
The LO, QCD and ${\rm EW_{\rm SDK}}$ predictions calculated in this work all depend on the same four-fermion operators that we have considered. By construction, no new Wilson coefficients can enter via NLO corrections. In the previous section, we demonstrated how higher-order corrections appreciably modify the relevant linear combinations of Wilson coefficients to which the data are sensitive, in particular, how they can lift and/or rotate LO flat directions. However, this analysis does not quantify whether the corrections bring genuinely new information in a global sense, beyond a rotation and/or dilatation in the space of constrained directions. 

Consider, for example, a process that is sensitive to some number of independent linear combinations of Wilson coefficients at LO. Whilst higher-order corrections may significantly modify these directions, the NLO results may still only depend on the same number of independent directions. The predictions are more accurate/precise, and may indeed better constrain specific UV scenarios, but they do not provide new information in the model-independent global approach in which one considers all possible coefficients to exist simultaneously.
In this case, the relevant question to ask is whether the total number of independent linear combinations of Wilson coefficients on which the data depend changes from LO to NLO.  

To this end, we analyse the Fisher Information (FI) matrix, whose definition and relevant properties are briefly summarised in~\cref{sec:fisherAPP}. Here, it is sufficient to note that it is a square matrix in the space of parameters-of-interest, which in this case, are the relevant Wilson coefficients affecting the process at hand, $C_i=\{C_1,\dots,C_n\}$. The eigenvectors, $\hat{e}_i$, of the FI matrix correspond to linearly independent directions in the Wilson coefficient space that are constrained by the data. The degree to which they are constrained is represented by their eigenvalues, $\lambda_i$, which give the inverse of their variance. The expected 1-$\sigma$ statistical uncertainty on each linear combination of coefficients for $\Lambda=1$ TeV is therefore, $1/\sqrt{\mathcal{L}\lambda_i}$, where $\mathcal{L}$ is the integrated luminosity. Null eigenvalues, instead, are flat directions that cannot be constrained by the data. Since we only consider statistical uncertainties in our analysis, the $\lambda_i$ are proportional to the integrated luminosity, $\mathcal{L}$. Given that we are mainly interested in relative sensitivities, we will work with the FI scaled by $\mathcal{L}$, which has units of cross-section.

For each of the processes considered, we individually construct and diagonalise the FI matrix for the differential distributions previously presented, namely the transverse momentum of the top quarks, $p_T^{t}$, in top-quark pair production at the LHC or at 10~TeV muon collider, and the dilepton invariant mass, $m_{\ell \ell}$, for the Drell-Yan process at the LHC. 

It is important to note that this is a numerical method, and so flat directions can only be identified within numerical uncertainties. We observe that the potential to lift flat directions is process-dependent. Whilst the impact of EW corrections may not be very significant in some cases, such as in the case of top-quark pair production, it can still yield interesting outcomes in processes like Drell-Yan, as will be seen in the upcoming discussion. 

To reliably identify flat directions in the FI matrix, we estimate the uncertainty of the FI eigenvalues by generating toy samples of MC predictions, drawn from normal distributions with standard deviations defined by the MC uncertainties. We then evaluate the standard deviation of the FI eigenvalues over these generated samples, using this as a measure of their uncertainty.
Eigenvalues that are consistent with zero within two standard deviations are considered indicative of flat directions. However, achieving precise predictions can be challenging due to the eigenvalues spanning several orders of magnitude, which complicates the accurate determination of flat directions amidst significant uncertainties.

The three rows of~\cref{fig:flat_fisher} show the result of this analysis for the three processes studied in the previous section. In each row, the left panel displays the eigenvalues of the luminosity-normalised FI matrix evaluated at different perturbative orders along with their $2\sigma$ uncertainty evaluated as described above. Error bars crossing zero are those that extend to the lower axis, indicating that the associated eigenvector is a flat direction. The remaining three panels display the predictions of the first four eigenvectors for the kinematic distribution used to construct the FI matrix, normalised to the SM. These are given for coefficient values corresponding to unit-normalised eigenvectors with $\Lambda=1$ TeV, again at different perturbative orders. As we will discuss below, they can be useful to highlight the lifting of flat directions due to higher-order corrections.
\begin{figure}[ht]
    \centering
    \includegraphics[width=\textwidth,trim={0 0.8cm 2cm 0cm},clip]{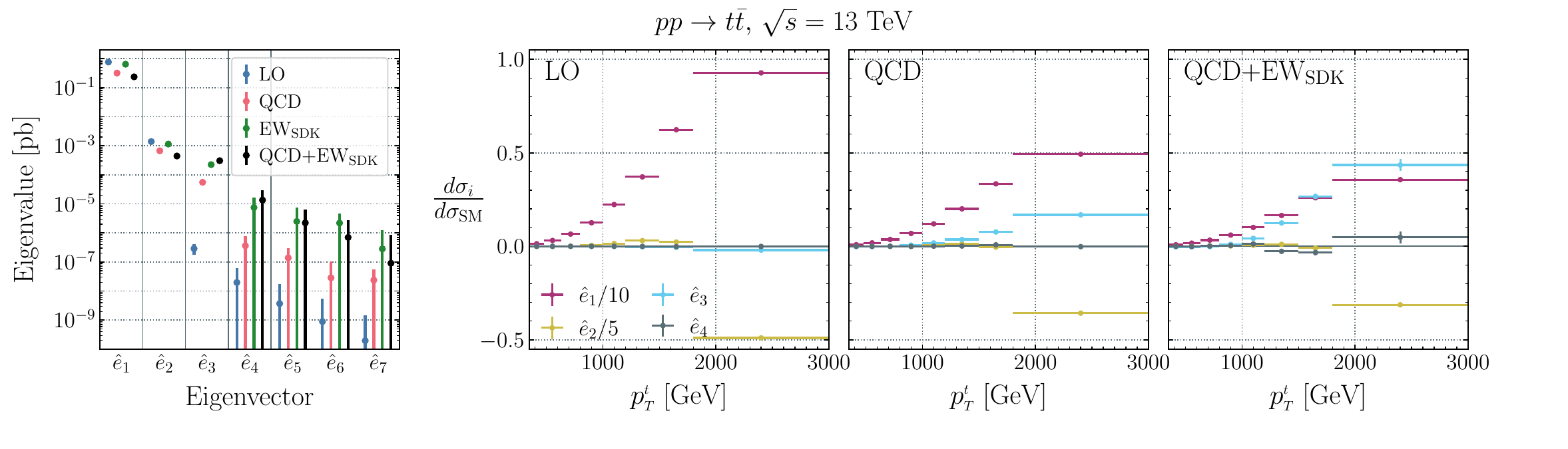}
    \\[3mm]
    \includegraphics[width=\textwidth,trim={0 0.8cm 2cm 0cm},clip]{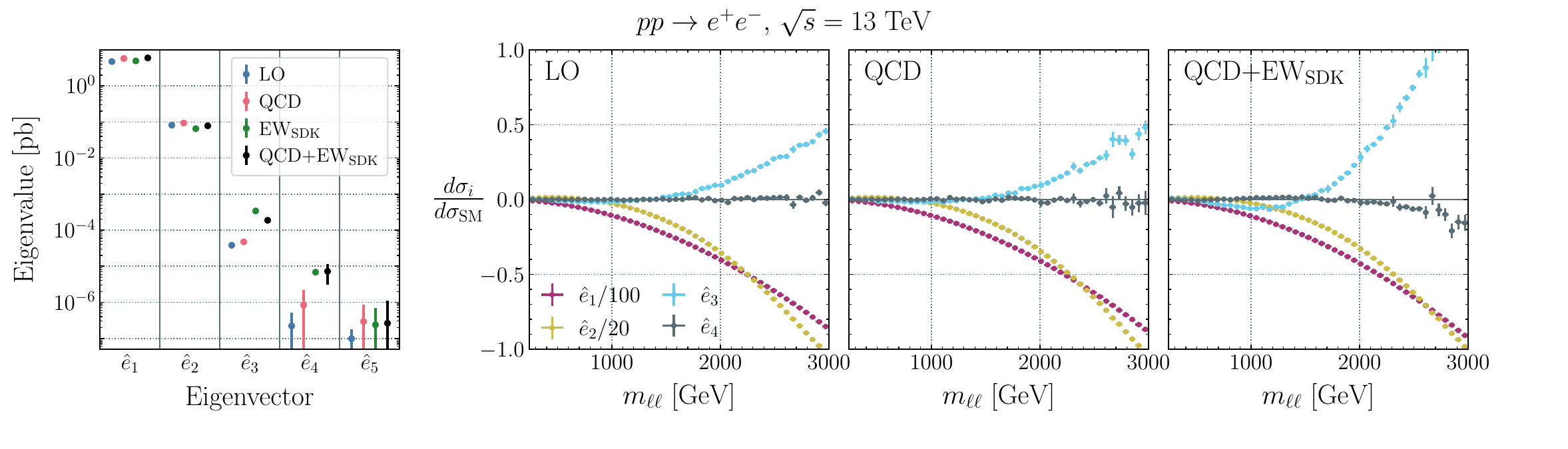}
    \\[3mm]
    \includegraphics[width=\textwidth,trim={0 0.8cm 2.3cm 0cm},clip]{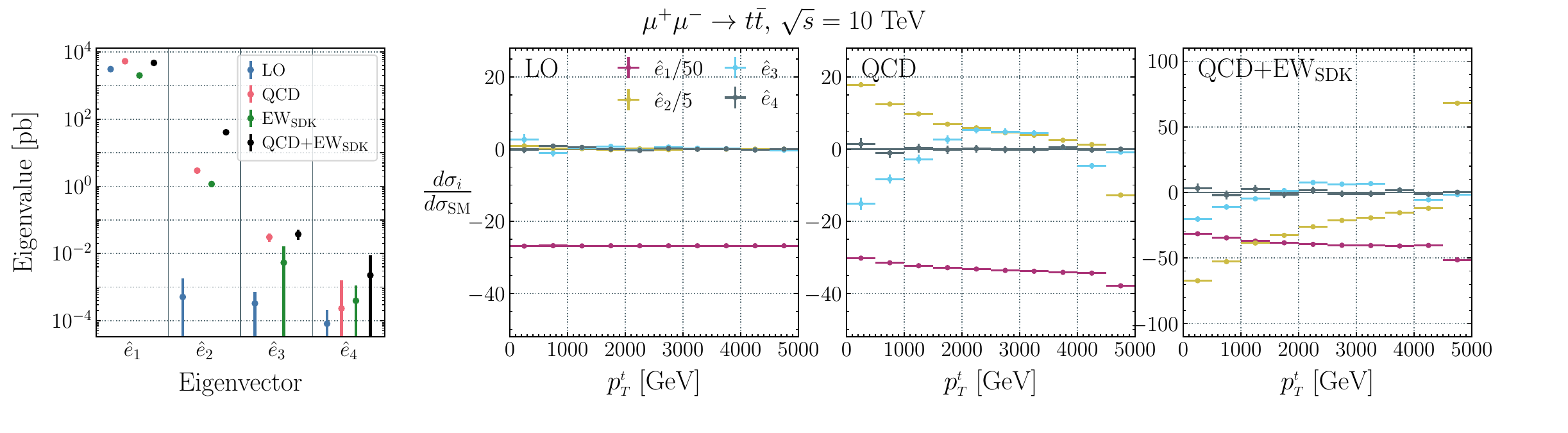}
    \vspace{-4mm}
    \caption{Eigensystems of expected Fisher information matrices derived from the differential distributions presented in~\cref{sec:LHCtt}--\cref{sec:muCtt}: \textbf{top:} $p_{\sss T}^{t}$ in $t\bar{t}$ production at the LHC, \textbf{middle:} $m_{\ell \ell}$ in the Drell-Yan process at the LHC, \textbf{bottom:} $p_{\sss T}^{t}$ in $t\bar{t}$ production at a 10~TeV muon collider. The left panels display the eigenvalues of the system with 2$\sigma$ MC uncertainties for LO, QCD, EW$_{\text{SDK}}$, and QCD+EW$_{\text{SDK}}$ predictions, coloured in blue, red, green, and black, respectively. The three other panels illustrate the ratio of SMEFT predictions to the SM for the first four eigenvectors at the perturbative orders specified in the upper left corner of each panel.\label{fig:flat_fisher}}
\end{figure}

\paragraph{Top-quark pair production at the LHC} 
The results for this process are shown in the upper row of~\cref{fig:flat_fisher}. This process differs from the other two in being dominated by the $gg$ initial state, which does not receive SMEFT corrections from four-fermion operators. Moreover, it also involves coloured particles in both initial and final states, leading to significant QCD corrections in general.
\begin{table}[ht]
    \centering
\begin{tabular}{|c|c|c|c|c|c|c|c|c|}
\hline
$\hat{e}_i$&$C_{\scriptscriptstyle Qq}^{\scriptscriptstyle 1,8}$&$C_{\scriptscriptstyle Qq}^{\scriptscriptstyle 3,8}$&$C_{\scriptscriptstyle Qu}^{\scriptscriptstyle 8}$&$C_{\scriptscriptstyle Qd}^{\scriptscriptstyle 8}$&$C_{\scriptscriptstyle td}^{\scriptscriptstyle 8}$&$C_{\scriptscriptstyle tq}^{\scriptscriptstyle 8}$&$C_{\scriptscriptstyle tu}^{\scriptscriptstyle 8}$&$\lambda_i$ [pb]\tabularnewline\hline
\multicolumn{9}{|l|}{LO}\tabularnewline\hline
$\hat{e}_1$&0.56&0.15&0.36&0.21&0.21&0.56&0.36&$7.6\times 10^{-1}$\tabularnewline\hline
$\hat{e}_2$&-0.12&0.69&0.29&-0.41&-0.41&-0.12&0.29&$1.4\times 10^{-3}$\tabularnewline\hline
$\hat{e}_3$&-0.56&0.32&0.18&0.51&0.49&-0.13&0.19&$2.9\times 10^{-7}$\tabularnewline\hline
\multicolumn{9}{|l|}{QCD}\tabularnewline\hline
$\hat{e}_1$&0.55&0.14&0.36&0.21&0.21&0.58&0.35&$3.2\times 10^{-1}$\tabularnewline\hline
$\hat{e}_2$&-0.09&0.73&0.22&-0.36&-0.40&-0.14&0.31&$6.7\times 10^{-4}$\tabularnewline\hline
$\hat{e}_3$&0.57&0.01&-0.10&-0.18&0.38&-0.67&0.20&$5.6\times 10^{-5}$\tabularnewline\hline
\multicolumn{9}{|l|}{EW$_{\text{SDK}}$}\tabularnewline\hline
$\hat{e}_1$&0.54&0.17&0.37&0.22&0.22&0.56&0.38&$6.4\times 10^{-1}$\tabularnewline\hline
$\hat{e}_2$&-0.17&0.63&0.29&-0.42&-0.42&-0.13&0.36&$1.1\times 10^{-3}$\tabularnewline\hline
$\hat{e}_3$&-0.69&-0.16&0.13&0.16&0.42&0.03&0.53&$2.3\times 10^{-4}$\tabularnewline\hline
\multicolumn{9}{|l|}{QCD+EW$_{\text{SDK}}$}\tabularnewline\hline
$\hat{e}_1$&0.51&0.17&0.38&0.22&0.23&0.57&0.38&$2.4\times 10^{-1}$\tabularnewline\hline
$\hat{e}_2$&0.01&0.67&0.17&-0.39&-0.50&-0.16&0.30&$4.5\times 10^{-4}$\tabularnewline\hline
$\hat{e}_3$&-0.48&0.02&0.16&0.04&0.51&-0.31&0.62&$3.1\times 10^{-4}$\tabularnewline\hline
\end{tabular}
    \caption{Composition of the first three eigenvectors of the luminosity-normalised Fisher Information matrix in the space of colour-octet four-fermion operators. The eigenvectors are constructed from the $p_T^{t}$ distribution in $t\bar{t}$ production at the LHC, as shown in the top row of~\cref{fig:flat_fisher}. They are shown at different perturbative orders alongside their eigenvalues in units of pb.}
    \label{tab:eigs} 
\end{table}
At LO, there are clearly two non-trivial constraints over the 7-dimensional space of coefficients, which can be seen to predict significant energy growing effects in the second panel from the left. A third direction appears to be constrained, but its associated eigenvalue is nearly four orders of magnitude smaller than the next best constrained direction. With the full 3 ab$^{-1}$ of LHC luminosity, the constraint on this direction is equivalent to a new physics scale of 700 GeV, assuming Wilson coefficients of $\mathcal{O}(1)$. Given that this analysis only considers statistical uncertainties, we can safely assume that this will be a flat direction for all intents and purposes. The remaining directions are approximately flat, reflecting the fact that data are not able to constrain them. Higher-order corrections, either QCD or EW$_{\text{SDK}}$ separately, or both combined, are able to lift the third eigen-direction which we argued to be initially flat at LO, as seen on the most left panel in the upper row of~\cref{fig:flat_fisher}. The complete QCD+EW$_\text{SDK}$ predictions allow for it to be constrained almost as well as the second eigenvalue.

~\Cref{tab:eigs} shows the composition of the constrained eigenvectors at the different perturbative orders. As expected, at LO, the two eigenvectors are orthogonal linear combinations of the vectorial coupling combinations $C^u_{VV}$ and $C^d_{VV}$. The first is essentially unchanged when including QCD corrections, in fact, it stays approximately the same regardless of the perturbative order. The second eigenvector, by contrast, varies significantly across different perturbative orders. As a consequence, $\hat{e}_1$, and especially $\hat{e}_2$, are no longer simple linear combinations of $C^u_{VV}$ and $C^d_{VV}$, but instead acquire components from all other coupling combinations. This is the 7-dimensional analogue of the flat direction analysis presented in~\cref{sec:flattt}. As discussed in~\cref{sec:LHCtt}, QCD corrections tend to be larger for the SM than for the four-fermion operator contributions, leading to reduced sensitivity. This is reflected in the smaller eigenvalues of the two constrained eigenvectors compared to those at LO.

In the presence of higher-order corrections, a third direction is probed. The eigenvalue for this third direction is significantly smaller than those of the first two eigen-directions in the presence of only QCD or EW$_{\text{SDK}}$ corrections, but becomes similar to $\lambda_2$ for QCD+EW$_{\text{SDK}}$, showing a comparable sensitivity. We observe that the remaining eigenvalues are consistent with zero based on our analysis. The existence of potential additional flat directions that may have been lifted cannot be confidently determined with the level of numerical accuracy employed. Assessing their statistical significance would necessitate a substantial investment in computational resources and time. Given that our MC uncertainty exceeds the precision achievable in foreseeable experiments, pursuing further numerical investigation is considered outside the current scope of this work.

\paragraph{The Drell-Yan process at the LHC} The results for this process are plotted in the middle row of~\cref{fig:flat_fisher}, for which a space of five Wilson coefficients are studied. The left panel shows the eigenvalues of the luminosity-normalised FI matrix at different orders in perturbation theory. At LO, we observe the existence of three constrained directions, whilst the last two eigenvalues are consistent with zero, and are therefore identified as flat directions. 

The four-fermion operators at hand can mediate interactions involving the right-handed up, right-handed down, or left-handed doublet quark currents, leading to amplitudes that grow identically with energy ($\propto s$). These can only be distinguished once the matrix elements are convoluted with the parton luminosities of different quark-antiquark initial states. Since we have assumed flavour universal Wilson coefficients, these correspond to the $u\bar{u}$, $d\bar{d}$ and sea quark-antiquark luminosities, each of which is expected to yield a different shape in $m_{\ell\ell}$ when convoluted with the four-fermion operator matrix elements. At LO, and at $\ord(\Lambda^{-2})$ in the SMEFT expansion, the three constrained directions should correspond to some linear combinations of these partonic contributions, made distinguishable by the fact that the underlying parton luminosities vary differently with the partonic centre-of-mass energy. 
The second panel from the left shows the dilepton invariant mass distributions of the first four eigenvectors.
The fact that $\hat{e}_{\sss 1-3}$ display a clear growth with energy is consistent with the fact that the invariant mass distribution is able to constrain them. The fourth eigenvector is, instead, flat and seems to predict no effect in this observable, within MC errors. This behaviour persists when considering NLO QCD-accurate predictions shown in the third panel from the left. However, when including EW$_{\text{SDK}}$ corrections, the fourth eigenvector shows an energy growth, as seen in the rightmost panel, indicating that this direction can now be constrained. This exemplifies how EW$_{\text{SDK}}$ corrections can potentially lift globally flat directions, enhancing our ability to constrain the Wilson coefficient space, compared to LO- or NLO QCD-accurate SMEFT predictions.
This can be understood from a matrix-element point of view, where EW corrections introduce an additional form of energy dependence proportional to $\log(s/\MW^2)$, or $\log^2(s/\MW^2)$, with respect to the Born amplitude, enabling the data to distinguish a new combination of coefficients.

~\Cref{tab:eigs_DY} reports the composition of the constrained eigenvectors in our analysis. One can see that $\hat{e}_{1-3}$ are essentially unchanged from LO to NLO QCD, reinforcing the fact that QCD corrections do not alter the picture besides providing slightly stronger constraints. EW$_{\text{SDK}}$ corrections, instead, rotate the space of constrained directions significantly modifying $\hat{e}_{3}$. Additionally, EW corrections introduce a sensitivity, albeit weak, to a fourth direction in the parameter space. The full prediction at QCD+EW$_{\text{SDK}}$ accuracy resembles the eigensystem constrained by the EW$_{\text{SDK}}$ predictions. We see that $\hat{e}_4$ at this order is more closely aligned with $\hat{e}_3$ eigenvector of the LO/QCD predictions. The newly constrained direction from EW$_{\text{SDK}}$ corrections corresponds to $\hat{e}_3$ of the full prediction, and it is  better constrained than the $\hat{e}_3$ eigenvector at LO/QCD. This underlines the fact that including EW corrections leads to new and non-trivial constraining power in this process.
\begin{table}[ht]
    \centering
\begin{tabular}{|c|c|c|c|c|c|c|}
\hline
$\hat{e}_i$&$C_{\scriptscriptstyle ql}^{\scriptscriptstyle (-)}$&$C_{\scriptscriptstyle ql}^{\scriptscriptstyle (3)}$&$C_{\scriptscriptstyle qe}$&$C_{\scriptscriptstyle ul}$&$C_{\scriptscriptstyle ue}$&$\lambda_i$ [pb]\tabularnewline\hline
\multicolumn{7}{|l|}{LO}\tabularnewline\hline
$\hat{e}_1$&0.27&-0.79&0.20&0.23&0.46&$4.8\times 10^{0}$\tabularnewline\hline
$\hat{e}_2$&0.74&0.53&0.04&0.21&0.35&$8.2\times 10^{-2}$\tabularnewline\hline
$\hat{e}_3$&-0.44&0.19&0.15&0.86&0.10&$3.8\times 10^{-5}$\tabularnewline\hline
\multicolumn{7}{|l|}{QCD}\tabularnewline\hline
$\hat{e}_1$&0.27&-0.79&0.20&0.23&0.46&$5.8\times 10^{0}$\tabularnewline\hline
$\hat{e}_2$&0.74&0.53&0.04&0.21&0.35&$9.4\times 10^{-2}$\tabularnewline\hline
$\hat{e}_3$&-0.45&0.20&0.19&0.84&0.11&$4.8\times 10^{-5}$\tabularnewline\hline
\multicolumn{7}{|l|}{EW$_{\text{SDK}}$}\tabularnewline\hline
$\hat{e}_1$&0.22&-0.80&0.19&0.22&0.47&$4.9\times 10^{0}$\tabularnewline\hline
$\hat{e}_2$&0.60&0.55&0.05&0.24&0.52&$6.5\times 10^{-2}$\tabularnewline\hline
$\hat{e}_3$&0.69&-0.17&0.03&0.16&-0.69&$3.4\times 10^{-4}$\tabularnewline\hline
$\hat{e}_4$&-0.33&0.09&0.00&0.93&-0.14&$6.8\times 10^{-6}$\tabularnewline\hline
\multicolumn{7}{|l|}{QCD+EW$_{\text{SDK}}$}\tabularnewline\hline
$\hat{e}_1$&0.23&-0.80&0.19&0.22&0.47&$6.0\times 10^{0}$\tabularnewline\hline
$\hat{e}_2$&0.63&0.55&0.05&0.24&0.49&$7.8\times 10^{-2}$\tabularnewline\hline
$\hat{e}_3$&0.60&-0.17&0.05&0.27&-0.73&$1.9\times 10^{-4}$\tabularnewline\hline
$\hat{e}_4$&-0.43&0.13&0.14&0.88&-0.04&$7.3\times 10^{-6}$\tabularnewline\hline
\end{tabular}
    \caption{Same as~\cref{tab:eigs}, showing the constrained eigenvectors in the space of four-fermion operators constructed from the $m_{\ell\ell}$ distribution in the Drell-Yan process at the LHC, as shown in the middle row of~\cref{fig:flat_fisher}.}
     \label{tab:eigs_DY} 
\end{table}
\paragraph{Top-quark pair production at a muon collider} We carry out this study using the $p^{t}_{T}$ distributions, as shown in the bottom row of~\cref{fig:flat_fisher}. Here, out of four Wilson coefficients, only one direction can be probed at LO, which can be seen to correspond to an overall rescaling of the cross-section in the second panel from the left, with no dependence on the kinematic variable. This is expected from the fact that the muon collider operates at fixed energy, and these operators mediate amplitudes proportional to Mandelstam $s$ only, with no non-trivial angular dependence at LO. This can be seen in~\cref{fig:muC_distr_tl} in the main text, as well as~\cref{fig:muC_distr_te_Qe_qlm} in~\cref{sec:muC_distr_remaining}. We observe that QCD and EW corrections, separately, can lift the LO degeneracy, with one and two additional directions being constrained when including EW and QCD corrections, respectively, as shown in the leftmost panel. The combined QCD+EW$_{\text{SDK}}$ result maintains three non-trivially constrained directions, the latter two of which are better constrained than in the individual QCD and EW$_\text{SDK}$ cases. The two rightmost panels of the corresponding plot in~\cref{fig:flat_fisher} show how the perturbative corrections lead to non-trivial shapes in the kinematic variable for all constrained eigenvectors, whilst the fourth remains flat within 2$\sigma$ of the MC uncertainties. 

~\Cref{tab:eigs_mumutt} shows the composition of the constrained eigenvectors in our analysis. 
\begin{table}[ht]
    \centering
\begin{tabular}{|c|c|c|c|c|c|}
\hline
$\hat{e}_i$&$[C_{\scriptscriptstyle Qe}]_{\scriptscriptstyle 22}$&$[C_{\scriptscriptstyle Ql}^{\scriptscriptstyle (-)}]_{\scriptscriptstyle 22}$&$[C_{\scriptscriptstyle te}]_{\scriptscriptstyle 22}$&$[C_{\scriptscriptstyle tl}]_{\scriptscriptstyle 22}$&$\lambda_i$ [pb]\tabularnewline\hline
\multicolumn{6}{|l|}{LO}\tabularnewline\hline
$\hat{e}_1$&0.14&0.78&0.54&0.27&$3.1\times 10^{3}$\tabularnewline\hline
\multicolumn{6}{|l|}{QCD}\tabularnewline\hline
$\hat{e}_1$&0.17&0.72&0.63&0.24&$5.3\times 10^{3}$\tabularnewline\hline
$\hat{e}_2$&0.09&-0.66&0.74&-0.06&$2.9\times 10^{0}$\tabularnewline\hline
$\hat{e}_3$&-0.58&-0.14&0.01&0.80&$3.1\times 10^{-2}$\tabularnewline\hline
\multicolumn{6}{|l|}{EW$_{\text{SDK}}$}\tabularnewline\hline
$\hat{e}_1$&0.16&0.51&0.78&0.31&$2.0\times 10^{3}$\tabularnewline\hline
$\hat{e}_2$&0.03&0.85&-0.52&-0.12&$1.2\times 10^{0}$\tabularnewline\hline
\multicolumn{6}{|l|}{QCD+EW$_{\text{SDK}}$}\tabularnewline\hline
$\hat{e}_1$&0.19&0.49&0.81&0.25&$4.7\times 10^{3}$\tabularnewline\hline
$\hat{e}_2$&-0.05&0.85&-0.52&0.07&$4.1\times 10^{1}$\tabularnewline\hline
$\hat{e}_3$&-0.59&-0.12&-0.03&0.80&$3.8\times 10^{-2}$\tabularnewline\hline
\end{tabular}
    \caption{Same as~\cref{tab:eigs}, showing the constrained eigenvectors in the space of four-fermion operators constructed from the $p_T^{t}$ distribution in $t\bar{t}$ production at a 10 TeV muon collider, as shown in the bottom row of~\cref{fig:flat_fisher}.}
    \label{tab:eigs_mumutt}  
\end{table}
We see that the decomposition of the best constrained direction, $\hat{e}_1$, is modified by higher-order corrections, in particular the EW ones. The eigenvalue $\lambda_1$ remains of the same order. Instead, the additionally constrained directions when including QCD corrections are quite different from the additional direction when including only EW$_{\text{SDK}}$ corrections, indicating that the latter predictions do probe an independent part of the parameter space. The resulting space of constrained directions appears to be a combination of the directions probed by the two individual corrections. The second eigenvector, $\hat{e}_2$, is somewhat aligned with the new direction probed by EW corrections, whilst the third is similar to $\hat{e}_3$ of the QCD predictions.  The full prediction provide the best overall sensitivity. In particular, $\lambda_{2}$ improves by a factor of 14 (34) w.r.t. the individual QCD (EW$_{\text{SDK}}$) predictions. 

By comparing the eigenvalues listed in~\cref{tab:eigs} and~\cref{tab:eigs_DY} against~\cref{tab:eigs_mumutt}, it is manifest that the capabilities of a muon collider in constraining the operators considered are much stronger than the case of the LHC, even without considering its ability to accumulate larger luminosity. Indeed, as already mentioned, the eigenvalues correspond to the inverse of the square of the $1\sigma$ uncertainty for every inverse $\rm pb$ of accumulated luminosity. The origin of this difference is twofold. First, the larger the energy, the larger the effect of EWSL and therefore the possibility of inducing new independent directions. Second, and this is the most significant aspect, whilst in the case of the LHC, the regions at high $p_T^t$ or high $m_{\ell\ell}$ correspond to the tails of the distributions, with very few events, in the case of a muon collider, the high $p_T^t$ region corresponds to the bulk of the cross-section, where EWSL are large and almost all the events are present.

To summarise, our Fisher information analysis has quantified how higher-order corrections can constrain new, independent linear combinations of Wilson coefficients at $\mathcal{O}(\Lambda^{-2})$ in the SMEFT expansion. In other words, their inclusion can allow data to access genuinely new information in the SMEFT parameter space. This was shown for all three processes considered in this study. For the Drell-Yan process at the LHC, the entire sensitivity is driven by the Sudakov EW corrections, whilst in top-quark pair production at the LHC and at a muon collider, both QCD corrections and EW$_{\text{SDK}}$ are responsible for the new information and consequently constraining directions in the SMEFT parameter space. However, in the case of the muon collider, the resulting space of constrained directions appears similar to a combination of the individual eigensystems of the two correction types, with an overall improved sensitivity. Moreover, since the bulk of the cross-section is associated to large values of $s$ in a muon collider, all the eigenvalues ($1\sigma$ uncertainties) are much larger (smaller) than at the LHC, both at LO, as well as when including higher-order corrections. 

Finally, we remind the reader that this analysis considers only the linear, or interference, contributions of the Wilson coefficients. Including quadratic terms introduces a dependence of the Fisher information on the specific point in parameter space under consideration, rendering general statements more difficult. A more comprehensive study, which incorporates these additional EFT quadratic contributions, is beyond the scope of this work.

\section{Summary and conclusions}
\label{sec:conclusions}
We investigated NLO EW corrections within the SMEFT in the high-energy limit, where these corrections are most significant and, conveniently, simpler to compute. At one loop, EW corrections are predominantly driven by Sudakov logarithms, referred to as EWSL throughout this paper. In SM, NLO EW corrections have been extensively studied, with their logarithmic components extractable for arbitrary processes using the algorithm developed by Denner and Pozzorini. This algorithm has been revisited and automated in {\aNLO}, enabling efficient computation of these corrections.

In this work, we have applied the {\denpoz} algorithm to the SMEFT. After reviewing NLO EW corrections, both in general and specifically in the high-energy limit, we have examined the principal theoretical aspects, initially focusing on amplitudes and subsequently on physical cross-sections. 

The {\denpoz} algorithm is applicable to all processes provided that their LO amplitude is not suppressed by powers of \(M_W / \sqrt{s}\), meaning it does not vanish in the \(s \to \infty\) limit. Whilst such cases are relatively rare in the SM, they are more prevalent in the SMEFT, particularly in scenarios where effective operators containing the Higgs doublet generate \(\mathcal{O}(v)\) Feynman rules when the Higgs field is non-dynamical. 
By restricting our focus to SMEFT amplitudes that are not mass-suppressed, we show that the computation of Sudakov logarithms in the SMEFT is structurally similar to that in the SM. We also provide an explicit counterexample for the top-quark chromomagnetic dipole operator in~\cref{sec:ctg}.

We have detailed and validated our computational setup. Subsequently, we conducted full-NLO accuracy computations, comprising exact NLO QCD and NLO EW corrections in the Sudakov approximation, for selected representative SMEFT processes: top-quark pair production and the Drell-Yan process at the LHC, as well as top-quark pair production at a future 10~TeV muon collider, incorporating four-fermion operators. 

As expected, NLO EW corrections become increasingly significant at high energies and, in some cases, surpass the magnitude of NLO QCD effects. For the Drell-Yan process, we observed a partial cancellation between the typically positive NLO QCD corrections and the generally negative NLO EW corrections. Although the structure of NLO EW corrections in the SMEFT is similar to that in the SM, their impact is not. The effects depend on the specific operator, highlighting the limitations of using a simple \(K\)-factor approach. Rescaling SMEFT cross-sections by SM \(K\)-factors fails to capture the relevant physical effects, and we generally advise against this practice.

Moreover, we demonstrated the impact of higher-order corrections on the linear combinations of Wilson coefficients probed by the processes considered in this work. We found that both QCD and EW corrections can alter the linear combinations probed, thereby modifying the flat directions in the parameter space. This effect was explored for all three processes discussed in this study. EW corrections were found to have a non-negligible impact, often more significant than that observed with QCD corrections alone. 

To determine whether the higher-order corrections provide genuinely new information rather than merely rotating the probed directions, we performed an analysis based on the Fisher information matrix. Our results show that higher-order corrections indeed lift some of the flat directions in the parameter space of the Wilson coefficients, which are present at LO. Whilst QCD corrections alone lift some flat directions for all three processes, we found that for the Drell-Yan process, EW corrections lift an additional flat direction. 

This analysis further strengthens our argument against the use of a simplistic \(K\)-factor approach for obtaining higher-order EW corrections in the SMEFT.
This work presents the first automated extraction of NLO EW corrections, albeit restricted to the high-energy limit, for selected relevant processes in the SMEFT. Our implementation, carried out within \textit{\aNLO}, is computationally efficient and numerically stable. The corresponding code will be publicly released, with the aim of facilitating the inclusion of NLO EW effects in future SMEFT analyses in a convenient and effective manner.

Beyond its intrinsic interest, we regard this work as a preliminary step towards the evaluation and automation of exact NLO EW corrections for arbitrary processes in the SMEFT, a task that is currently advancing within the community.

\section*{Acknowledgements}
We are grateful to Hua-Sheng Shao and Valentin Hirschi for their helpful discussions. H.F., C.S., and E.V.\ are supported by the European Union’s Horizon 2020 research and innovation programme under the EFT4NP project (grant agreement no.\@ 949451) and by a Royal Society University Research Fellowship through grant URF/R1/201553. K.M.\ is supported by an Ernest Rutherford Fellowship from the STFC, Grant No.\ ST/X004155/1. The work of M.Z.\ has been partly supported by the ``Programma per Giovani Ricercatori Rita Levi Montalcini'' granted by the Italian Ministero dell’Universit\`a e della Ricerca (MUR). D.P.~and M.Z. acknowledge financial support by the MUR through the PRIN2022 Grant 2022EZ3S3F, funded by the European Union – NextGenerationEU. D.P. acknowledges support from the DESY computing infrastructures provided through the DESY theory group. 

\appendix
\section{Validation of the {\denpoz} algorithm in the SMEFT}
\label{sec:validations}
In this section, we present the numerical results produced to validate the application of the {\denpoz} algorithm in the SMEFT. We focus on the subset of processes and operators studied in this work, rather than addressing the general case, since, as discussed in~\cref{sec:MSDP}, the {\denpoz} algorithm cannot be applied universally to arbitrary processes and operators.  

The MC implementation of the {\denpoz} algorithm, as formulated in Refs.~\cite{Denner:2000jv, Denner:2001gw} for the SM case, has already been considered in Ref.~\cite{Pagani:2021vyk} for its automation within {\aNLO}. 
Given the discussion of~\cref{sec:theo}, and particularly the one of~\cref{sec:dSMEWinSMEFT}, we expect the algorithmic procedure for the calculation of EWSL to be identical in the SM and in the SMEFT. In other words, the computation of the quantity $\delta^{\SM}_{\rm EW}$ remains unchanged between the two cases, the only difference being the tree-level amplitudes that are fed into the algorithm. 

For the sake of clarity and completeness, in this section, we explicitly demonstrate the agreement between the use of the {\denpoz} algorithm and the results obtained by taking the $\MW^2/s \to 0$ limit of an exact analytical calculation of the corresponding one-loop Feynman diagrams in the SMEFT. Our validation has been designed to be the simplest and most minimal comparison necessary to confirm the validity of the {\denpoz} algorithm in the SMEFT, for amplitudes that grow maximally with energy.

The evaluation of NLO cross-sections requires several components, which can be broadly classified as: \textit{(i)} tree-level amplitudes squared, which also constitute the LO contributions; \textit{(ii)} virtual amplitudes, {\it  i.e.}, loops diagrams, interfered with tree-level amplitudes, which are both UV and IR divergent; \textit{(iii)} UV counterterms interfered with the tree-level amplitudes to cancel the UV divergences; \textit{(iv)} real-emission amplitudes squared, which are themselves IR divergent and cancel the remaining IR divergences. 

Our comparison involves virtual one-loop amplitudes and their UV counterterms, which we evaluate using the implementation of the {\denpoz} algorithm in {\aNLO}~\cite{Pagani:2021vyk}. On the other hand, the analytical approach corresponds to explicitly computing the one-loop Feynman diagrams and taking the high-energy limit. We note that there is no need to validate both the dimension-six and the dimension-six-squared terms of the NLO cross-section. By inspecting~\cref{eq:NLO2-6limit,eq:NLO2-8limit}, we observe that once the validity of the {\denpoz} algorithm is confirmed for \(\NLO^{(8)}_2\), it is also implicitly confirmed for \(\NLO^{(6)}_2\). This is because the ingredients in the latter are the same as those in the former, with the addition of SM terms, for which the validity of the {\denpoz} algorithm has already been established.

Following these considerations, we selected two representative partonic processes: 
\begin{equation}
    u \bar u \to t \bar t\,, \hspace{1cm} u \bar u \to e^- e^+\,, 
\end{equation}
and, for each, calculated the high-energy limit of the interference of the dimension-six tree-level amplitude \(\MzNP\) with its one-loop EW correction  \(\MoNP\), denoted in~\cref{eq:NLO2-8limit} as:
\begin{equation}
    \lim_{\MW^2/s\to 0} \NLO^{(8)}_2\Big|_{\rm virt.}. \label{eq:for_validation}
\end{equation}
To perform our validation, the calculation was carried out using two different approaches:

\paragraph{The analytical approach} 
We perform the calculation of the virtual part of \(\MoNP\), which includes all the one-loop diagrams contributing to the process, using {\tt FeynArts}~\cite{Hahn:2000kx} and {\tt FeynCalc}~\cite{Kublbeck:1990xc, Shtabovenko:2016sxi, Shtabovenko:2020gxv}. The loop integrals are evaluated analytically with {\tt PackageX}~\cite{Patel:2015tea, Patel:2016fam}. The calculation is carried out in the HV scheme~\cite{tHooft:1972tcz} for \dimreg, with \(\gamma^5\) treated according to the BMHV prescription~\cite{Breitenlohner:1975hg}. UV counterterms contributing to  \(\MoNP\) are taken from the literature~\cite{Denner:1991kt}, and we confirmed the cancellation of UV poles. 
After completing the full calculation and interfering the amplitudes with the tree-level ones \(\MzNP\), we take the high-energy limit. For simplicity, all \(r_{kl}\) invariants are assumed to of the same order of \(s\), consistent with the original derivation of the {\denpoz} algorithm~\cite{Denner:2000jv, Denner:2001gw}. We retain the EWSL of the form \(\log^2(|r_{kl}|/M^2)\) and \(\log(|r_{kl}|/M^2)\), as in~\cref{eq:generallogs}, but neglect those of the form  \(\log(|r_{mn}|/|r_{pq}|)\) unless they multiply the aforementioned logarithms.

\paragraph{The {\denpoz} algorithm approach}  
We evaluate the quantity described in~\cref{eq:for_validation} using the {\denpoz} algorithm, as implemented in {\aNLO} and detailed in Ref.~\cite{Pagani:2021vyk}.

Notably, the implementation of Ref.~\cite{Pagani:2021vyk} preserves logarithms among invariants, \emph{i.e.}, \(\log(|r_{mn}|/|r_{pq}|)\). However, and as previously mentioned, our analytical approach omits these terms. Since the predictions for phenomenological results presented in~\cref{sec:results} account for such logarithms, we have decided to retain them in this approach. This choice is deemed appropriate, as we anticipate, \emph{a priori}, a minimal impact from these logarithmic contributions based on our choice of a fixed scattering angle, which ensures that \(|t| \simeq s\). A more detailed discussion of this point will be provided later in this section. 

We remind the reader that, for the purpose of this comparison, it is not necessary to perform any combination with real emissions, and it is sufficient to validate at the level of IR-divergent amplitudes. At the same time, we emphasise that the results presented in this section, whilst sufficient for validating the high-energy limit of NLO EW corrections in the SMEFT, are \textit{not} directly relevant for physical observables\footnote{~Phenomenological results presented in the following sections will employ the $\SDKw$ scheme introduced in~\cref{sec:sdkweak}, which is IR-finite, includes real emission contributions, and is suitable for physical predictions.}. 
The reason is that we use the {\denpoz} algorithm for the evaluation of the \textit{amplitudes}, an approach that has been dubbed as the `SDK approach' in Ref.~\cite{Pagani:2021vyk}. This approach inherently produces IR-divergent quantities and, whilst it correctly approximates the virtual contribution to NLO EW cross-sections, it is unsuitable for physical observables. 

Finally, since our comparison here involves IR-divergent amplitudes, a dependence on the regularisation scale is expected. In accordance with the choice already made in Refs.~\cite{Denner:2000jv, Denner:2001gw}, we set \(\mu = \sqrt{s}\).

Whilst in the original {\denpoz} algorithm the IR divergences are regulated by a fictitious photon mass (soft limit) or the fermion masses for the light flavours (collinear limit), in modern calculations, such regularisation is performed, similarly to the UV case, via dimensional regularisation in $d=4-2\epsilon$ dimensions. The corresponding formulae can be found for example in Refs.~\cite{Pagani:2021vyk, Lindert:2023fcu}. In the following discussion, it is understood that in the comparison between the analytical and the {\denpoz} algorithm, only the finite term, which is proportional to $\epsilon^0$, is considered, \emph{i.e}, we do not consider the $1/\epsilon$ poles.

In~\cref{fig:validations_4f}, we examine the partonic processes $u \bar{u} \to t \bar{t}$ and $u \bar{u} \to e^- e^+$, plotting the $\text{LO}|_{\text{VD}}$ squared SMEFT amplitude (we assign the ${\text{`VD'}}$ tag to the quantities used in this validation):
\begin{equation}
   \text{LO}|_{\text{VD}}\equiv\sum_{\text{col.}} \sum_{\text{hel.}} |\MzNP|^2, \label{eq:LOplot}
\end{equation}
which is shown by the solid lines in the main panel. We also plot the analytical $\text{NLO}|_{\text{VD}}$ squared amplitude in the high-energy limit:
\begin{equation}
    \text{NLO}|_{\text{VD}}\equiv\sum_{\text{col.}} \sum_{\text{hel.}} \left[ |\MzNP|^2 + \lim_{\MW^2/s\to 0} 2\Re\left(\MzNP\left(\MoNP\right)^*\right) \right], \label{eq:NLOplot}
\end{equation}
as dashed lines in the main panel. For clarity, instead of plotting the amplitude directly, we rescale it by $\Lambda^4/s^2$ to remove the leading dependence on $s$ and $\Lambda$ (see~\cref{eq:masssupprSMEFT}). The ratios of~\cref{eq:NLOplot} to~\cref{eq:LOplot} denoted as $\text{R}_{\text{VD}}$ can be simply written as:
\begin{equation}
    \text{R}_{\text{VD}}\equiv\frac{\text{NLO}|_{\text{VD}}}{\text{LO}|_{\text{VD}}}\label{eq:RVD}
\end{equation}
and are plotted in the first inset. The differences between the $\text{NLO}|_{\text{VD}}$ of~\cref{eq:NLOplot} (a quantity evaluated manually starting from one-loop amplitudes and taking the the high-energy limit) and the corresponding quantity utilising the {\denpoz} algorithm as implemented in {\aNLO} according to the SDK approach in Ref.~\cite{Pagani:2021vyk}, denoted as $\text{NLO}^{\text{DP}}|_{\text{VD}}$, is quantified as follows:
\begin{equation}
    \text{D}_{\text{VD}} \equiv \frac{\text{NLO}|_{\text{VD}} - \text{NLO}^{\text{DP}}|_{\text{VD}}}{\text{LO}|_{\text{VD}}}, \label{eq:DVD}
\end{equation}
and displayed in the second inset. We plot the aforementioned quantities as a function of $\sqrt{s}$ for a fixed value of the ratio $t/s$, corresponding to a final-state scattering angle of $75^\circ$. Different colours correspond to contributions from different operators, with the corresponding Wilson coefficient set to $C/\Lambda^2 = 1/\text{TeV}^2$, whilst all other coefficients are set to zero.
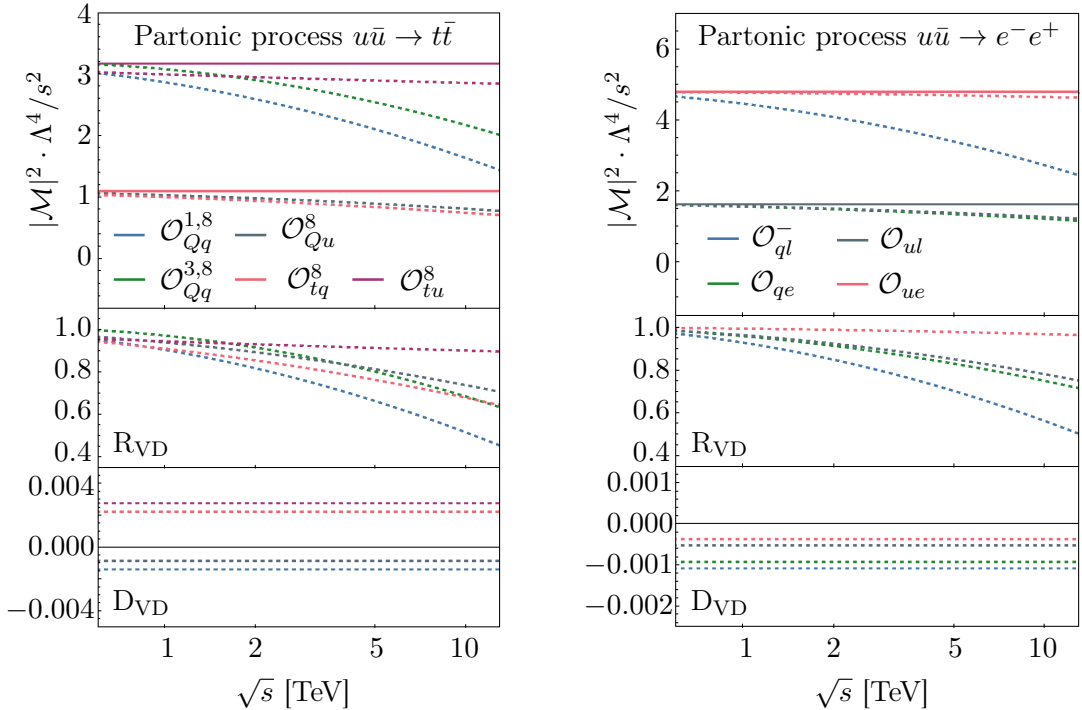
\begin{figure}[ht]
    \centering
    \begin{tikzpicture}
    \draw (0, 0) node[inner sep= 0] {\includegraphics[width=.495\textwidth]{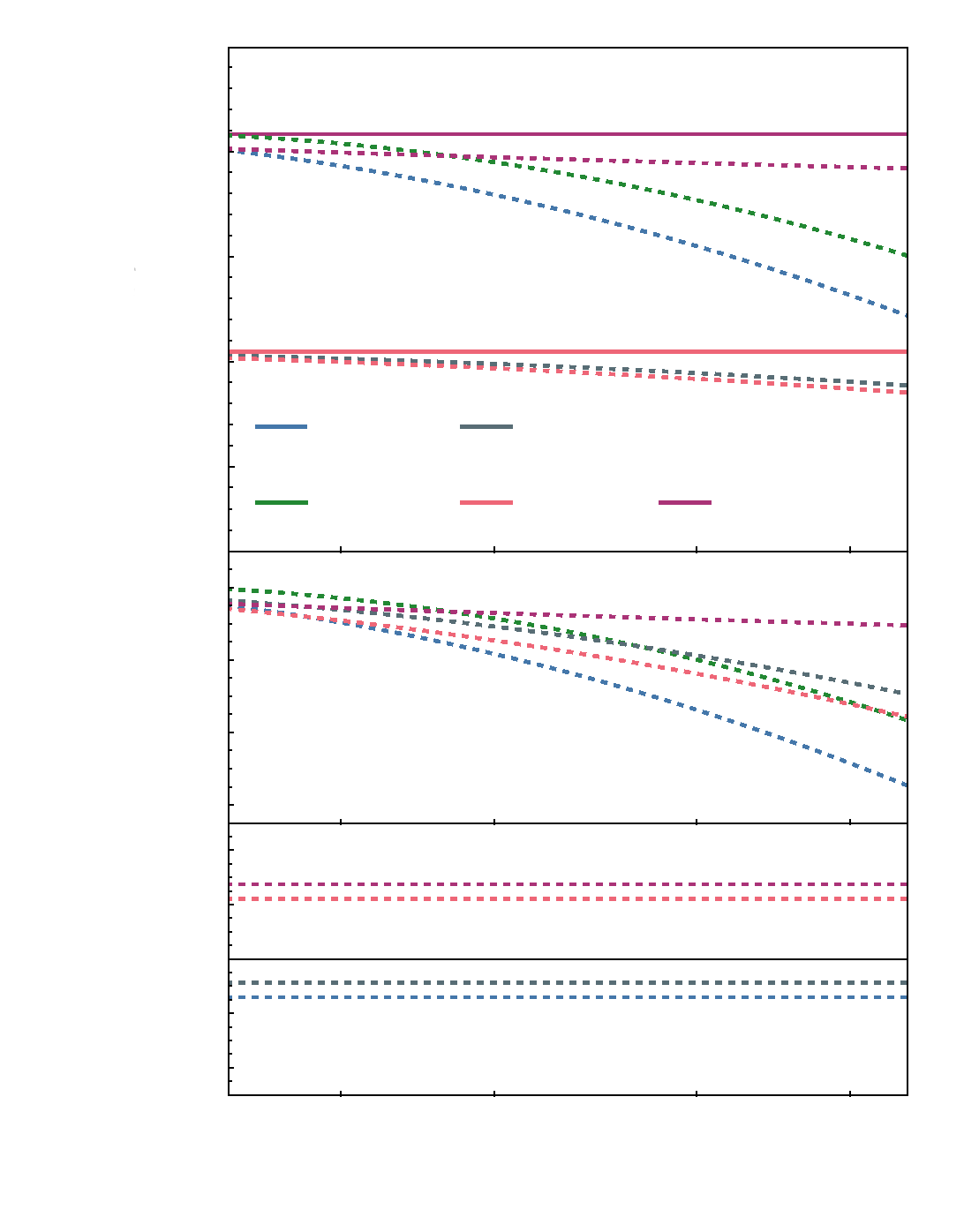}};
    \draw (0.6, 4.1) node {Partonic process $u \bar u \to t \bar t$};
    \draw (-2.15, 4.4) node {$4$};
    \draw (-2.15, 3.6) node {$3$};
    \draw (-2.15, 2.8) node {$2$};
    \draw (-2.15, 2.0) node {$1$};
    \draw (-2.15, 1.2) node {$0$}; 
    \draw (-0.8, 1.5) node {$\mathcal O_{Qq}^{1,8}$};
    \draw (-0.8, 0.85) node {$\mathcal O_{Qq}^{3,8}$};
    \draw (0.8, 1.5) node {$\mathcal O_{Qu}^8$};
    \draw (0.8, 0.85) node {$\mathcal O_{tq}^8$};
    \draw (2.3, 0.85) node {$\mathcal O_{tu}^8$};
    \draw (-2.7, 2.6) node [rotate=90] {$|\mathcal M|^2 \cdot \Lambda^4 / s^2$};
    \draw (-2.30, 0.25) node {$1.0$};
    \draw (-2.30, -0.3) node {$0.8$};
    \draw (-2.30, -0.85) node {$0.6$};
    \draw (-2.30, -1.40) node {$0.4$}; 
    \draw (-1.4, -1.3) node [rotate=0] {$\text{R}_{\text{VD}}$};
    \draw (-2.45, -1.85) node {$0.004$};
    \draw (-2.45, -2.65) node {$0.000$};
    \draw (-2.60, -3.5) node {$-0.004$};
    \draw (-1.4, -3.4) node [rotate=0] {$\text{D}_{\text{VD}}$};
    \draw (-1.05, -4.0) node {$1$};
    \draw (0.1, -4.0) node {$2$};
    \draw (1.7, -4.0) node {$5$};
    \draw (2.85, -4.0) node {$10$};
    \draw (0.6, -4.6) node {$\sqrt s$ [TeV]};
    \end{tikzpicture}
    \begin{tikzpicture}
    \draw (0, 0) node[inner sep= 0] {\includegraphics[width=.495\textwidth]{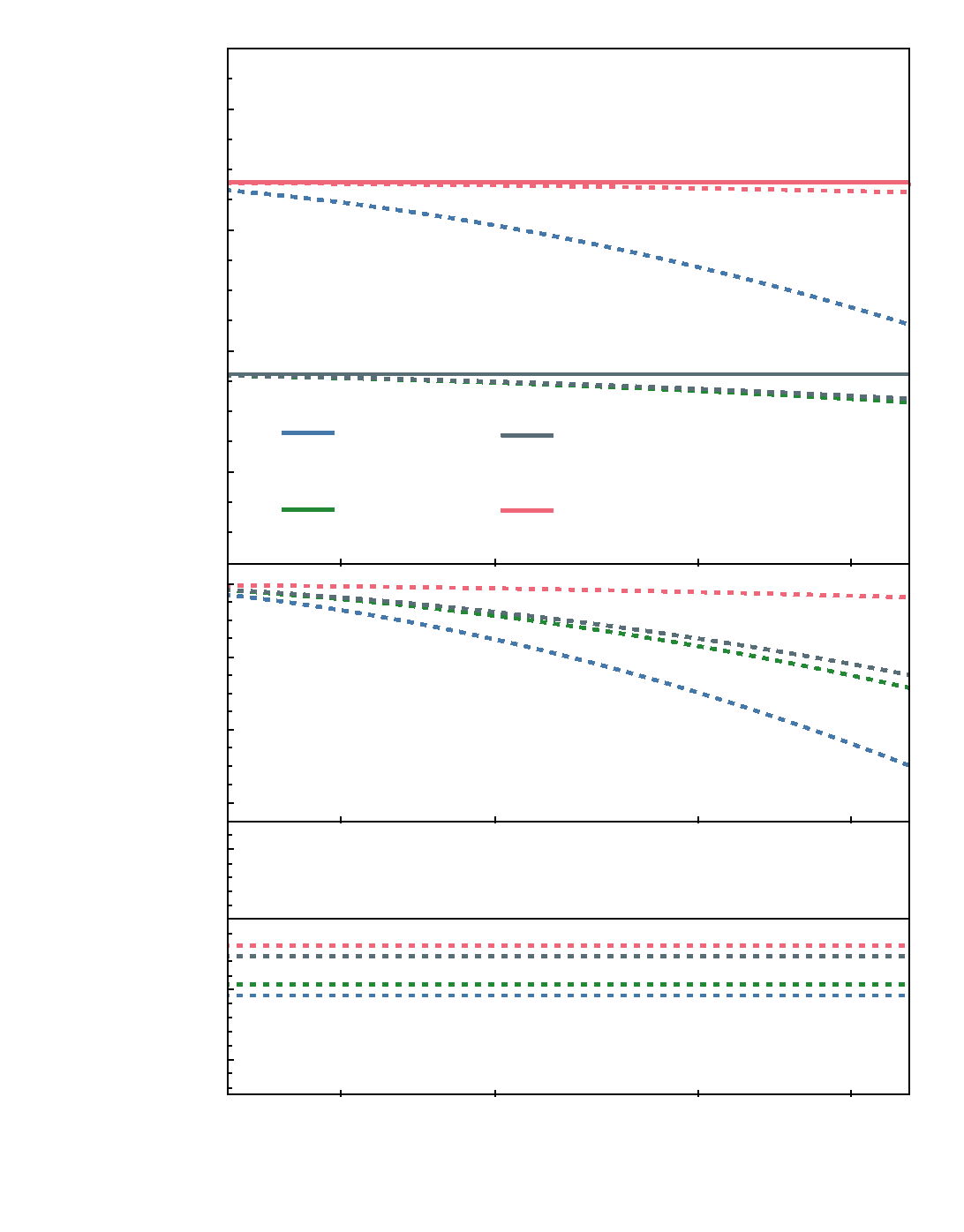}};
    \draw (0.7, 4.1) node {Partonic process $u \bar u \to e^- e^+$};
    \draw (-0.7, 1.4) node {$\mathcal O_{ql}^-$};
    \draw (-0.7, 0.8) node {$\mathcal O_{qe}$};
    \draw (1.0, 1.4) node {$\mathcal O_{ul}$};
    \draw (1.0, 0.8) node {$\mathcal O_{u e}$}; 
    \draw (-2.15, 3.95) node {$6$};
    \draw (-2.15, 3.00) node {$4$};
    \draw (-2.15, 2.05) node {$2$};
    \draw (-2.15, 1.10) node {$0$}; 
    \draw (-2.7, 2.6) node [rotate=90] {$|\mathcal M|^2 \cdot \Lambda^4 / s^2$};
    \draw (-2.30, 0.25) node {$1.0$};
    \draw (-2.30, -0.3) node {$0.8$};
    \draw (-2.30, -0.85) node {$0.6$};
    \draw (-2.30, -1.40) node {$0.4$};   
    \draw (-1.4, -1.3) node [rotate=0] {$\text{R}_{\text{VD}}$};
    \draw (-2.45, -1.8) node {$0.001$};
    \draw (-2.45, -2.35) node {$0.000$};
    \draw (-2.60, -2.9) node {$-0.001$};
    \draw (-2.60, -3.45) node {$-0.002$};
    \draw (-1.4, -3.4) node [rotate=0] {$\text{D}_{\text{VD}}$};
    \draw (-1.05, -4.0) node {$1$};
    \draw (0.1, -4.0) node {$2$};
    \draw (1.7, -4.0) node {$5$};
    \draw (2.85, -4.0) node {$10$};
    \draw (0.6, -4.6) node {$\sqrt s$ [TeV]};
    \end{tikzpicture}    
    \vspace{-8mm}
    \caption{Results of our validations for the partonic processes $u \bar{u} \to t \bar{t}$ (left) and $u \bar{u} \to e^- e^+$ (right) are presented. The main plot, generically labelled as $|\mathcal{M}|^2 \cdot \Lambda^4 / s^2$, features the quantities $\text{LO}|_{\text{VD}}$ (solid) and $\text{NLO}|_{\text{VD}}$ (dashed), as detailed in~\cref{eq:LOplot,,eq:NLOplot}, respectively. This considers individual contributions of the relevant four-fermion dimension-six insertions as discussed in~\cref{sec:OPandNOT}, one at a time. The first inset depicts the ratio $\text{R}_{\text{VD}}$, as defined in~\cref{eq:RVD}, whilst the second inset provides a comparison between the two approaches utilised, $\text{D}_{\text{VD}}$, as quantified in~\cref{eq:DVD} and discussed in the accompanying text.}
\label{fig:validations_4f}
\end{figure}
As can be seen in~\cref{fig:validations_4f}, the EWSL induce negative corrections of order $10$ to $50 \%$ 
of the LO at partonic centre-of-mass energies between $1$ and $10$ TeV. The first inset of both plots shows that the relative corrections scale with $s$ as $\alpha \log^2(s / M^2)$ and $\alpha \log(s / M^2)$, and strongly depend on the specific operator that is considered.   
Indeed, the dominant helicity configurations are different for each operator and the EW corrections are sensitive to the helicities of the external legs. For example, it is worth noting the comparatively small corrections arising from the operator $\mathcal{O}_{ue}$ in the $u \bar{u} \to e^- e^+$ partonic process, owing to its purely right-handed configuration.
Independently of $\sqrt s$, and consequently of the ratio $\text{R}_{\text{VD}}$, we observe in both plots that the quantity $\text{D}_{\text{VD}}$ in the second inset of both plots in~\cref{fig:validations_4f} is constant and always at the sub-percent level for all operators. This feature unambiguously confirms that all EWSL are correctly captured by the the {\denpoz} algorithm and that its implementation into {\aNLO} can be safely utilised. 
We give more details concerning this check in the following.

Expanding in powers of $\MW^2/s$, in the high-energy limit ($\MW^2/s\to0$), the quantity $\text{R}_{\text{VD}}$ can be written as follows:
\begin{equation}
\text{R}_{\text{VD}}=\frac{\alpha}{4\pi s_{\rm w}^2}\left(A_2 \log^2 \frac{s}{\MW^2}+ A_1 \log \frac{s}{\MW^2} + B +\ord\left({\frac{\MW^2}{s}}\right) \right)\,, \label{eq:RVCexp}
\end{equation}
where the $A_2,A_1$ and $B$  terms are, in general, functions of the Mandelstam invariants $s,t$ and $u$\footnote{~In the cases considered here, since all diagrams entering the amplitudes with four-fermion operators have the same topology, these functions are in fact constants. As can be seen in Section 6 of Ref.~\cite{Denner:2000jv}, in the SM for four-fermion neutral-current processes, these function are constant, but for instance they are not in the case of the production of neutral gauge-boson pairs in $l^+l^-$ annihilation.}. Moreover, these three terms are also generally process- and helicity-dependent and they are exactly of  $\ord(1)$, namely of order $(\MW^2/s)^0$ in the $\MW^2/s$ expansion. In other words,~\cref{eq:RVCexp} explicitly shows the division of NLO EW corrections into logarithmically enhanced contributions ($A_2$, {\it i.e.}~the double logarithms, and $A_1$, {\it i.e.}~the single ones), the constant terms $B$, and the mass-suppressed corrections, namely, the remaining terms of $\ord\left(\MW^2/s\right)$. 
The {\denpoz} algorithm in its original formulation, without taking into account the $\Delta^{s\TO r_{kl}}$ terms introduced in Ref.~\cite{Pagani:2021vyk}, consists of the exact calculation of $A_2$ and $A_1$. Any other terms, including $B$, are discarded. On the contrary, when the high-energy limit is performed for our analytical calculation, spurious terms from $B$ are retained. In particular, we ignore the subtraction of possible (single or double) logarithms of the form $\log(M^2/\MW^2)$ in the expansion of the poly-logarithms emerging from the exact loop calculation. Since such terms do {\it not} depend on $s$ and the largest possible value for $M$ is $m_t$, if EWSL are correctly calculated ($A_2$ and $A_1$) the quantity $\text{D}_{\text{VD}}$ defined in~\cref{eq:DVD} should satisfy the following relation:
\begin{equation}
    \ord(\text{D}_{\text{VD}})=\frac{\alpha}{4\pi s_{\rm w}^2} \log^2 \frac{\mt^2}{\MW^2}\simeq 0.2\%, \label{eq:estimate}
\end{equation}
and therefore manifest as a horizontal line, for any operator, in~\cref{fig:validations_4f}. This is precisely what is observed in this figure, which validates our implementation.

For simplicity, in the previous argument, we have assumed that $s\simeq |t| \simeq |u|$, as in the original derivation of the {\denpoz} algorithm. Otherwise, logarithms involving the ratios among these three invariants may in general be much larger than the one considered in~\cref{eq:estimate} and, as discussed before, they are included in the {\denpoz} algorithm validation approach but not in the analytical one. However, in~\cref{fig:validations_4f}, we have chosen a final-state scattering angle of $75^\circ$, which implies  $|t|\simeq s/2$ and therefore the aforementioned logarithms  are of the same order of the one in~\cref{eq:estimate}\footnote{~In the high-energy limit, neglecting any mass, for a scattering angle of $75^\circ$ we obtain $t/s=(\sqrt{6}-\sqrt{2}-4)/8\simeq-0.4$ and $u/s=(-\sqrt{6}+\sqrt{2}-4)/8\simeq-0.6$.  This means that $\log(s/|u|)<\log(s/|t|)<\log(\mt^2/\MW^2)$.}.

In principle, even if $s\gg |t|, |u|$, the previous discussion could be repeated taking into account two steps: the expansion in powers of $\MW^2/s$ and then in powers of $|t|/s$, $|u|/s$, {with either removing the contribution of $\Delta^{s\TO r_{kl}}$ from the {\denpoz} algorithm (as implemented in Ref.~\cite{Pagani:2021vyk}) results or retaining all the logarithms among invariants in the analytical ones.
That said, we remind the reader that all the terms that are not logarithmically enhanced are beyond the desired accuracy in this work and can only be controlled in an exact calculation of NLO EW corrections; neither of the DP algorithm approach nor the analytical one discussed here is expected to return the correct result if this level of accuracy is required. 

\section{Mass-suppressed processes in the SMEFT} \label{sec:ctg}
As discussed in Sec.~\ref{sec:MSDP}, the {\denpoz} algorithm is not expected to be valid for processes that feature mass-suppressed amplitudes at the tree-level. As an example of such situation, which -- unlike the SM -- is quite common in the SMEFT, we have highlighted the case of the process: 
\begin{equation}
    u \, \bar u \to t \bar t\, \label{ctg_proc}\, ,
\end{equation}
and the associated tree-level amplitude, $\MzNP$, induced by the top-quark chromomagnetic operator $\mathcal{O}_{tG}$, defined in~\cref{eq:ctg_lagrangian}. The EW corrections, $\MoNP$, to the amplitude $\MzNP$ involve the diagrams depicted in~\cref{fig:ctg_diagsinmain}. In Sec.~\ref{sec:MSDP}, we have emphasised why these diagrams, in particular the one on the right,  can lead to  contributions that cannot be correctly captured in the high-energy limit by the {\denpoz} algorithm. In this Appendix, we explicitly report the calculation of the high-energy limit of  $\MoNP$, and we pinpoint the terms that are omitted by the {\denpoz} algorithm. We emphasise once again that we are not asserting a failure of the {\denpoz} algorithm. Rather, this calculation is performed not only within the context of the SMEFT but also for a tree-level mass-suppressed amplitude and therefore lies beyond the algorithm's range of applicability. Since this type of amplitude is rather ubiquitous when considering higher-dimensional operators, our results serve to discourage the generic use of the {\denpoz} algorithm for SMEFT calculations.

The Feynman diagrams of~\cref{fig:ctg_diagsinmain} lead to contributions of $\ord(\mt^2/v^2)$ within $\MoNP$ for the process in~\cref{ctg_proc}, 
considering a single insertion of the $\mathcal{O}_{tG}$ operator. Additional corrections of the same order arise from the UV counterterm for the vertex.
In particular, the Feynman rule $\mathcal F$  associated to the $t\bar t g$ vertex in $\MzNP$, which arises from $\mathcal{O}_{tG}$ after electroweak symmetry breaking, is of the form:
\beq
\mathcal F^{\rm tree} \propto v \, C_{tG}\,,
\eeq
and therefore, the associated Feynman rule for the UV counterterm at one-loop for the same vertex reads 
\beq
\mathcal F^{\rm CT}= \mathcal F^{\rm tree}(\delta v+ \delta\psi_t + \delta C_{tG} )\,. 
\eeq

The quantity $\delta\psi_t $ is the counterterm of the top-quark wave function, $\delta v$ is the UV counterterm of  the vacuum expectation value and $\delta C_{tG}$ is the UV counterterm of the Wilson coefficient $C_{tG}$. Since we are interested in the $\ord(\mt^2/v^2)$ piece of the $\ord(\alpha)$ EW corrections, no counterterms associated to the gluon field are present. We consider on-shell renormalisation for all quantities besides $C_{tG}$, which is treated in the $\overline{\text{MS}}$ scheme and we set the renormalisation scale $\mu_R = \sqrt{s}$. Our counterterms have all been extracted independently and agree with the results in the literature~\cite{Denner:1991kt,Jenkins:2013wua}.

Defining $\widehat \MoNP$ as the unrenormalised amplitude, we obtain
\begin{align}
\MoNP &= \widehat \MoNP + \MzNP(\delta v+  \delta\psi_t + \delta C_{tG} )\,
\end{align}
which, in the high-energy limit and considering only the $\ord(\mt^2/v^2)$ component of the $\ord(\alpha)$ EW corrections, reads:
\begin{align}
   \lim_{s\to\infty} \MoNP\Big|_{\ord(\mt^2/v^2)}&= \MzNP \, \frac{y_t^2}{32 \pi^2} \Big( \underbrace{-\frac{3}{\varepsilon}  - 3 \log \frac{s}{m_t^2}}_{\text{loops~in~}\widehat \MoNP} \, \underbrace{+\, \frac{15}{2\varepsilon}}_{\delta C_{tG}} \nonumber \\
    &\hspace{3cm} \underbrace{-\frac{3}{2 \varepsilon}  - \frac{3}{2} \log \frac{s}{m_t^2}}_{\delta\psi_t} \ \underbrace{-\frac{3}{\varepsilon}  - 3 \log \frac{s}{v^2}}_{\delta v}  \, \Big) \nonumber \\
    &\simeq  \MzNP \, \frac{y_t^2}{32 \pi^2} \Big( \frac{15}{2}\log \frac{s}{\MW^2} \Big),\label{ctg_virt}
\end{align}
where we have dropped finite terms that do not grow with energy. As expected, the final result is finite and in the last line of~\cref{ctg_virt}, we have rewritten it in terms of $\log (s/\MW^2)$, dropping terms such as $\log (m_t^2/\MW^2)$ and $\log (v^2/\MW^2)$ as done in the {\denpoz} algorithm.

Let us now see the result we would have obtained had we performed the same calculation using the {\denpoz} algorithm as we do for the four-fermion operators (which do not induce mass-suppressed contributions) in this work. The {\denpoz} algorithm captures the logarithms stemming from field and parameter renormalisation in the second line of~\cref{ctg_virt}, but not the mass-singular logarithm $-3\log (s/m_t^2) \simeq -3\log (s/\MW^2) $ in the first line. It is easy to see that this is a logarithmic divergence of collinear origin, arising from the bubble diagrams on the right of Figure \ref{fig:ctg_diagsinmain}, when the Higgs and the top-quark momenta, respectively $p_h$ and $p_t$, are aligned and of similar size: $p_h = x \, p_t$, with $x \sim 1$.

Following Appendix A of Ref.~\cite{Denner:2001gw}, the collinear mass-singularity of a bubble diagram written generically as
\begin{equation}
    I_{\text{coll}} = \int \frac{- i \, (4\pi \mu^\varepsilon)^2\, N(\ell^\nu)}{(\ell^2-m_1^2)((p-\ell)^2-m_2^2)} \, \frac{d^d \ell}{(2 \pi)^d}\,,
\end{equation}
is given by
\begin{equation}
    I_{\text{coll}} = \log \frac{\mu^2}{m^2} \int_0^1 N(x \, p^\nu) \, dx\,, \label{Icoll}
\end{equation}
where $m \equiv \text{max}(m_1, m_2)$. 

For the class of bubble diagrams as the right one in~\cref{fig:ctg_diagsinmain},~\cref{Icoll} leads to the collinear logarithm
\begin{equation}
    I_{\text{coll}} = \MzNP \Big( -\frac{3 y_t^2}{32 \pi^2} \log \frac{\mu^2}{m_t^2} \Big),
\end{equation}
which is indeed what we see in the first line of~\cref{ctg_virt}. The collinear contribution $I_{\text{coll}}$ factorises the Born matrix element $\MzNP$ and a mass-singular logarithm. However, such logarithm is {\it not}  captured by the {\denpoz} algorithm, whose derivation relies on Ward identities based on the symmetry of the unbroken SU(2) $\times$ U(1) SM Lagrangian, and especially relies on the absence of mass-suppressed contributions.

We notice that following Refs.~\cite{Denner:2000jv, Denner:2001gw}, and in particular the argument around Eq.~(2.9) of Ref.~\cite{Denner:2001gw}, one would obtain
\begin{equation}
    I_{\text{coll}} = 0 + \mathcal O(v),
\end{equation}
which in the limit $\MW^2/s \to 0$, is formally correct since $\MzNP\to0$. However, although formally correct, such a result is not very useful.
\raggedbottom
\section{Additional distributions}
\label{app:AdditionalDistr}
This appendix contains additional differential distribution plots omitted from the main text.
\subsection{Top-quark pair production at the LHC} 
\label{sec:tt_distr_remaining}
\Cref{fig:tt_distr_tu8_td8_Qu8,fig:tt_distr_tq8_Qq83} of this section display the distributions for the top-quark pair production at the LHC, similar to~\cref{fig:tt_distr_Qd8}, but for the remaining four-fermion operators in~\cref{eq:Qq38}.
\begin{figure}[H]
    \centering
    \includegraphics[width=0.5\textwidth]{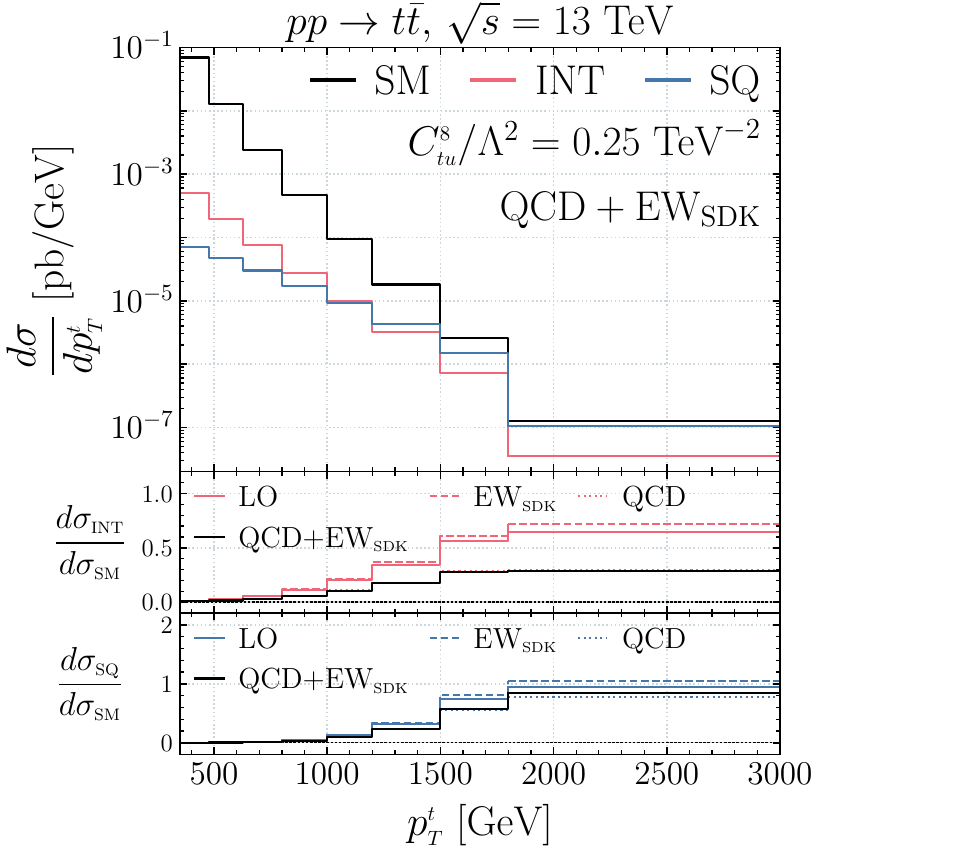}
    \hspace{-1.0cm}
    \includegraphics[width=0.5\textwidth]{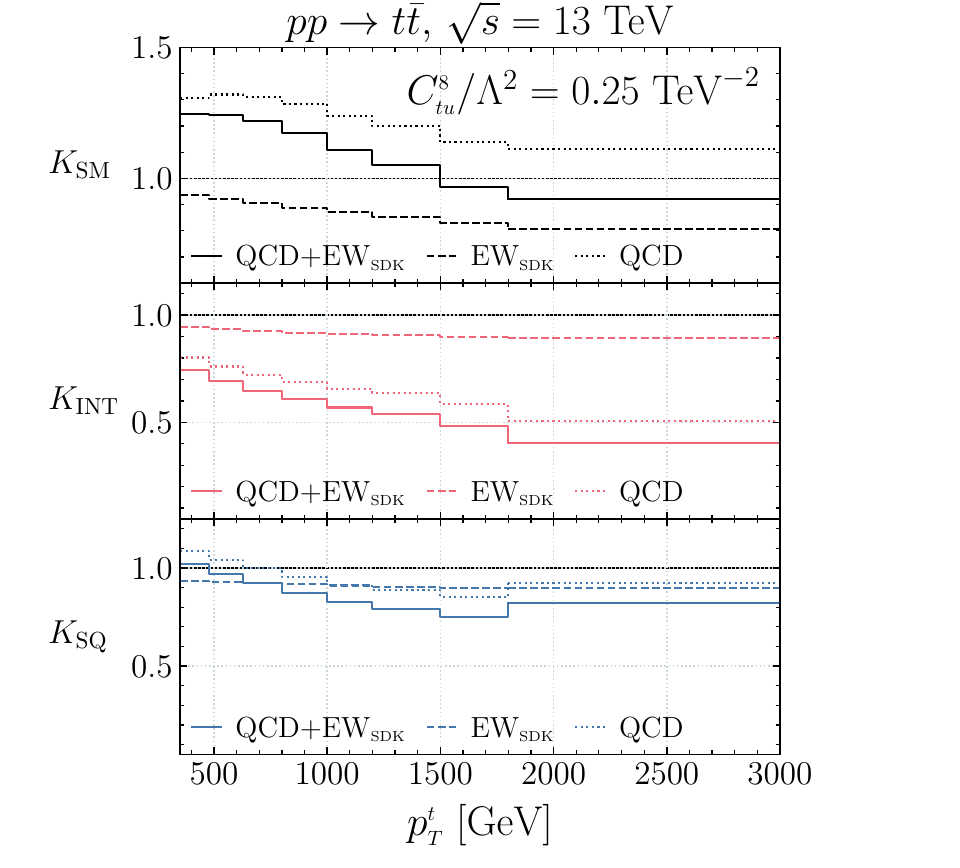}\\
    \vspace{2mm}
    \includegraphics[width=0.5\textwidth]{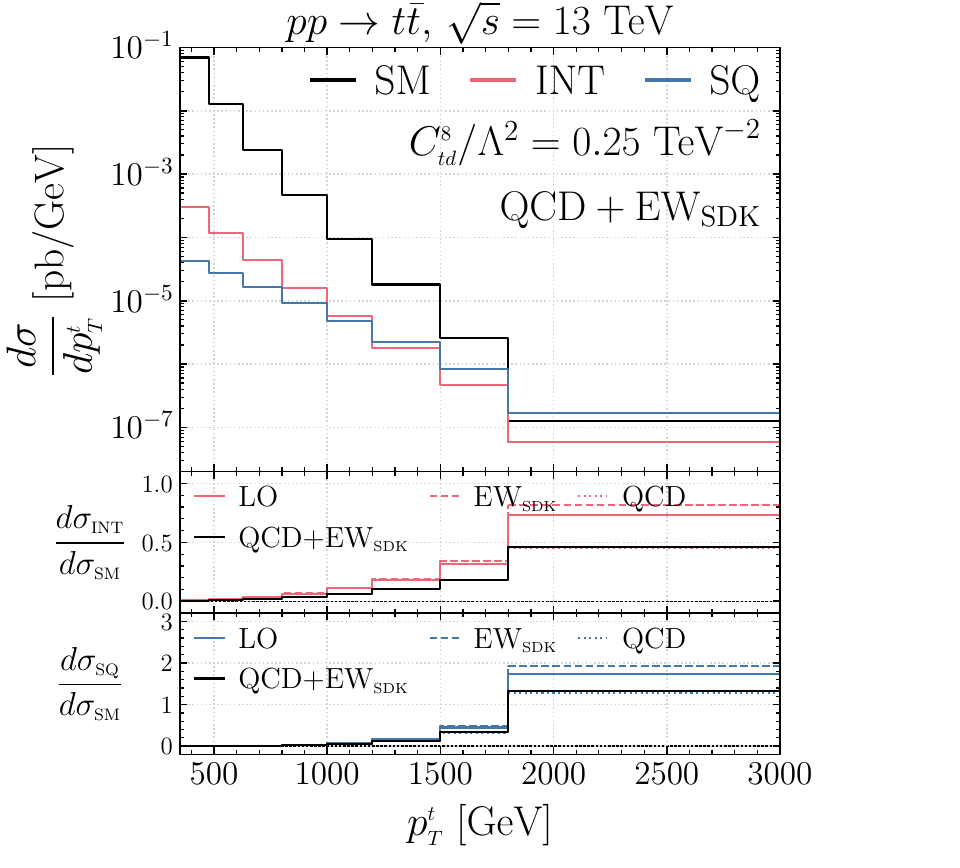}
    \hspace{-1.0cm}
    \includegraphics[width=0.5\textwidth]{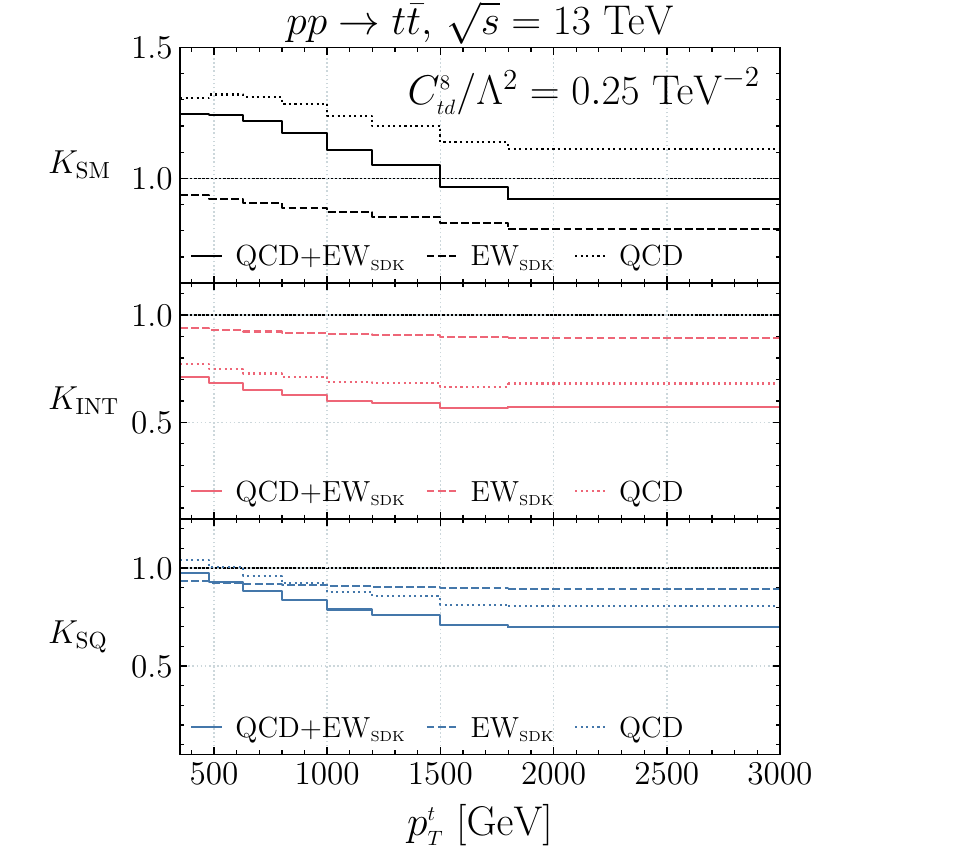}\\
    \vspace{2mm}
    \includegraphics[width=0.5\textwidth]{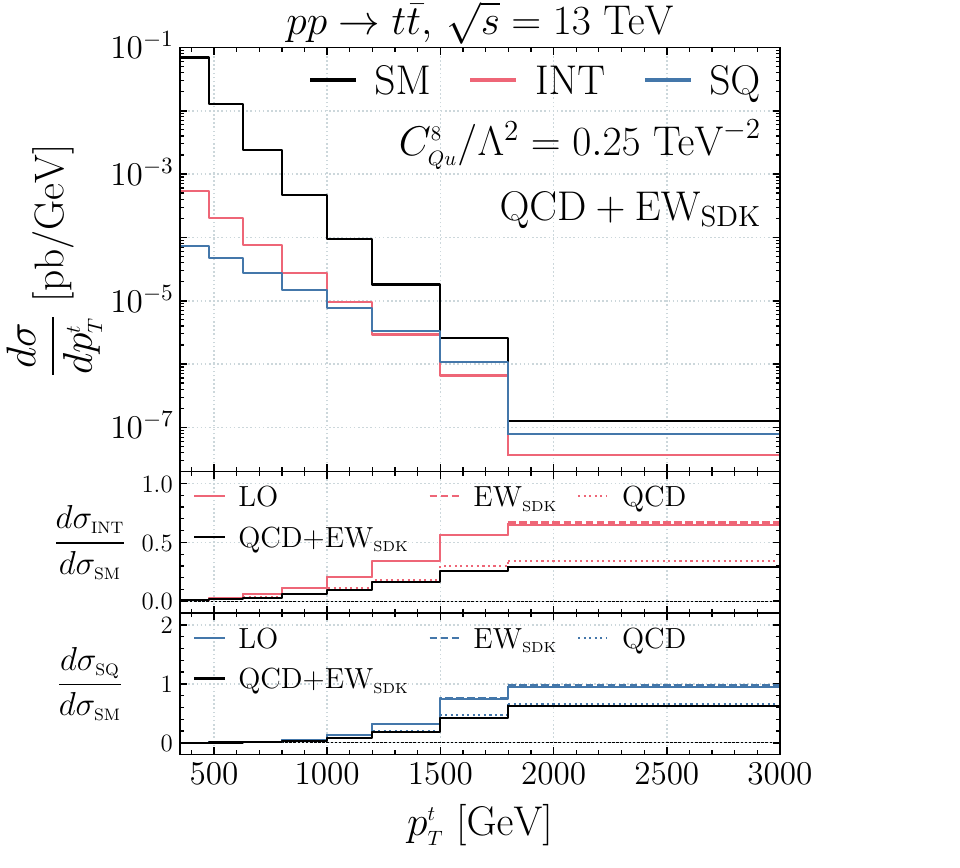}
    \hspace{-1.0cm}
    \includegraphics[width=0.5\textwidth]{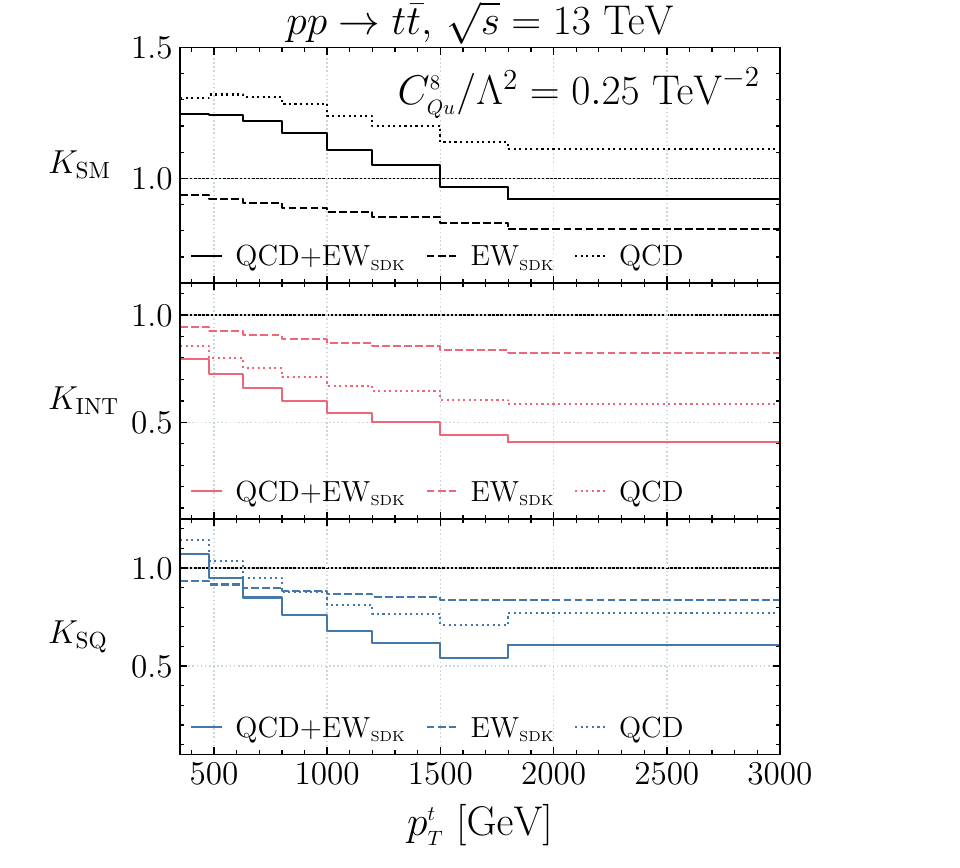}\\
    \vspace{-4mm}
   \caption{Same as~\cref{fig:tt_distr_Qd8} for the operators $\mathcal O_{tu}^8$ (top row), $\mathcal O_{td}^8$ (middle row) and $\mathcal O_{Qu}^8$ (bottom row).}
    \label{fig:tt_distr_tu8_td8_Qu8}
\end{figure}
\begin{figure}[H]
    \centering
    \includegraphics[width=0.5\textwidth]{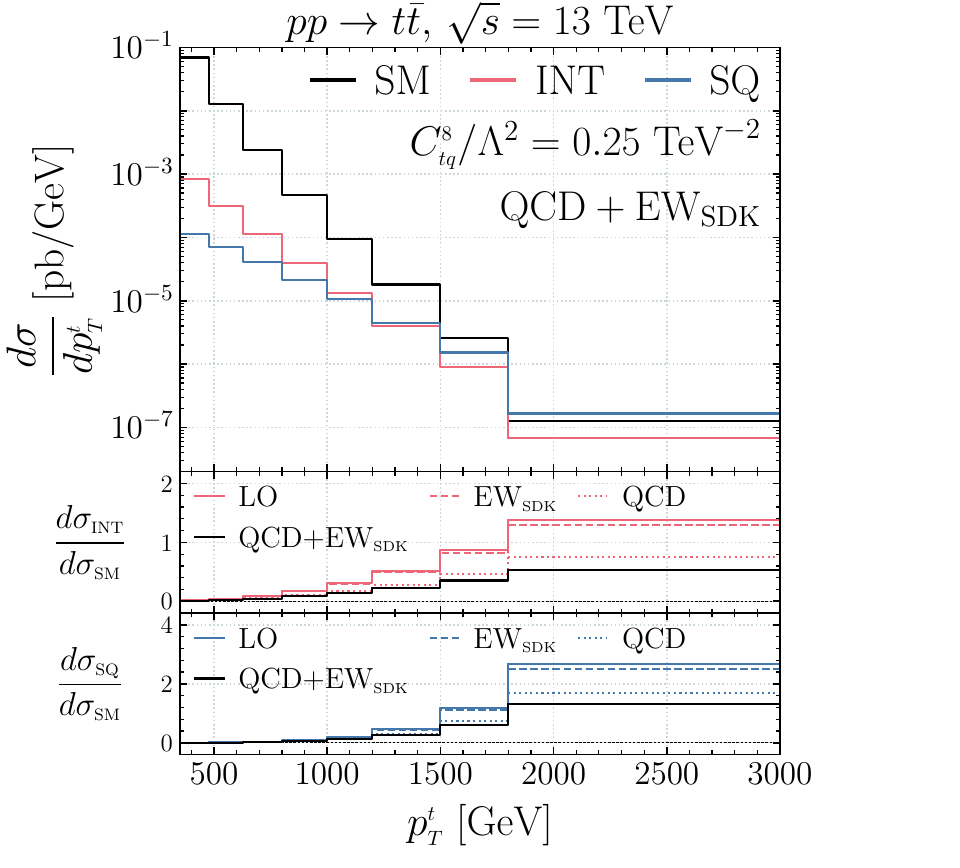}
    \hspace{-1.0cm}
    \includegraphics[width=0.5\textwidth]{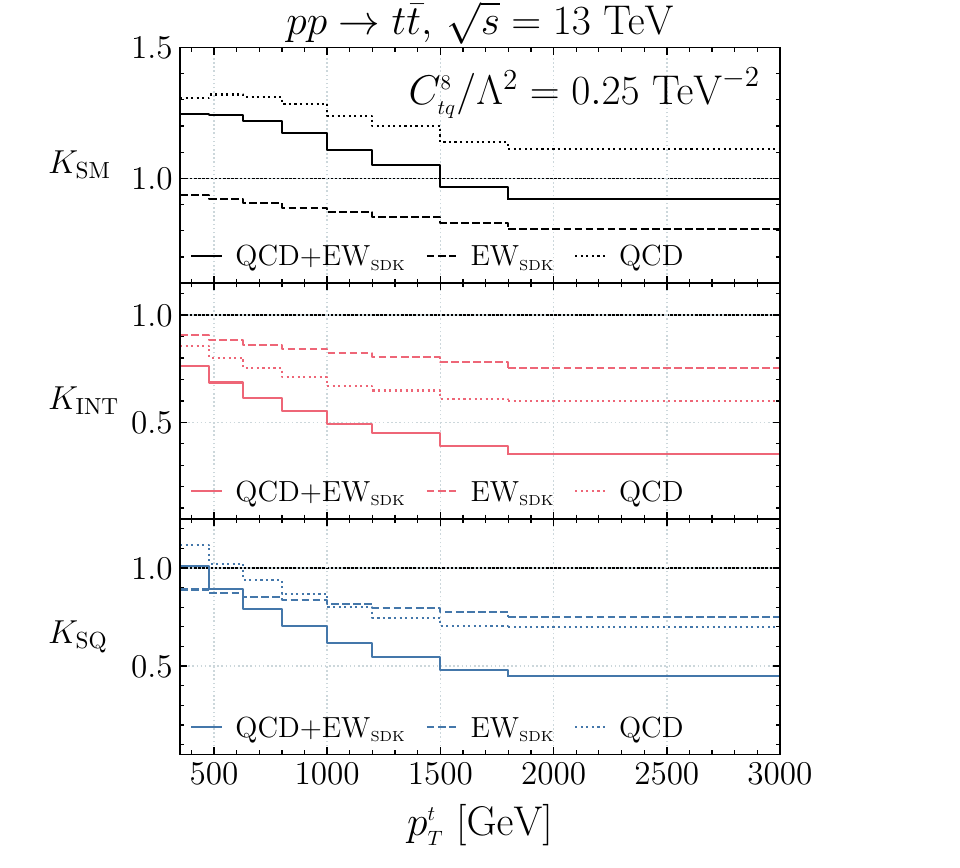}\\
    \vspace{2mm}
    \includegraphics[width=0.5\textwidth]{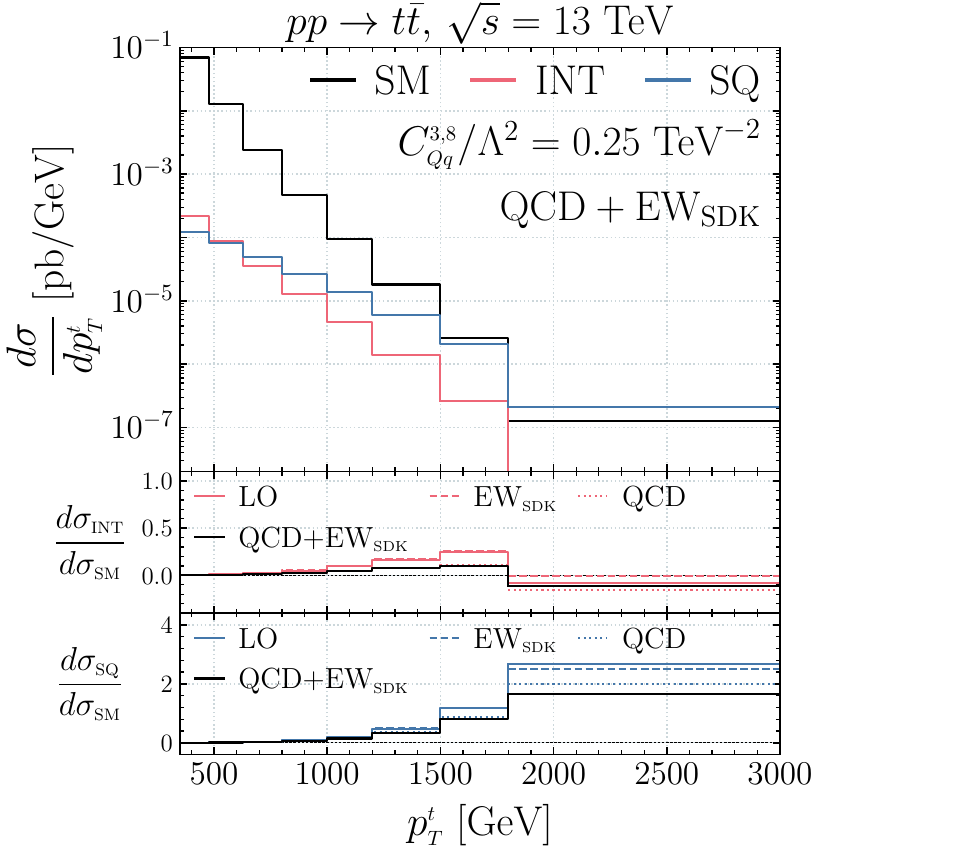}
    \hspace{-1.0cm}
    \includegraphics[width=0.5\textwidth]{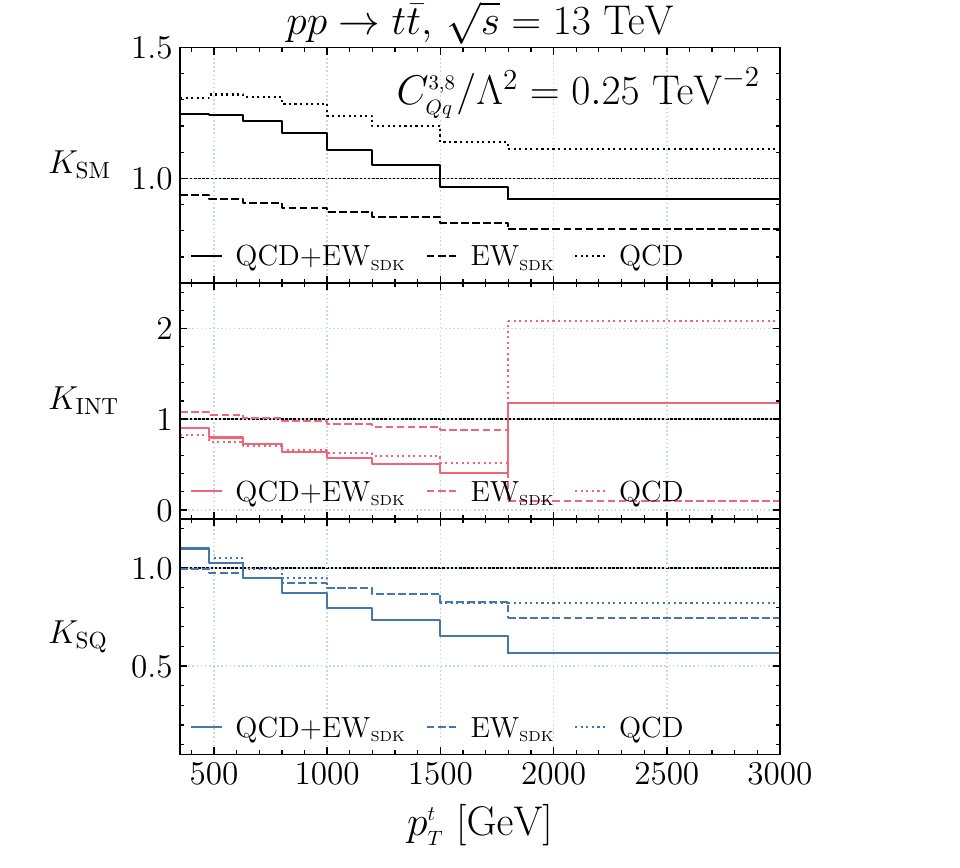}\\
    \vspace{-4mm}
   \caption{Same as~\cref{fig:tt_distr_Qd8} for the operators $\mathcal O_{tq}^8$ (top row) and $\mathcal O_{Qq}^{8,3}$ (bottom row).}
    \label{fig:tt_distr_tq8_Qq83}
\end{figure}
\subsection{The Drell-Yan process at the LHC} \label{sec:DY_distr_remaining}
\Cref{fig:DY_distr_ue_ul_ql3} displays the remaining distributions of the electron-positron pair production at the LHC from the four-fermion operators in~\cref{eq:Qte} (and considering the rotation in~\cref{eq:13rotation}).
\begin{figure}[htb!]
    \centering
    \includegraphics[width=0.5\textwidth]{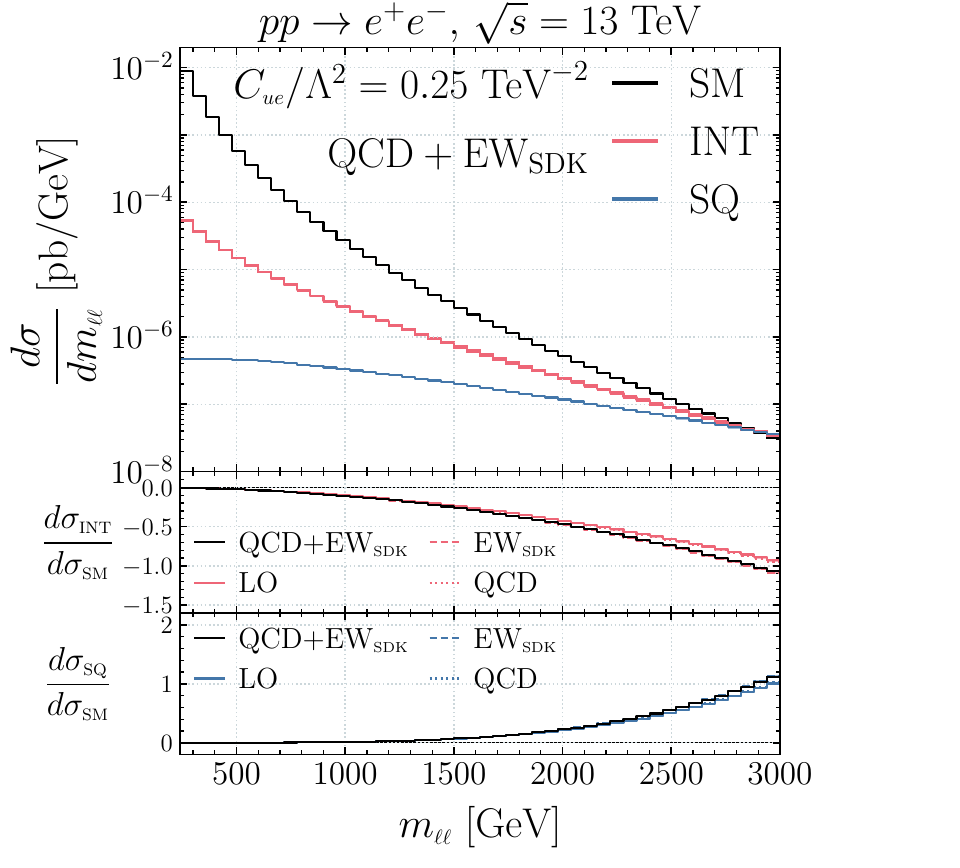}
    \hspace{-1.0cm}
    \includegraphics[width=0.5\textwidth]{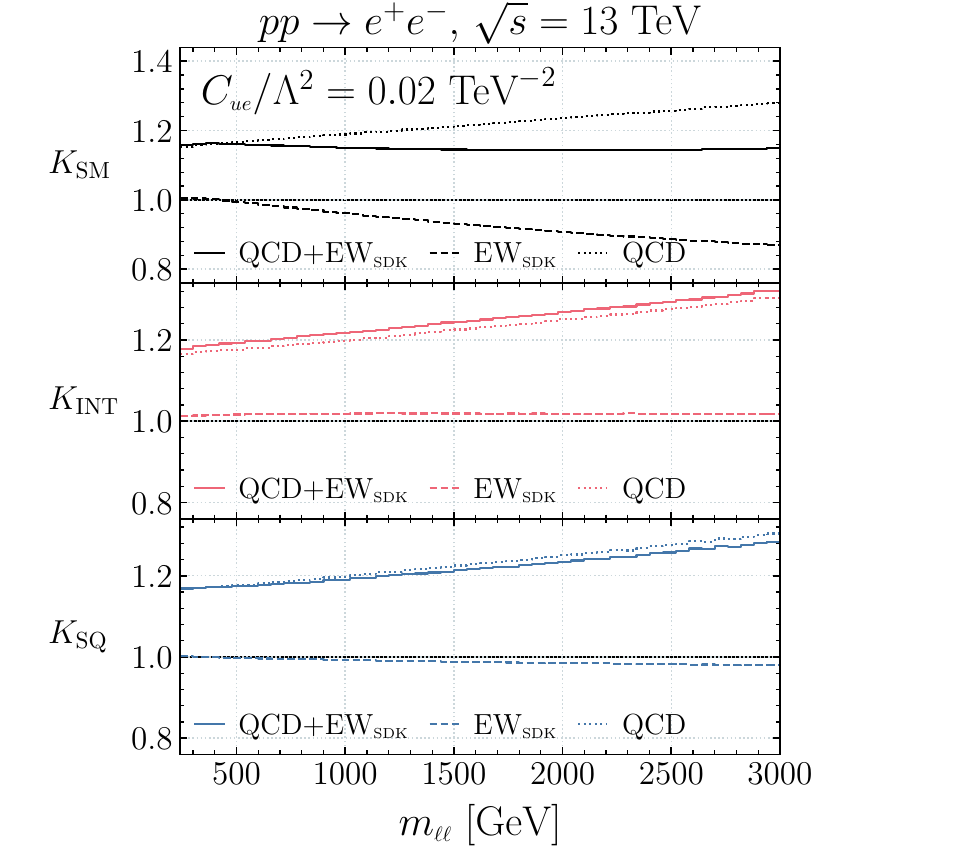}\\
    \vspace{2mm}
    \includegraphics[width=0.5\textwidth]{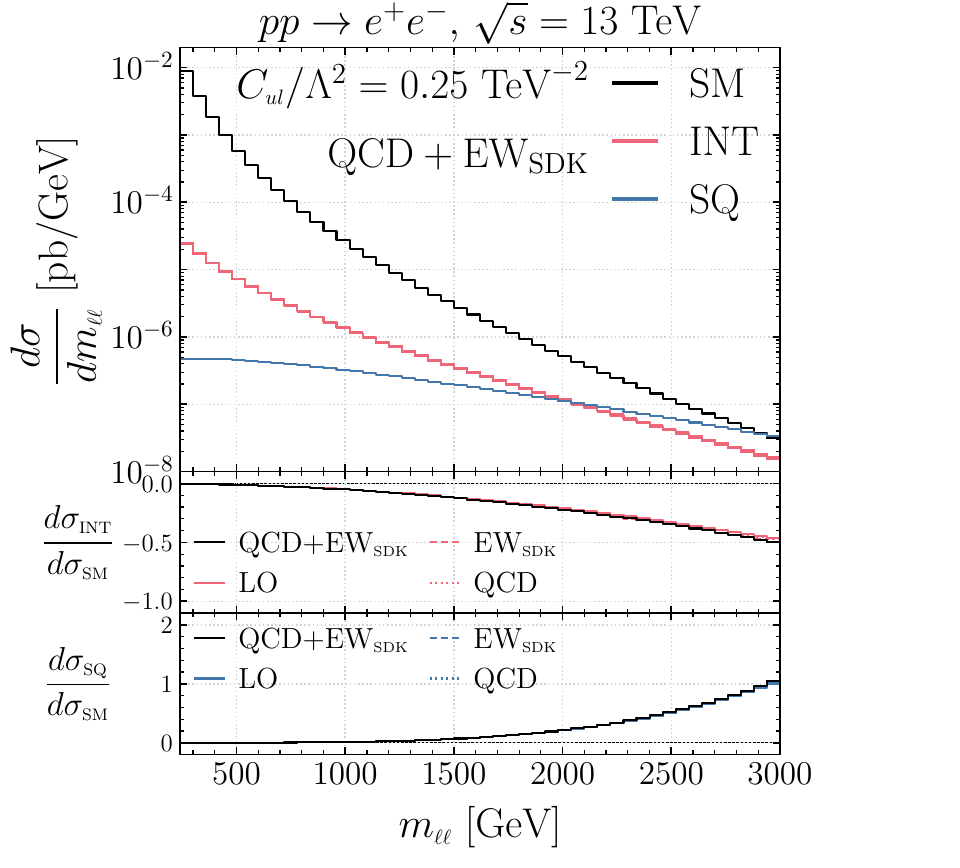}
    \hspace{-1.0cm}
    \includegraphics[width=0.5\textwidth]{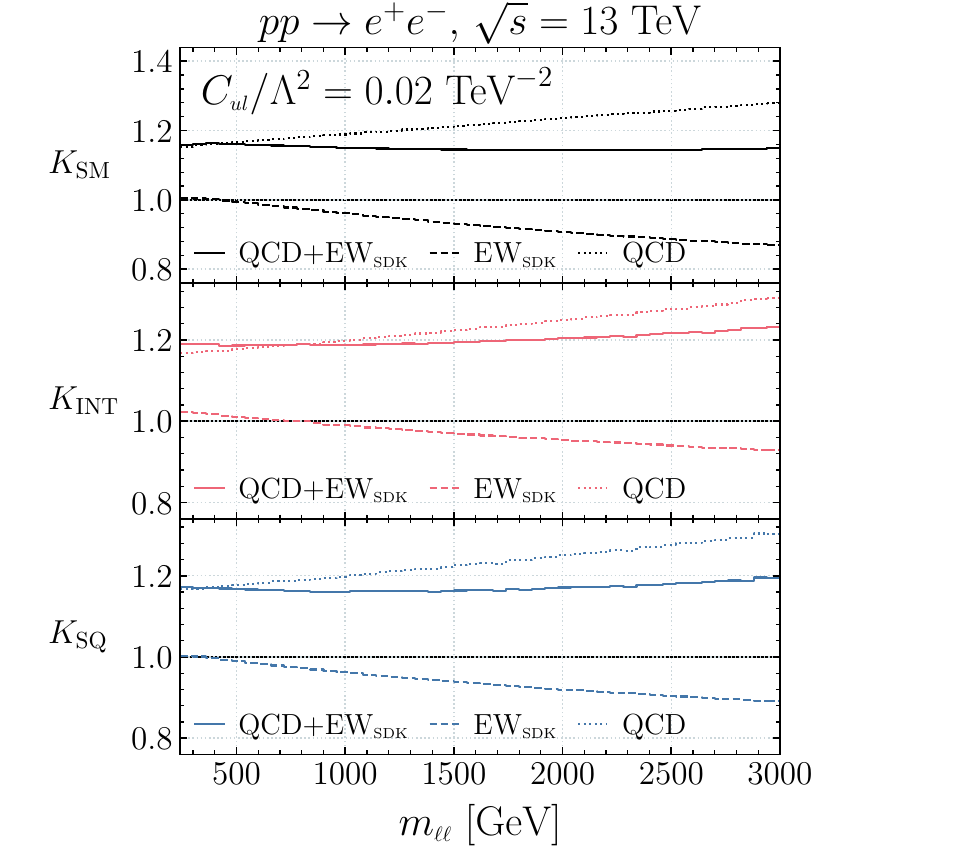}\\
    \vspace{2mm}
    \includegraphics[width=0.5\textwidth]{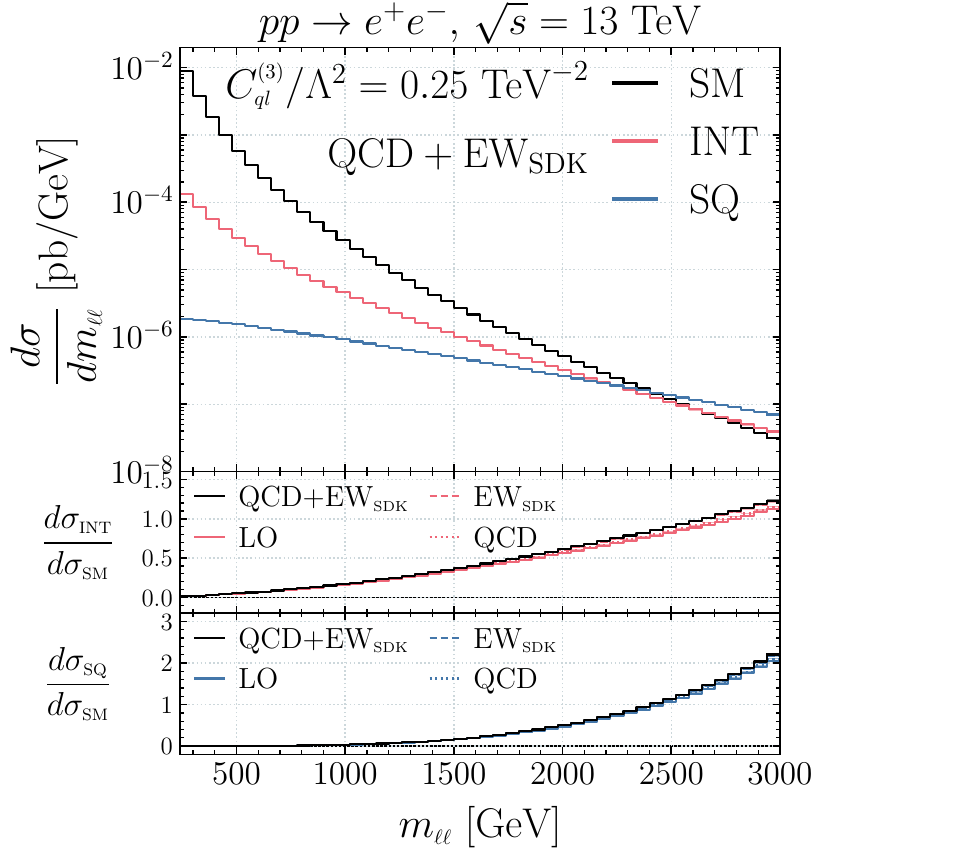}
    \hspace{-1.0cm}
    \includegraphics[width=0.5\textwidth]{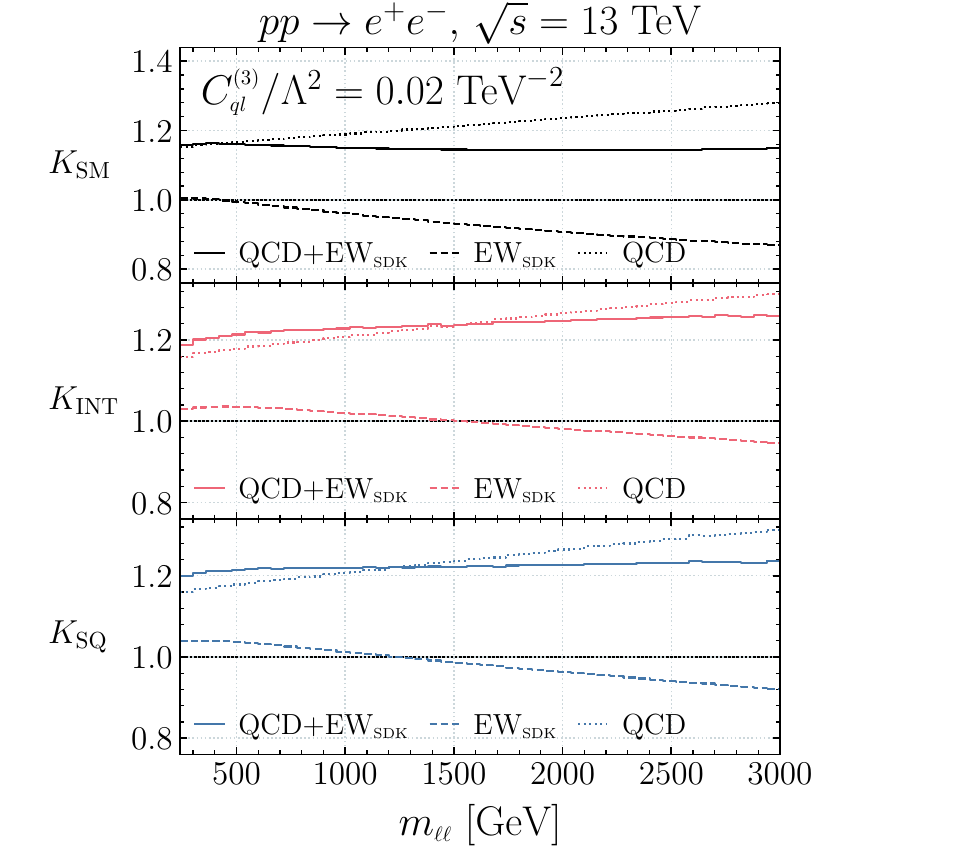}\\
    \vspace{-4mm}
   \caption{Same as~\cref{fig:tt_distr_Qd8}, differential in $m_{\ell\ell}$, for the operators $\mathcal O_{ue}$ (top row), $\mathcal O_{u\ell}$ (middle row), and $\mathcal O_{q\ell}^3$ (bottom row).}
   \label{fig:DY_distr_ue_ul_ql3}
\end{figure}
\subsection{Top-quark pair production at a muon collider} \label{sec:muC_distr_remaining}
\Cref{fig:muC_distr_te_Qe_qlm} shows the distributions for the top-quark pair production at a $10 \, \text{TeV}$ muon collider, similar to~\cref{fig:muC_distr_tl}, but for the remaining four-fermion operators in~\cref{eq:Que} (and considering the rotation in~\cref{eq:13rotation1}).
\begin{figure}[htb!]
    \centering
    \includegraphics[width=0.5\textwidth]{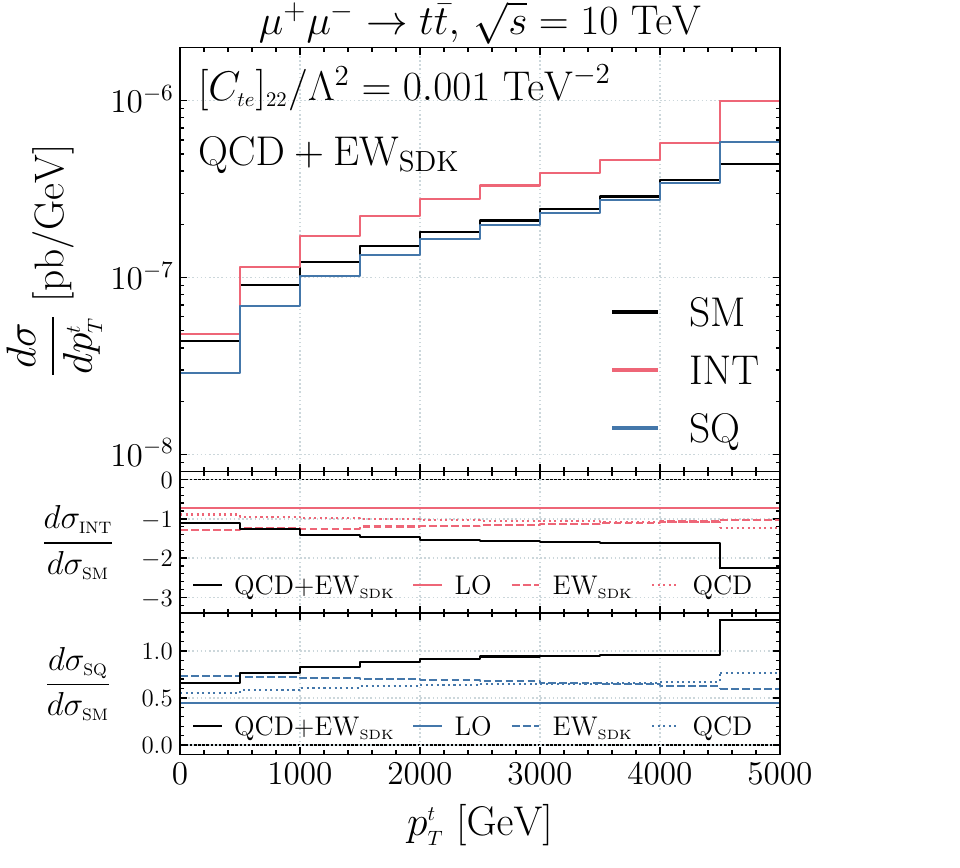}
    \hspace{-1.0cm}
    \includegraphics[width=0.5\textwidth]{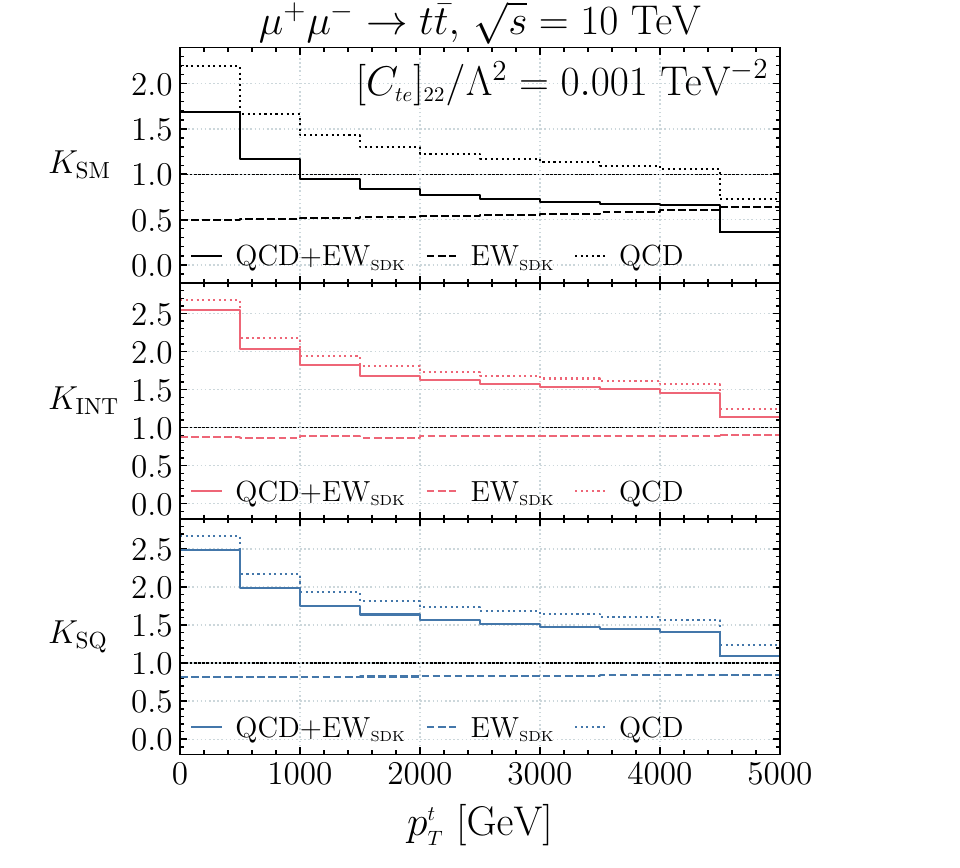}\\
    \vspace{2mm}
    \includegraphics[width=0.5\textwidth]{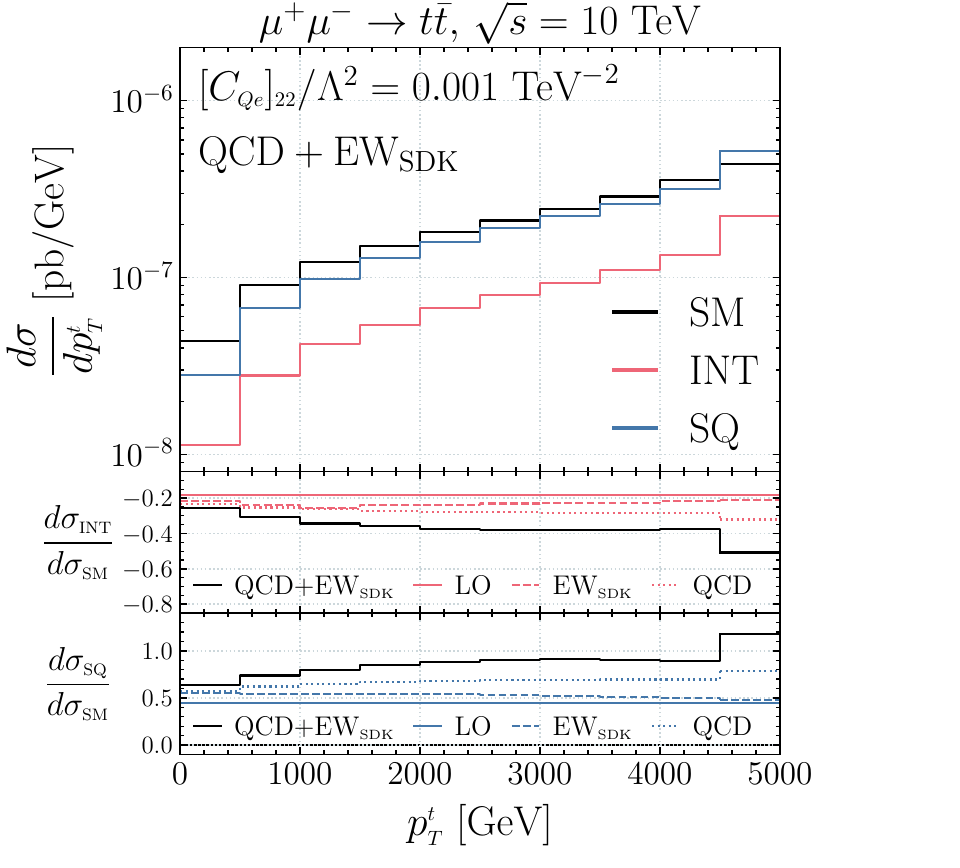}
    \hspace{-1.0cm}
    \includegraphics[width=0.5\textwidth]{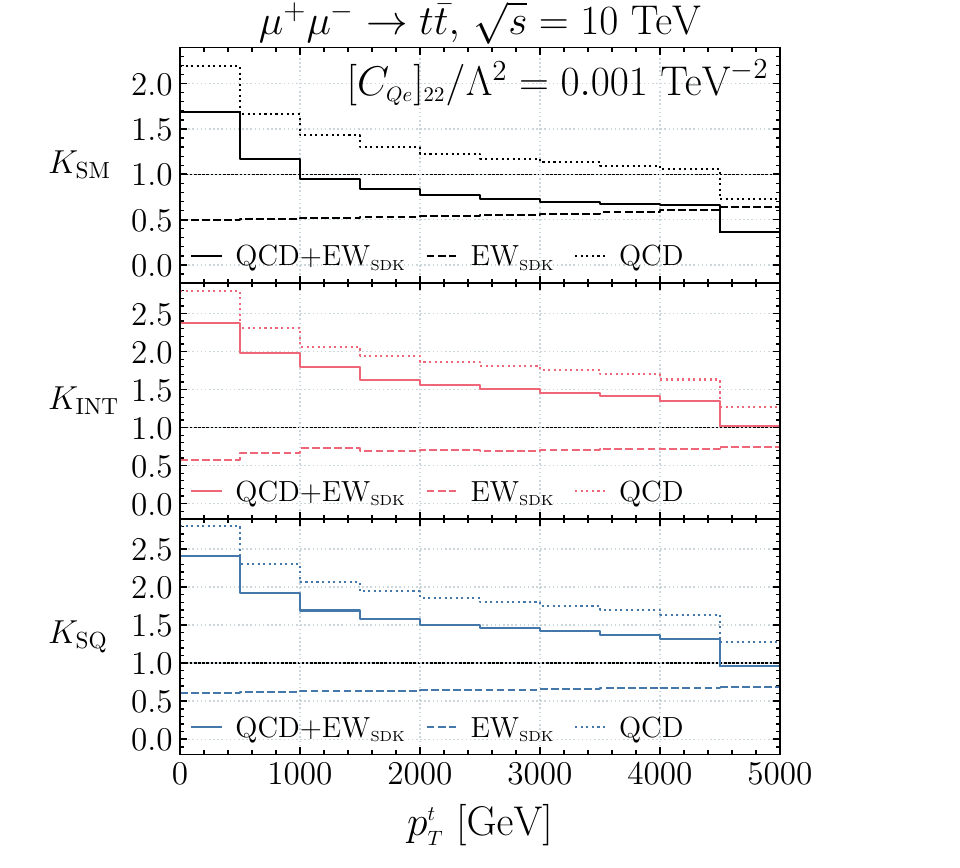}\\
    \includegraphics[width=0.5\textwidth]{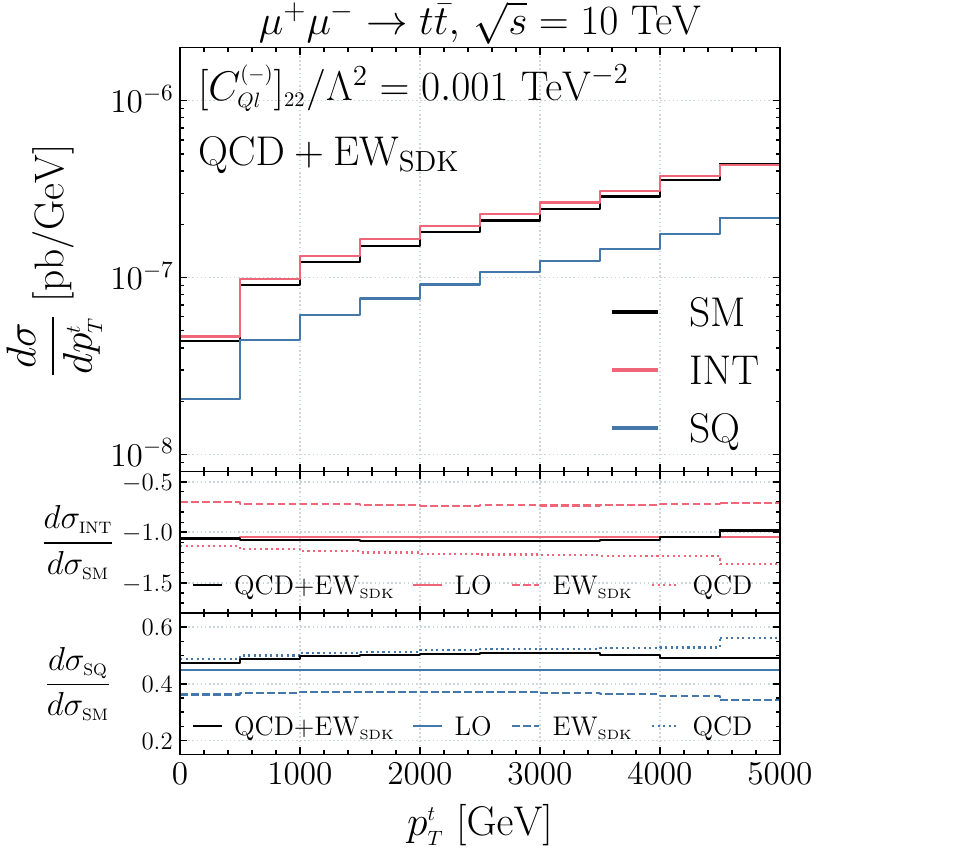}
    \hspace{-1.0cm}
    \includegraphics[width=0.5\textwidth]{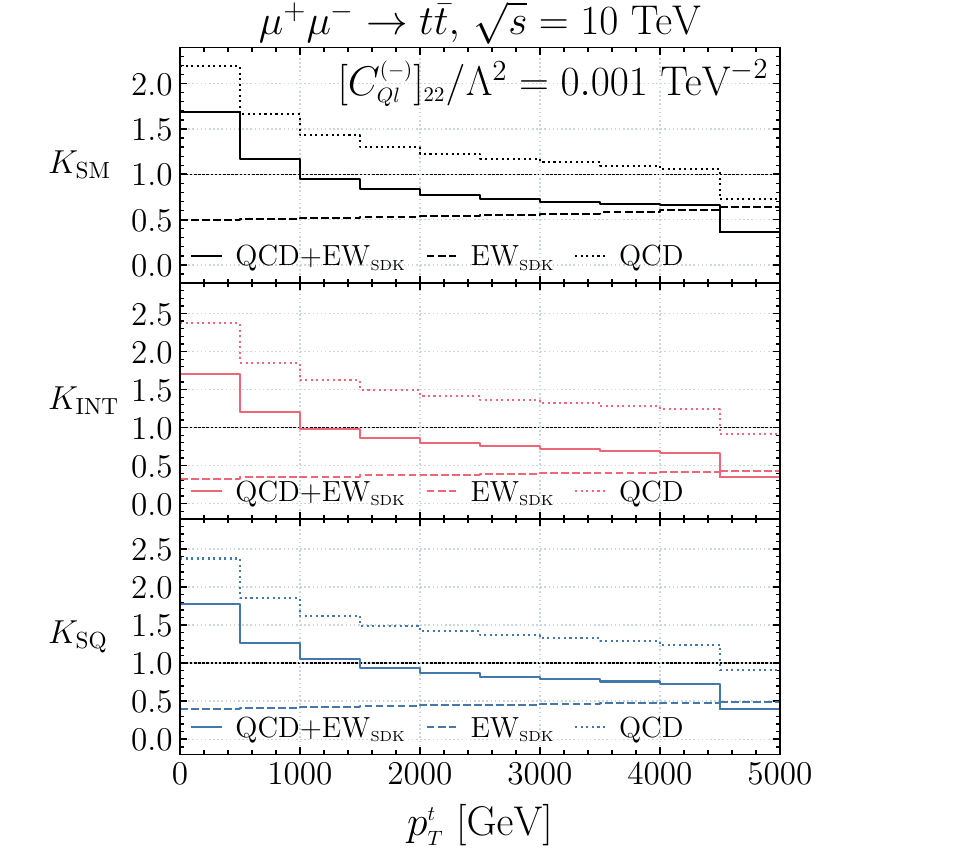}\\
    \vspace{-4mm}
   \caption{Same as~\cref{fig:muC_distr_tl} for the operators $\mathcal O_{te}$ (top row), $\mathcal O_{Qe}$ (middle row), and $\mathcal O_{Q\ell}^{(-)}$ (bottom row).}
   \label{fig:muC_distr_te_Qe_qlm}
\end{figure}
\clearpage
\section{Fisher information and flat directions} 
\label{sec:fisherAPP}
One way to study globally flat directions with respect to kinematic distributions is to define an expected Fisher Information (FI) matrix:
\begin{align}
F_{ij} = E\left[
\frac{\partial^2 \log L(\bm{c}|\bm{x})}{\partial c_i\partial c_j}
\Bigg| \bm{c}
\right],
\end{align}
where $E$ denotes the expectation value and $L$ is the likelihood function over an $n$-dimensional Wilson coefficient space described by the vector, $\mathbf{c}=\{c_1,c_2,\dots c_n\}$, assuming the observation of a generic dataset, $\bm{x}$. A simple treatment is to assume a multivariate Gaussian likelihood, which yields a $\chi^2$ log-likelihood:
\begin{align}
    \chi^2(\bm{c}) &= (\bm{x} - \bm{\mu}(\bm{c}))^{\sss \top}\cdot\,\bm{V}^{-1}\cdot
                    \,(\bm{x} - \bm{\mu}(\bm{c}))\,,
\end{align}
where $\bm{\mu}(\bm{c})$ denotes the vector of predictions for the data points, $\bm{x}$, as a function of the coefficients, and $\bm{V}$ is the covariance matrix of the data. In the linear approximation, we have:
\begin{align}
\label{eqn:def_H}
    \bm{\mu}(\bm{c})&=\bm{\mu}^{\rm SM}+\bm{H}\cdot\bm{c}\,,
\end{align}
where $\bm{\mu}^{\rm SM}$ are the SM predictions, and the $\chi^2$ can be written in terms of the FI and the best fit point, $\hat{\bm{c}}$:
\begin{align}
\begin{split}
        \chi^2(\bm{c}) &= \chi^2_{\sss \text{min.}} + (\bm{c} - \hat{\bm{c}})^{\sss \top}\cdot \bm{F}\cdot(\bm{c} - \hat{\bm{c}})\,,\\
    \chi^2_{\sss \text{min.}} &\equiv  \chi^2_{\rm SM} - \hat{\bm{c}}^{\sss \top}\cdot\bm{F}\cdot \hat{\bm{c}},
\end{split}
\end{align}
with
\begin{align}
\bm{F} = (\bm{H}^{\sss \top}\cdot\bm{V}^{-1}\cdot \bm{H}),
\end{align}
and $\chi^2_{\sss \text{min.}}$ is the $\chi^2$ value at the best-fit point which generally improves on the SM.

One can see that the FI measures the curvature of the log-likelihood in the vicinity of the best-fit point and hence quantifies the sensitivity of the data to a given direction in parameter space. The eigensystem of the FI matrix describes independent directions that are constrained by a given dataset. Well-constrained directions have large eigenvalues and eigenvectors with zero eigenvalue correspond to flat directions that cannot be probed by the input data.

We can apply this to study the impact of approximate EW corrections in the SMEFT by constructing the expected FI matrices corresponding to the hypothetical measurement of the distributions plotted in the previous sections, assuming the SM prediction is observed and neglecting SMEFT contributions at $\mathcal{O}(\Lambda^{-4})$. For a given integrated luminosity, $\mathcal{L}$, the number of events observed in each bin is
\begin{align}
    N_a = \mathcal{L}\left(\sigma_a^{\rm SM} + \sigma_{a,i} \,\frac{c_i}{\Lambda^2}\right),
\end{align}
where $\sigma_a^{\rm SM}$ is the SM contribution to the cross-section in that bin and $\sigma_{a,i}$ is the linear contribution from the $c_i$ coefficient. We therefore have, from~\cref{eqn:def_H},
\begin{align}
    H_{ai} = \mathcal{L}\frac{\sigma_{a,i}}{\Lambda^2}.
\end{align}
Taking only statistical uncertainties for simplicity, 
the covariance matrix for the number of observed events is diagonal,
\begin{align}
    V_{ab} = \delta_{ab}\,\mathcal{L}\, \sigma_{a}^{\rm SM},
\end{align}
where no sum over the indices is implied, and the FI is
\begin{align}
    F_{ij} = \mathcal{L}\sum_{a=1}^{n_{\text{bins}}}\frac{H_{ai} H_{aj}}{\sigma^{\rm SM}_a}.
\end{align}
Since we are mainly interested in relative differences in sensitivity, we will work with the luminosity-normalised FI, $\bm{F}/\mathcal{L}$, which has units of cross-section. Moreover, since eigenvectors are only defined up to an overall sign and normalisation, we use a convention in which they have unit norm and a positive inner-product with the unit vector in the $(1,1,\dots, 1)$ direction.

The Fisher information generally pertains only to the linear or interference contributions of Wilson coefficients. Whilst it can be constructed for any likelihood function, it is not possible to make as general statements when including, \emph{e.g.}, quadratic terms in the dependences of~\cref{eqn:def_H}.
This is because the Fisher information, being a second derivative, begins to explicitly depend on the best fit point in parameter space, given the observed data. In the case that one observes exactly the SM prediction, one recovers the FI from the linearised approximation and no further information is added by the quadratic dependences.

\bibliographystyle{JHEP}
\bibliography{bibliography}

\end{document}